\documentclass{article}

\usepackage[preprint]{neurips_2025}

\usepackage[utf8]{inputenc}
\usepackage[T1]{fontenc}
\usepackage[english]{babel}
\usepackage{microtype}
\usepackage{url}
\usepackage{booktabs}
\usepackage{tabularx}
\usepackage{array}
\usepackage{ragged2e}
\usepackage{graphicx}
\graphicspath{{images/}{./images/}{./}}
\usepackage{xcolor}
\usepackage{colortbl}
\usepackage{placeins}
\usepackage{amssymb}
\usepackage{textcomp}
\usepackage{rotating}
\usepackage[hidelinks]{hyperref}

\setcitestyle{authoryear,round,aysep={,},yysep={;}}

\definecolor{hdrnavy}{HTML}{1F3864}
\definecolor{hdrblue}{HTML}{2E5496}
\definecolor{cellpale}{HTML}{EEF1F5}
\definecolor{hdrpale}{HTML}{C9DAF8}
\definecolor{hdrgrey}{HTML}{EFEFEF}
\definecolor{hdrblue2}{HTML}{CFE2F3}
\definecolor{hdrtan}{HTML}{EFE7CF}
\definecolor{markyellow}{HTML}{FFF2A8}
\definecolor{rulegrey}{HTML}{9AA5B1}

\arrayrulecolor{rulegrey}
\newcommand{\thc}[2][hdrnavy]{\cellcolor{#1}\textcolor{white}{\textbf{#2}}}
\newcommand{\thcl}[1]{\cellcolor{hdrpale}\textbf{#1}}
\newcommand{\thcg}[1]{\cellcolor{hdrgrey}\textbf{#1}}
\newcommand{\thcb}[1]{\cellcolor{hdrblue2}\textbf{#1}}
\newcommand{\thct}[1]{\cellcolor{hdrtan}\textbf{#1}}
\newcommand{\pc}{\cellcolor{cellpale}}

\newcolumntype{L}[1]{>{\RaggedRight\arraybackslash}p{#1}}

\makeatletter
\newcommand{\blfootnote}[1]{%
  \begingroup
    \renewcommand{\thefootnote}{}\footnote{#1}%
    \addtocounter{footnote}{-1}%
  \endgroup
}
\makeatother

\makeatletter
\renewcommand{\paragraph}{%
  \@startsection{paragraph}{4}{\z@}%
                {-1.2ex \@plus -0.4ex \@minus -0.2ex}%
                { 0.4ex \@plus  0.2ex}%
                {\normalsize\bfseries\raggedright}%
}
\makeatother

\newcommand{\appsection}[1]{%
  \refstepcounter{section}%
  \section*{Appendix \thesection\ -\ #1}%
  \addcontentsline{toc}{section}{Appendix \thesection\ -\ #1}%
}

\newcommand{\cae}[2]{{\par\leftskip=\dimexpr #1pt*88/100\relax\relax #2\par}}

\title{Understanding as an Explicit and Assessable Component of Frontier AI
Safety Decisions}
\author{%
  Stephen Barrett\textsuperscript{1} \quad
  Robin Bloomfield\textsuperscript{2} \quad
Alexandra Chirilă\textsuperscript{1} \\[2pt]
  \bfseries Mamoon Masud\textsuperscript{1} \quad
  David Meredith Hardy\textsuperscript{1} \quad
  Phillip Mulvana\textsuperscript{1} \\[6pt]
  \normalsize\textsuperscript{1}Arcadia Impact AI Governance Taskforce \\
  \normalsize\textsuperscript{2}City St George's, University of London
}

\begin{document}

\maketitle

\blfootnote{Correspondence to: Stephen Barrett
\textless{}\href{mailto:steve.barrett@arcadiaimpact.org}{steve.barrett@arcadiaimpact.org}\textgreater{}}

\phantomsection
\addcontentsline{toc}{section}{Abstract}
\begin{abstract}

Decision makers need sufficient understanding to make good decisions about training or deploying frontier AI systems. However, such decisions are increasingly made under time-pressure, and this combined with the use of AI generated artefact creation, can mean that the existence of safety cases and system cards may no longer demonstrate that sufficient understanding exists. Our provisional methodology for making understanding explicit and assessable requires the production of an explicit description of 4 objects of understanding (decision, decision-frame, safety justification, system-in-context) and a justification for the adequacy of this understanding. In addition, the methodology provides a mechanism for describing and evaluating the adequacy of the decision-maker representation of this understanding.  It builds on recent developments in safety cases using the Assurance 2.0 framework to operationalise the philosophical basis of understanding from Elgin and Arendt. To assess the methodology we trialled two different scenarios.  One scenario, which we investigated through role-based analysis, concerned the risk of scheming in the deployment of an AI coding agent in a robotics company and the other scenario was for the higher uncertainty, more decision-critical argument of ‘If Anyone Builds It, Everyone Dies’ (Yudkowsky and Soares). The trial's central finding, for these two scenarios, is that the methodology could be applied and was found to be generative: we found the analyses that justify sufficiency of understanding (internal coherence, tethering, felicitous falsehoods, external coherence) drives the engineering. 

\end{abstract}

\clearpage
\section*{Executive Summary}
\addcontentsline{toc}{section}{Executive Summary}

\textbf{What is the problem with current frontier AI deployment
decision-making?}

\begin{itemize}
  \item Decision makers need sufficient understanding for the decision they
        make to be  adequate  and to be held accountable.
  \item Safety and assurance cases and other justification documentation now risk becoming detached from the
        understanding needed for responsible engineering and governance
        decisions.
  \item There are two issues. Firstly, the production and evaluation of
        critical socio-technical systems increasingly face the challenge of
        pressure for increased tempo leading to reduced scrutiny. Secondly,
        there is growing use of AI-generated artefacts that may produce outputs
        that appear coherent without supporting genuine human understanding.
\end{itemize}

\textbf{What is needed to remedy this problem?}

\begin{itemize}
  \item Understanding needs to be made explicit and assessable.
\end{itemize}

\textbf{How can understanding be made explicit and assessable?}

\begin{itemize}
  \item There are a number of objects of understanding that need to be made
        explicit.
  \item The sufficiency of understanding of these objects needs to be justified
        and this sufficiency needs to be made explicit and communicable to
        others, including auditors.
  \item The understanding also needs to have been sufficiently mentally
        internalised by the decision-maker such that the decision-maker can
        wield this understanding, and the sufficiency of this internalisation
        is something that also needs to be made explicit and assessable.
\end{itemize}

\textbf{What are the objects of understanding?}

\begin{itemize}
  \item Safety justification: The claims, argument and evidence that provide
        justification that the system is safe to deploy.
  \item System-in-context: Description of the system to be deployed, and the
        environment within which it operates.
  \item Decision-in-frame: Description of the decision and related context such
        that a decision-maker can bear accountability for the outcome.
  \item Frame-selection: Description of the selected decision-frame, and
        rationale for its selection.
\end{itemize}

\textbf{How can the sufficiency of understanding be made explicit?}

\begin{itemize}
  \item Assurance 2.0 \citep{bloomfieldAssurance20Manifesto2021} is a rigorous and systematic approach to developing,
        presenting, and examining assurance cases to support indefeasible
        confidence in safety or other critical properties.
  \item Sufficiency of understanding is documented in an Understanding Basis
        (UB), developed using Assurance 2.0,which has the following elements.
  \item Internal coherence: Demonstration that the safety justification
        argument hangs together, is logically valid, sound and indefeasible
        and supports the top level claim with sufficient confidence.
  \item Sufficiency of tethering: Demonstration that the argument is based on
        evidence and fact.
  \item Felicitous falsehoods which are true enough: Demonstration that
        idealisations, abstractions, simplifications and models that are used
        are true enough for the decision in hand.
  \item External coherence: Demonstration of sufficient consensus with other
        sources of information external to the task in hand, and demonstration
        where there is a lack of consensus (i.e. where there is surprise) that
        this is sufficiently explained. 
\end{itemize}

\textbf{How can the decision-maker's internalised level of understanding be
made assessable?}

\begin{itemize}
  \item Assessment of a decision-maker's internalised understanding is made
        explicit and assessable in a Personal Understanding Statement (PUS).
  \item A Personal Understanding Statement (PUS) is structured as a small assurance case so that the personal
        understanding can be made explicit, challenged and revisable. It makes
        a top-level claim of the form ``My understanding is sufficient to make
        decision D, on the basis of UB.''
  \item It comprises a decomposition over four sub-claims over the 4 objects of
        understanding.
  \item For each of the four objects of understanding sufficiency of grasp is
        evidenced through demonstration of five capabilities. The 5
        capabilities include the ability to reason, explain, predict, challenge
        and revise.
  \item The decision-maker also makes explicit what is not grasped and whether
        that matters
\end{itemize}

\textbf{Does the methodology work?}

\begin{itemize}
  \item We showed that the 4 objects of understanding can be made explicit
  \item We showed that the sufficiency of externalised understanding can be
        made explicit and assessable using the additional information provided
        in an Understanding Basis (internal coherence, tethering, felicitous
        falsehoods, external coherence).
  \item We showed that a PUS could plausibly be created and that tests could be
        devised for assessing the decisionmaker's internalised understanding.
        This was successfully achieved for each of the 4 objects of
        understanding (Frame selection, decision-in-frame, system-in-context,
        safety justification) and for each of 5 capabilities (Explain, Reason,
        Predict, Challenge, Revise).
  \item However, we did not yet attempt to take the tests devised in the PUS,
        to determine their practicality and efficacy in establishing the level
        of internalised understanding.
\end{itemize}

\textbf{What was discovered about the methodology?}

\begin{itemize}
  \item The methodology is generative and drove the evolution of the decision,
        the system and the safety justification.
  \item Understanding develops dynamically. The 4 objects of understanding are
        interdependent. Difficulties in building sufficient understanding of
        one object might be resolved by changing another.
\end{itemize}

\textbf{What existing features of Assurance 2.0
\citep{bloomfieldAssurance20Manifesto2021} are supportive of making
understanding explicit and assessable?}

\begin{itemize}
  \item Check for sufficiency of internal coherence is supported by Assurance
        2.0's positive view, negative view and residual risk/confidence view.
  \item Sufficiency of tethering is supported using techniques for weighing of
        evidence
  \item CAE argumentation allows for identification of felicitous falsehoods.
\end{itemize}

\textbf{What augmentations to current safety case practice were deemed most
valuable?}

\begin{itemize}
  \item Search for candidate felicitous falsehoods, and assessment of whether
        they are `true enough'.
  \item Search for external coherence, that may not otherwise be evident or
        made explicit and assessable in an Assurance 2.0 style case.
\end{itemize}

\textbf{How does the ability to make understanding explicit and assessable vary
as a function of uncertainty?}

\begin{itemize}
  \item Felicitous falsehoods are present at every level of uncertainty.
  \item Tethering persists at every level of uncertainty. What changes is the
        nature of the tethering. Where epistemic uncertainty is low, tethering
        may take the form of direct measurements on the system, whilst at the
        other extreme, it may be structural and theoretical arguments that need
        to be tethered.
\end{itemize}

\textbf{How does the ability to make understanding explicit and assessable vary
as a function of decision criticality?}

\begin{itemize}
  \item There is a massive increase in rigour and evidence required to move
        from everyday reliability to the very high system reliability that is
        achieved, for example, in aviation
  \item Where decision-criticality is high there is a need for highly detailed
        supporting argumentation and justification for sufficiency of
        understanding. The depth and complexity likely entail development of understanding that is distributed across people and artefacts.
\end{itemize}

\textbf{What other findings were there?}

\begin{itemize}
   \item Customers of frontier AI would benefit from the availability of component safety cases (for the general purpose frontier AI component), that they could use in building a control safety argument. However, frontier AI developers do not currently make these available.
  \item Attacks on the assurance infrastructure itself can represent a risk
        pathway.
  \item An inability to understand highly advanced AI may itself be a useful
        thing to understand.
\end{itemize}

\section*{Contribution Statement}
\addcontentsline{toc}{section}{Contribution Statement}

\textbf{Author Contributions}: (alphabetical listings within categories)

\textbf{Conceptualization}: Stephen Barrett, Robin Bloomfield

\textbf{Provisional Methodology}: Robin Bloomfield

\textbf{Potential updates to provisional methodology}:  Stephen Barrett, Robin Bloomfield, Alexandra Chirilă,
Mamoon Masud, David Meredith Hardy, Phillip Mulvana

\textbf{Investigation}: Stephen Barrett, Robin Bloomfield, Alexandra Chirilă,
Mamoon Masud, David Meredith Hardy, Phillip Mulvana

\textbf{Writing}: Stephen Barrett, Robin Bloomfield, Alexandra Chirilă, Mamoon
Masud, David Meredith Hardy, Phillip Mulvana

\textbf{Review}: Stephen Barrett, Robin Bloomfield, Alexandra Chirilă,
Mamoon Masud, David Meredith Hardy, Phillip Mulvana

\textbf{Lead editor, Supervision}: Stephen Barrett

\textbf{Project Administration}: Stephen Barrett, Francesca Gomez, Ben R. Smith

\section*{Acknowledgements}
\addcontentsline{toc}{section}{Acknowledgements}

The work was undertaken as part of Arcadia Impact's AI Governance Taskforce.

\textbf{Disclosure on use of generative AI}. Any use of generative AI in this
manuscript adheres to ethical guidelines for use and acknowledgement of generative AI in academic
research. Each author has made a substantial contribution to the work, which
has been thoroughly vetted for accuracy, and assumes responsibility for the
integrity of their contributions.

\section*{Abbreviations and terms}
\addcontentsline{toc}{section}{\textbf{Abbreviations and terms}}

\begin{center}
\begin{tabularx}{\textwidth}{|L{1.27in}|X|}
\hline
ASI       & Artificial Super Intelligence \\ \hline
CAE       & Claims Argument Evidence framework \citep{DeclareGuidanceCAE} \\ \hline
DB        & Decision Basis \\ \hline
DU        & Decision-in-frame \\ \hline
DU+       & Decision frame selection \\ \hline
FF       & Felicitous Falsehood \\ \hline
GDM       & Google DeepMind \\ \hline
HAZOP       & Hazard and Operability Study \\ \hline
IDE       & Integrated Development Environment \\ \hline
IABIED    & If Anyone Builds It, Everyone Dies (Argument presented in
             \citep{yudkowskyIfAnyoneBuilds2025}) \\ \hline
NPSA      & National Protective Security Authority (UK) \\ \hline
NPT       & Non-Proliferation Treaty \\ \hline
PR       & Pull Request \\ \hline
PUS       & Personal Understanding Statement \\ \hline
RL       & Reinforcement Learning \\ \hline
RobotCorp & Fictional robotics company \\ \hline
SC        & System-in-context \\ \hline
SJ        & Safety Justification \\ \hline
STPA       & Systems Theoretic Process Analysis \\ \hline
TLC        & Top Level Claim \\ \hline
UB        & Understanding Basis \\ \hline
\end{tabularx}
\end{center}

\FloatBarrier

\clearpage
\phantomsection
\addcontentsline{toc}{section}{Contents}
{\hypersetup{linkcolor=black}
\makeatletter\renewcommand{\@dotsep}{2}\makeatother
\tableofcontents
}

\clearpage

\section{Introduction}
\label{sec:introduction}

Policy-makers and other risk owners need methodologies to provide a
sufficiently well justified decision to deploy a frontier AI. We assert that a
well-justified decision to deploy, or not, a frontier AI cannot be made unless
there is sufficient understanding. In this paper we explore how to make
understanding explicit, assessable and sufficient.

\subsection{The problem}
\label{sec:the-problem}

Safety cases are one method for demonstrating and communicating understanding
via the arguments about the safety or otherwise of deploying a system and
safety cases are seen as a viable approach for making decisions about
deployment of frontier AI systems \citep{SafetyCasesAISI}, \citep{phuongEvaluatingFrontierModels2025}, \citep{buhlAlignmentSafetyCase2025}. Historically,
creation of a good safety case has been sufficient evidence to demonstrate the
understanding exists amongst those people that create such cases. However,
safety and assurance cases now risk becoming detached from the understanding
needed for responsible engineering and governance decisions \citep{fingletonNuclearRegulatoryReview2025}.

This is because of two issues. Firstly, the production and evaluation of
critical socio-technical systems increasingly face the challenges of pressures
for increased tempo leading to reduced scrutiny \citep{khlaafFissionAlgorithmsUndermining2025}. Secondly, there is growing use
of AI-generated artefacts that may produce outputs that appear coherent without
supporting genuine human understanding \citep{bloomfieldUnderstandingReframingAutomation2026}.

Conventional assurance approaches focus on producing and documenting claims,
arguments, evidence, and confidence while development processes focus on
verification and validation artefacts. Neither make explicit what relevant
actors understand, how that understanding is formed, how it is challenged, or
how it changes when new evidence or defeaters arise.

But what does it mean to understand something? The philosopher Catherine Elgin
states: ``At a first approximation, an understanding is an epistemic commitment
to a comprehensive, systematically linked body of information that is grounded
in fact, is duly responsive to reasons or evidence, and enables nontrivial
inference, argument, and perhaps action regarding the topic the information
pertains to'' \citep{elginTrueEnough2017}.

In the context of frontier AI deployment decisions, the decision-maker needs
sufficient understanding of a number of objects: the decision, the decision
frame, the system-in-context and the justification for sufficiency of safety.
The sufficiency of understanding of these objects needs to be justified and
this sufficiency needs to be made explicit and communicable to others,
including auditors. In addition the understanding needs to have been
sufficiently mentally internalised by the decision-maker such that the
decision-maker can wield this understanding, and the sufficiency of this
internalisation is something that also needs to be made explicit and
assessable.

This new understanding-related challenge is relevant for the deployment of any
safety critical system. In this study we explore and exemplify the research
questions in the context of decisions to deploy frontier AI models in systems
with safety implications. The understanding challenge is particularly acute for
frontier AI systems for the following reasons:

\begin{itemize}
  \item \textbf{High tempo of AI system development} and pressure to release
        systems.
  \item \textbf{Loss of control}: Loss of control is the concern that humanity
        loses control over AI systems. If a human (humanity) is to retain
        control over whether an AI system is deployed or whether an AI should
        be turned off, the human decision-maker (controller) needs to
        understand the risks and benefits of the AI system.
  \item \textbf{Concerns with using AI to `solve' AI Safety}: The assumption
        and de facto plan of some AI developers has previously been stated as
        relying on AI to `solve' AI safety \citep{IntroducingSuperalignment2023}. This assumption is
        questionable, if this approach does not also yield, or is not augmented
        by suitable methods, to enable the requisite understanding by human
        decision-makers.
\end{itemize}

In \citep{bloomfieldUnderstandingReframingAutomation2026} Elgin's conceptual
basis for making claims about understanding is used in order to devise a
framework for making understanding explicit, assessable and open to challenge.
The new contributions of this paper include:

\begin{itemize}
  \item Providing further details of a potential methodology for making
        understanding explicit and assessable
  \item Assessment of this proposed methodology through role-based analyses.
  \item Stress testing of the proposed methodology for arguments where uncertainty and decision-criticality are particularly high.
\end{itemize}

The paper addresses the following research questions

\textbf{Q1}) Whether and how understanding could become an explicit,
assessable, and defensible component of decision making for frontier AI safety
cases?

\textbf{Q2}) How current safety case practice would need to be adapted or
augmented in order to support making understanding explicit, accessible and
defensible?

\textbf{Q3}) How the ability to make understanding explicit and assessable
varies as a function of uncertainty?

\textbf{Q4}) How the ability to make understanding explicit and assessable
varies as a function of decision criticality?

\subsection{Research project design}
\label{sec:research-project-design}

Our work began, by building on the work of one of this paper's authors
\citep{bloomfieldUnderstandingReframingAutomation2026} to provide a detailed
method for making understanding explicit and assessable. This is provided in
Section 2.

Whether the methodology is successful in making understanding explicit and
assessable is evaluated in Section 3.1. Here, we make use of role-based
analysis. Specifically, we considered a scenario that we expect to become
increasingly common, which is that of an organisation making a decision on
whether to make use of general purpose frontier AI for the purposes of
improving organisational competitiveness. We considered the scenario of a
fictional company, RobotCorp, creating safety-critical technology (robots
operating in co-working environments with humans) that wishes to use a frontier
AI coding agent to create product code. We considered harms materialising
through the risk pathway of AI agent scheming. This particular risk pathway was
selected because of the prior availability of a potential `component' safety case, that
had been provided by Google DeepMind \citep{phuongEvaluatingFrontierModels2025},
and because of the availability of a review of this case, the production of which was led by one of this paper's authors 
\citep{barrettLessonsExternalReview2026}.

In Section 3.2 we evaluate how the ability to make understanding explicit and
assessable varies as a function of uncertainty and decision-criticality
(answering Research questions 3 and 4), by considering the argument for `If
Anyone Builds It, Everyone Dies' \citep{yudkowskyIfAnyoneBuilds2025}, which has very different uncertainty and
decision-criticality characteristics compared to the RobotCorp case. The
overall findings of our work are documented in Section 3.3.

\setcounter{subsection}{0}
\section{Provisional methodology for making understanding explicit and assessable}
\label{sec:provisional-methodology}

We start from the perspective that a decision maker needs and wants sufficient
understanding to make a decision that is a good decision for which they
can be held accountable and which they can document and, if necessary,
explain in the future.

The decision is based on a collection of analyses and documents that we call the Understanding Basis, and for which the safety justification is a key component.
Through analysing, probing and discussion the decision maker develops a
sufficient grasp of the Understanding Basis to reach a conclusion.

There are four objects the decision-maker needs to understand. They need to
understand the safety justification and supporting analyses, they need to have
sufficient understanding of what the safety justification is about (the system
in context), of the decision itself and its context (what is it for?), and the
selection of the decision frame itself (why this decision rather than the
alternatives).

The decision maker typically wields their understanding to decide whether or not to accept the top-level claim of the safety justification and provides both a sentencing statement, to justify the decision, and a Personal Understanding Statement to make their grasp (and lack of grasp) explicit, challengeable, and revisable.

Development of the four objects of understanding is dynamic. For example, if
we find potential defeaters on the safety justification that need addressing we
might revise our understanding of the system (it may broaden the system scope
as other dependencies come to light). The decision itself may change, for
example only permitting deployment in less critical scenarios than was
originally envisaged, and we might revise the framing of the decision itself.
So we need a protocol that captures the dynamic aspects and the evolution of
understanding.

Understanding needs to be sufficient, so an equilibrium develops between the
four objects of understanding and we should have frugal and incremental methods
for developing that understanding.

Previous work \citep{bloomfieldUnderstandingReframingAutomation2026} has
defined the concepts of the two key artefacts, the \textbf{Understanding Basis
(UB)} and the \textbf{Personal Understanding Statement (PUS)} and these are
defined in the next sections.

In this project we detail and operationalise these concepts. We also introduce
the need for the Protocol and by applying the UB and PUS we demonstrate the
requirements for such a protocol and generalise this in our recommendations.

\subsection{Understanding Basis}
\label{sec:understanding-basis}

The Understanding Basis (UB) is a structured Assurance 2.0
\citep{bloomfieldAssurance20Manifesto2021} based justification addressing why
the available understanding is sufficient for a specific decision.

The Understanding Basis defines:

\begin{itemize}
  \item[$\bullet$] what epistemic commitments and account the decision relies on;
  \item[$\bullet$] how the justification for the decision is tethered to evidence;
  \item[$\bullet$] how idealisations and felicitous falsehoods are being used;
  \item[$\bullet$] how coherence is established and challenged;
  \item[$\bullet$] how fallibility and the possibility of rupture are handled.
\end{itemize}

The following tables provide guidance on operationalising these
philosophically-leaning concepts.

\subsubsection{Epistemic commitments}
\label{sec:epistemic-commitments}

Sufficiency of understanding of epistemic commitments is captured by the
Assurance 2.0 CAE-based safety case, which has both an explicit argument
structure (usually graphical) and a narrative. Compared with a ``classic''
case, an existing case may need to evolve, as shown below in Figure 2.1.

\begin{table}[htbp]
\centering
\begin{tabularx}{\textwidth}{|L{1.56in}|X|}
\hline
\thc{Commitment and account} & \thc{How it is operationalised} \\ \hline
\pc A tenable account with epistemic commitments
  & Provided by a safety case. Expressed through structured claims,
    assumptions, theories, and justification.

    \smallskip
    System description provided as part of the safety case context \\ \hline
\end{tabularx}
\end{table}

\begin{figure}[htbp]
\centering
\includegraphics[width=4.375in]{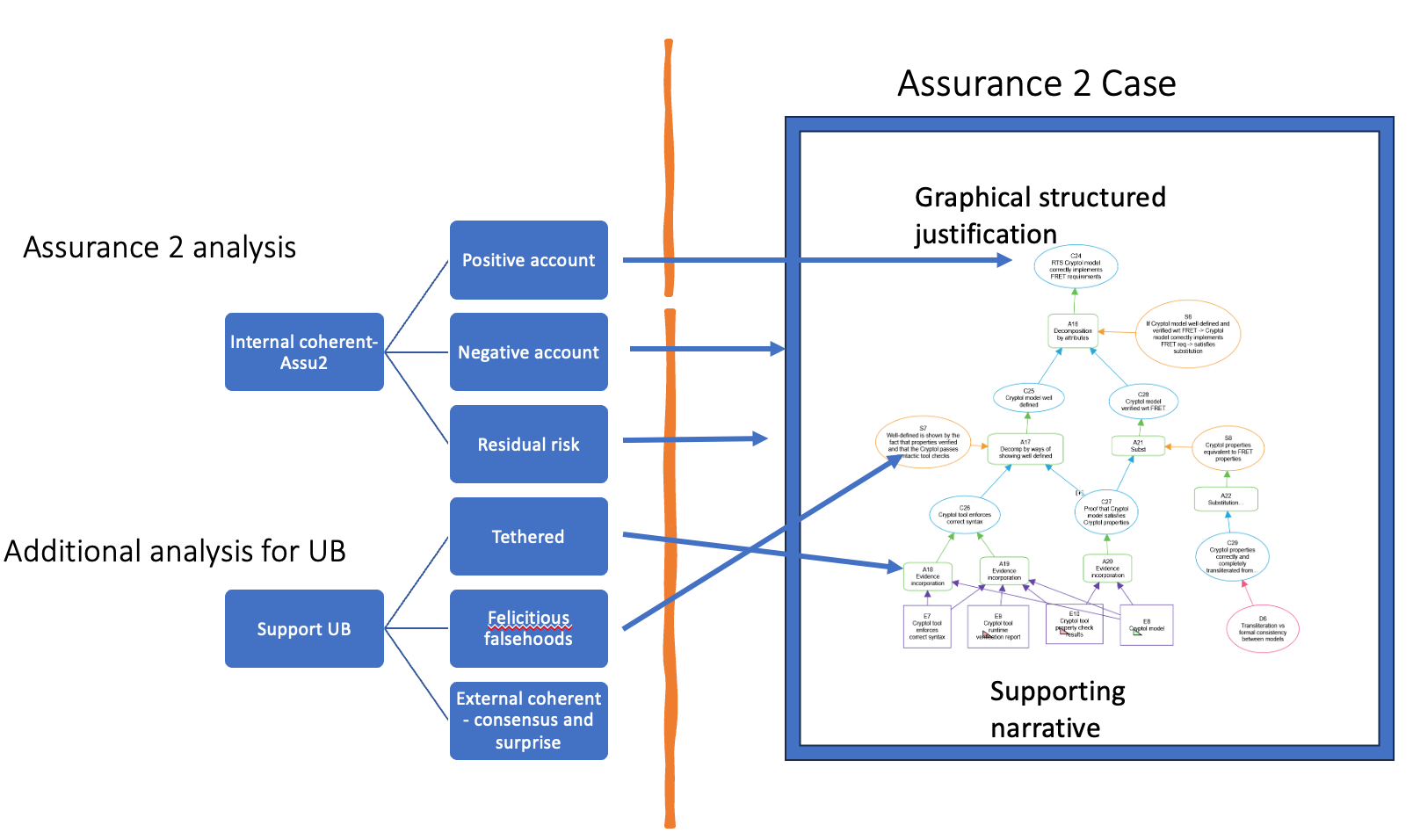}

\smallskip
Figure 2.1) Augmentation of safety case to support additional understanding claims
\end{figure}

\FloatBarrier

We augment the CAE-based safety case with specific claims to capture the new
items.

\subsubsection{Tethering}
\label{sec:tethering}

Tethering maps directly to an Assurance 2.0 safety justification:

\begin{table}[htbp]
\centering
\begin{tabularx}{\textwidth}{|L{1.56in}|X|}
\hline
\thc{Tethering} & \thc{How it is operationalised} \\ \hline
\pc Justification anchored in evidence
  & Achieved with evidential claims and confirmation theory, including explicit
    claims on evidence provenance and trustworthiness. \\ \hline
\end{tabularx}
\end{table}

\FloatBarrier

\subsubsection{Felicitous falsehoods}
\label{sec:felicitous-falsehoods}

Felicitous falsehoods are idealisations, models, or simplifications that are
not literally true but are ``true enough'' to reveal salient system structures
or risks. These map readily onto Assurance 2.0 concepts, but the actual safety
case may need augmenting to bring them out.

\begin{table}[htbp]
\centering
\begin{tabularx}{\textwidth}{|L{1.56in}|X|}
\hline
\thc{Felicitous falsehoods} & \thc{How it is operationalised} \\ \hline
\pc Idealisations that are ``true enough''
  & Appear in model assumptions and decomposition strategies, justified through
    side-claims and confidence assessments. \\ \hline
\end{tabularx}
\end{table}

\FloatBarrier

Felicitous falsehoods form a sub-category of the `idealisation' epistemic
commitment, where categories of epistemic commitment in a case include at
least:

\begin{itemize}
  \item Empirical claims
  \item Theoretical principles
  \item Assumptions
  \item Idealisations
    \begin{itemize}
      \item[$\circ$] Felicitous falsehoods
    \end{itemize}
  \item Modelling conventions
\end{itemize}

Note that a felicitous falsehood differs from a useful assumption, in that a
useful assumption is a proposition accepted as an epistemic commitment for the
purposes of reasoning, analysis, or decision. Its value lies in enabling
inquiry, whether or not it is ultimately true. In contrast, a felicitous
falsehood is a deliberately inaccurate representation whose departure from
literal truth increases understanding by exposing salient structure,
relationships, or mechanisms that would otherwise be obscured.

\subsubsection{Coherence}
\label{sec:coherence}

Coherence refers to consistency and mutual support among the elements of an
account. We operationalise it by distinguishing three types: internal,
external, and pathological.

\paragraph{Internal coherence}
\label{sec:internal-coherence}

Internal coherence is provided by the three evaluation viewpoints in Assurance
2.0, positive view, negative view and residual doubt. It addresses the
question: does the argument hang together and support the top claim with
sufficient confidence?

\begin{table}[htbp]
\centering
\begin{tabularx}{\textwidth}{|L{1.88in}|X|}
\hline
\thc{Internal coherence --- dimension} & \thc{How it is operationalised} \\ \hline
\pc Logical coherence --- positive view
  & Positive evaluation in Assurance 2.0, based on CAE blocks and
    argument-validity checks. \\ \hline
\pc Logical coherence --- negative view
  & The evaluation ``hangs together'' despite a dialectical challenge by exploration of potential defeaters.
    Concepts used: indefeasibility, residual doubt, rebuttals, defeater analysis. \\ \hline
\pc Confidence coherence --- residual-doubt view
  & A consistent approach to residual doubt. Concepts used: confidence
    arguments, confidence defeaters. \\ \hline
\end{tabularx}
\end{table}

\FloatBarrier

The following table defines some defeater prompts for the sufficiency of
internal coherence:

\begin{table}[htbp]
\centering
\begin{tabularx}{\textwidth}{|L{1.88in}|X|}
\hline
\thc{Dimension} & \thc{Defeater prompts} \\ \hline
\pc Narrative coherence
  & Is the narrative self-consistent and consistent with the technical
    argumentation? \\ \hline
\pc Whole-case coherence
  & Are the parts of the case in contradiction? Does evidence for one aspect of the case contradict,
    weaken, or strengthen other aspects of the case? \\ \hline
\pc Ontological coherence
  & Are the terms used in the case consistent and defined? Do terms have
    specialised meanings in different domains (e.g.\ safety and security)?
    \\ \hline
\end{tabularx}
\end{table}

\FloatBarrier

\paragraph{External coherence}
\label{sec:external-coherence}

External coherence relies on assessment of consensus and surprise. Consensus
concerns how the case relates to other cases and source of information.
Surprise relates to whether novelty and innovation are explicitly recognised
and explained. Social coherence concerns whether partial viewpoints of the case are
coherent. The first should form
part of a positive claim; the second is treated as a source of defeaters.

\begin{table}[htbp]
\centering
\begin{tabularx}{\textwidth}{|L{1.88in}|X|}
\hline
\thc{Dimension} & \thc{How it is operationalised} \\ \hline
\pc Consensus and surprise
  & Positive claim showing consensus with other relevant cases or approaches,
    and where differences exist, e.g.\ in the form of novelty or non-standard
    aspects that these have been identified and explained as being acceptable.
    \\ \hline
\end{tabularx}
\end{table}

\FloatBarrier

A search for social incoherence may be used as a defeater prompt.

\begin{table}[htbp]
\centering
\begin{tabularx}{\textwidth}{|L{1.88in}|X|}
\hline
\thc{Dimension} & \thc{Defeater prompt} \\ \hline
\pc Social coherence
  & Are the different viewpoints of the case (subsystem-specific,
    role-specific) coherent? Do viewpoints reduce coherence, and how? Are
    different uses of the case coherent? \\ \hline
\end{tabularx}
\end{table}

\FloatBarrier

\paragraph{Pathological coherence}
\label{sec:pathological-coherence}

Coherence has a dark side: organisational myths, post-hoc rationalisations, and
over-polished safety cases can be highly coherent. Coherence can be achieved by
ignoring or suppressing things that should have been considered. A search for
pathological coherence can be used as a defeater prompt.

\begin{table}[htbp]
\centering
\begin{tabularx}{\textwidth}{|L{1.88in}|X|}
\hline
\thc{Pathological coherence} & \thc{Defeater prompts} \\ \hline
\pc Coherence achieved by suppression
  & Are awkward facts being excluded? Are doubts acknowledged only locally but
    neutralised globally? Are alternative explanations considered seriously?
    Does the case become coherent only by narrowing the system boundary? Are
    dissenting expert views represented or smoothed away? Does the case resist
    or absorb anomalies too easily? \\ \hline
\end{tabularx}
\end{table}

\FloatBarrier

\subsubsection{Reflective equilibrium, fallibilism, and rupture}
\label{sec:reflective-equilibrium-fallibilism-and-rupture}

Reflective equilibrium makes understanding iterative and revisable, while
fallibilism ensures that openness to error is treated as a resource for inquiry
rather than a threat to it.

\begin{table}[htbp]
\centering
\begin{tabularx}{\textwidth}{|L{1.88in}|X|}
\hline
\thc{Concept} & \thc{How it is operationalised} \\ \hline
\pc Reflective equilibrium and fallibilism --- a state where claims, evidence,
    and assumptions are mutually supportive and integrated into a defensible
    ``fabric'' of understanding
  & Operationalised through defeater analysis, uncertainty management,
    dialectical challenge, and explicit documentation of reasoning. \\ \hline
\pc Rupture and re-evaluation --- recognition that established frameworks may
    fail in the face of radical novelty, requiring re-examination of
    assumptions, evidence, and judgement rather than forced coherence
  & Supported by the defeaters, described above, which are specifically
    designed to resist forced coherence, together with explicit claims about
    confidence and expected future behaviour. \\ \hline
\end{tabularx}
\end{table}

\FloatBarrier

Reflective equilibrium and fallibilism are integrated closely into the
Assurance 2.0 methodology; a case developed with Assurance 2.0 will show the
reflective-equilibrium process. In particular, the negative viewpoint of  Assurance 2.0 implements fallibilism and revision. For this
project we subsume this into coherence, and into claims about the negative
evaluation and its process. Whether reflective equilibrium \emph{as a state}
has been achieved is something to reflect on in the PUS.

Note that the Understanding Basis and its associated process should incorporate
designed friction,but we treat this as out of scope because the project is
short. Designed friction is instead developed within the PUS, where it
generates and tests grasp.

\subsubsection{Summary of the Understanding Basis}
\label{sec:summary-of-the-understanding-basis}

The Understanding Basis is developed from an Assurance 2.0 case, incorporating
some Assurance 2.0 analyses as well as additional claims, as described above.
Some aspects are addressed by introducing new classes of defeaters.

Note that an important part of the UB and any safety case is the understanding
of the system-in-context that is required to provide sufficient understanding
of what the safety justification is about.

\subsection{The Personal Understanding Statement (PUS)}
\label{sec:the-personal-understanding-statement-pus}

The UB is a structured account of the understanding and sufficiency of that
understanding that is embodied in a case. The Personal Understanding Statement
(PUS) is the complementary artefact produced by the person who must rely on
that understanding to act.

Where the UB asks ``what is the understanding embodied in this case?'', the PUS
asks ``can I, this decision-maker, genuinely grasp and stand behind that
account well enough to make this decision, and where I cannot, does it
matter?''

The PUS is given a formal spine by a CAE claim structure together with guided
elicitation prompts that generate the evidence for that spine. The formal spine
makes the PUS defensible; the prompts make it tractable to produce. Its
idealised expression is a sentencing statement, in which the decision-maker
expresses the judgement they are making on the case presented.

The UB might also contain explicit evidence of the UB developer's grasp,
enabling a decision-maker to defer to the understanding of the developer. We
build second-order understanding in as a named mode within the PUS (warranted
deference) rather than setting it aside.

\subsubsection{The PUS as a claim-based artefact}
\label{sec:the-pus-as-a-claim-based-artefact}

The PUS is structured as a small assurance case so that it can be made
explicit, challenged, and revised. Its top claim is decision-oriented:

\begin{center}
\textbf{PUS top claim: ``My understanding is sufficient to make decision D, on
the basis of UB.''}
\end{center}

This claim decomposes over four objects of understanding the decision-maker
must hold. The first three sit within the chosen decision frame; the fourth
concerns the choice of frame itself, and conditions the validity of the others.

\begin{itemize}
  \item[$\bullet$] \textbf{Safety justification (SJ)}: The claims, argument and
        evidence and supporting analysis that provide justification that the
        system is safe to deploy.
  \item[$\bullet$] \textbf{System-in-context (SC)}: Description of the system
        to be deployed, and the environment within which it operates.
  \item[$\bullet$] \textbf{Decision-in-frame (DU):} Description of the decision
        and related context such that a decision-maker can bear accountability
        for the outcome. Describes what the decision is for, its impact, and
        what follows if the wrong decision is made, given the frame.
  \item[$\bullet$] \textbf{Frame-selection (DU+):} Description of the selected
        decision-frame, and rationale for its selection. Describes why this
        decision-frame was selected rather than the alternatives.
\end{itemize}

Each object of understanding requires a sub-claim of the form ``I grasp the SJ
sufficiently for DU'', supported by (a) positive evidence of grasp and (b)
explicit self-defeaters, the things known not to be grasped. Making lack of
grasp a structural element lets the PUS claim sufficiency clearly. The fourth
sub-claim (frame-selection, DU+) is distinctive: sub-claims 1--3 can all be
fully satisfied and the PUS still fail, because they may answer the wrong
question. Frame-selection therefore conditions the validity of the others
rather than sitting alongside them, and is best treated as a gate on the
understanding verdict.

The safety justification and system-in-context form part of the Understanding
Basis. Whilst the Decision-in-frame and Frame-selection form part of the
Decision Basis.

\subsubsection{Direct grasp and warranted deference}
\label{sec:direct-grasp-and-warranted-deference}

Real safety justifications are composed from many partial understandings; a
decision-maker rarely reconstructs every argument personally. The PUS
distinguishes two named modes of grasp, and tags each sub-claim with the mode
relied upon.

\begin{table}[htbp]
\centering
\begin{tabularx}{\textwidth}{|L{1.45in}|X|}
\hline
\thc[hdrblue]{Mode of grasp} & \thc[hdrblue]{What it asserts and what it must
justify} \\ \hline
\pc Direct grasp
  & ``I can reconstruct, explain, and defend this part of the case myself.''
    Evidence is the person's own reasoning, narrative, or reconstruction.
    \\ \hline
\pc Warranted deference
  & ``I cannot reconstruct this part myself, but I grasp its role in the case
    and have defensible grounds to trust the grasp of those who produced it.''
    This mode carries its own obligation: the PUS states why the deference is
    warranted (based on the analysis and evidence in the UB). Warranted
    deference is itself an understanding claim, not an absence of one.
    \\ \hline
\end{tabularx}
\end{table}

\FloatBarrier

This is the project's treatment of distributed and second-order understanding.
The decision-maker need not grasp everything directly, but every instance of
deference must be visible and justified, so the boundary between what is
understood and what is trusted is clear.

\subsubsection{Evidencing grasp: the objects \texttimes\ capabilities matrix}
\label{sec:evidencing-grasp-the-objects-capabilities-matrix}

For each of the four objects of understanding (SJ, SC, DU, DU+), grasp is
evidenced through five capabilities. These are reframed as things the
decision-maker can personally do, rather than properties the analysis
demonstrates: ``I can predict how the case behaves if the boundary shifts'' is
a PUS claim; ``sensitivity analysis demonstrates prediction'' is a UB property.

\begin{table}[htbp]
\centering
\begin{tabularx}{\textwidth}{|L{1.08in}|X|}
\hline
\thc[hdrblue]{Capability} & \thc[hdrblue]{What the decision-maker should be
able to do, per object} \\ \hline
\pc Reason
  & Trace why the core claims follow from the evidence and theories; identify
    the set of particularly important arguments and theories at the core of the case.
    \\ \hline
\pc Explain
  & Render the formal structure as a coherent narrative in their own words,
    to themselves and to a challenger. \\ \hline
\pc Predict
  & Say how the case would behave under change: a shifted system boundary, a
    weakened item of evidence, a different use-case. \\ \hline
\pc Challenge
  & Generate plausible defeaters; spot felicitous falsehoods doing too much
    work; detect pathological coherence. \\ \hline
\pc Revise
  & Use confidence/claim trade-offs and chains of confidence to say how the
    case would have to change to support the decision, and whether such
    revision is feasible. \\ \hline
\end{tabularx}
\end{table}

\FloatBarrier

The matrix provides a coverage map. A reviewer who can reason and explain the
SJ but can only defer on a particular evaluation, while still being able to
predict the consequence if that evaluation were wrong, has a defensible, and
explicit understanding profile.

The five capabilities apply to the frame-selection too: Reason becomes
articulating the selection argument; Predict becomes saying how the decision
would change under a different frame; Challenge becomes generating the
alternatives in the decision space; Revise becomes saying what evidence would
trigger a change in the selection; and Explain concerns
articulating the selection to those who bear the consequences.

Designed friction produces and tests grasp with the intention of avoiding passive acceptance of a
coherent-looking artefact. Each mechanism is reframed as a test yielding
evidence for a specific capability.

\begin{table}[htbp]
\centering
\begin{tabularx}{\textwidth}{|L{2.57in}|X|}
\hline
\thc[hdrblue]{Friction test} & \thc[hdrblue]{Evidence it produces} \\ \hline
\pc Translate (part of) the formal CAE structure into a coherent narrative.
  & Evidence for Explain; exposes structure that is followed but not
    understood. \\ \hline
\pc Defend the case without automated assistance.
  & Evidence against hidden epistemic debt and shallow uptake. \\ \hline
\pc Engage a plausible but logically flawed automated output.
  & Evidence for Challenge; exposes theory opacity. \\ \hline
\pc HAZOP-style guideword sweep over UB nodes.
  & HAZOP (Hazard and Operability Study), is a systematic guide-word structured technique for identifying hazards. It can be used to systematically populate the residual-incomprehension register. \\ \hline
\end{tabularx}
\end{table}

\FloatBarrier

A possible ``challenge'' activity is to apply a fixed set of guidewords to each
UB node (each claim, item of evidence, assumption, and felicitous falsehood),
in the spirit of a HAZOP applied to a safety case.

\subsubsection{The residual-incomprehension register}
\label{sec:the-residual-incomprehension-register}

An important part of the PUS is the explicit record of what is not grasped. The
analogue, in the PUS of residual doubt and defeaters in the safety case.

Every entry carries two independent judgements:

\begin{enumerate}
  \item \textbf{Do I grasp this?} (not at all / its role only / partially /
        fully)
  \item \textbf{Does my decision depend on grasping it?}
\end{enumerate}

The second connects understanding to decision criticality and uncertainty, and
provides the principled route to declaring understanding sufficient despite
gaps: a gap the decision does not depend on is felicitous ignorance; a gap the
decision turns on is a genuine self-defeater that must be resolved, deferred
with a warrant, or treated as a reason not to decide. The register thus
distinguishes felicitous ignorance from negligence. These concepts are illustrated with some examples in the table below.

\begin{table}[htbp]
\centering
\begin{tabularx}{\textwidth}{|X|L{0.89in}|L{0.96in}|L{1.47in}|}
\hline
\thc[hdrblue]{UB element / item} & \thc[hdrblue]{Grasp} &
\thc[hdrblue]{Decision depends on it?} & \thc[hdrblue]{Resolution} \\ \hline
\pc (e.g.\ internal mechanism of evaluation X)
  & Role only
  & No --- only its result matters
  & Felicitous ignorance; recorded, no action \\ \hline
\pc (e.g.\ assumption behind a felicitous falsehood)
  & Partial
  & Yes
  & Self-defeater: resolve, defer with warrant, or do not decide \\ \hline
\end{tabularx}
\end{table}

\FloatBarrier

\subsubsection{Felt unease as a defeater-discovery heuristic}
\label{sec:felt-unease-as-a-defeater-discovery-heuristic}

The PUS records the decision-maker's overall sense of the adequacy of the UB,
including any felt unease or tension. This is given an explicit
role: unease is treated as a trigger for a more directed search for
pathological coherence, examples of which include an unexamined assumption, or an awkward fact that has
been smoothed away. Unease that survives investigation is promoted to a
defeater; unease that dissolves is recorded as resolved. Either way it is
explicitly addressed, rather than being left as soft introspection.

\subsubsection{Summary: the shape of a PUS}
\label{sec:summary-the-shape-of-a-pus}

\begin{enumerate}
  \item A \textbf{top claim} of sufficiency-for-decision DU on the basis of UB.
  \item \textbf{Four sub-claims} over the safety justification (SJ), the
        system-in-context (SC), the decision-in-frame (DU), and
        frame-selection (DU+), the last conditioning the validity of the
        first three.
  \item Each sub-claim \textbf{evidenced via the five-capability matrix}, with
        every item tagged direct grasp or warranted deference.
  \item A \textbf{residual-incomprehension register} carrying the `grasp'
        judgement and the key `does-it-matter' judgement.
  \item Generated and validated through \textbf{designed-friction tests} and a
        \textbf{HAZOP-style guideword sweep}, with \textbf{felt unease} as a
        defeater-discovery heuristic.
\end{enumerate}

This makes the PUS, like the UB, an explicit and revisable assurance artefact.

\subsection{Process and protocol}
\label{sec:process-and-protocol}

Since we are concerned with complex engineered systems, the development of the
Understanding Basis and the Decision Basis would be expected to be part of an
engineering workflow that will have been defined for the users of the
methodology in those organisations.

Rather than try and define a universal process we instead define the components
of such a workflow and their interaction that could be instantiated within a
particular organisation.

We define:

\begin{itemize}
  \item The basic stages of the process
  \item The objects that form the DB and UB
  \item The interaction between these objects that needs to be taken into
        account in a dynamic workflow which can support a frugal approach to the development of understanding,
\end{itemize}

The overall stages of the process are:

\textbf{Stage A: Scope and entry conditions.} Describes what must be in place
before the protocol starts: an existing safety case, a named decision with an
accountable decision-maker, and a stated decision context (purpose,
criticality, consequence of error).

\textbf{Stage B: Developing the UB (domain-expert workflow).} Uses the UB's
structure to define a set of steps: developing or converting the case in Assurance
2.0 form, running the three Assurance 2.0 analyses (positive, negative,
residual risk) for the baseline, then adding what a classic case lacks: felicitous
falsehoods as explicit blocks, tethering/provenance side-claims, and the new
defeater classes. Provide supporting descriptions: models and analyses of the
system and its context. The deliverable is the UB plus the domain expert's own
account of where the embodied understanding is thin.

\textbf{Stage C: Developing the PUS (decision-maker workflow).} The reflective
workflow: demonstrating grasp of the 4 objects of understanding, summarising
coverage via the objects x capabilities matrix; identifying residual aspects
that are not comprehended and whether this matters for the decision in hand.
Search for any felt unease as part of defeater discovery. The deliverable is
the populated PUS.

\textbf{Stage D: The UB$\leftrightarrow$PUS $\leftrightarrow$DB loop.} The
reflective-equilibrium process: here any concerns that are identified with the PUS, either results
in the UB being sent back for revision, the concern becoming an accepted deferral with
recorded warrant, or grounds for a decision-withheld verdict. This is equilibrium as a
workflow between two people or groups (decision-maker and domain expert) rather
than a property asserted of one document. As the understanding develops the top
level claim and decision itself may change as the UB and Decision Basis (DB)
come into equilibrium.

\textbf{Stage E: Acceptance and sentencing.} The exit conditions: when the UB
is sufficient to hand over, and when the understanding verdict can be
``sufficient'' / ``sufficient subject to reservations'' / ``insufficient'',
with the audit trail (completed matrix, register, deferral justifications) that
makes the verdict inspectable by a third party.

\textbf{Aggregation and scaling note.} The above stages are simplified. The UB
might in practice cover several domains and be composed of a number of
different safety justifications and there may be many actors each of whom
provide partial understanding that is synthesised into a whole.

\section{Trial application of the understanding methodology}
\label{sec:trial-application}

In the previous section we defined a proposed methodology for making
understanding explicit and assessable. In this section we present our results
of applying the methodology.

We made this investigation through role-based analyses.

Team members played the following two roles:

\begin{itemize}
  \item Domain expert (author of Understanding Basis)
  \item Decision-maker (person who decides, for example, whether the system can
        be deployed, and who is responsible for making sufficiency of their internalised understanding
        explicit using a Personal Understanding Statement (PUS)).
\end{itemize}

The methodology was trialled by an international, interdisciplinary team of 4 with
diverse backgrounds in safety engineering, decision strategy, autonomous
vehicles and philosophy. This team was supported by a research team leader and the
developer of the understanding based methodology. The work was undertaken over
12 weeks.

We chose to focus the role-based analysis around potential risks and failure
modes arising due to use of frontier AI in a fictional company, RobotCorp that
creates robotic systems. We decided to emulate a situation that is commonly
seen in safety engineering practice wherein an application developer wishing to
make use of a 3rd party provided component, has to create their own
application-level safety justification. Such an application safety
justification may make use of a component safety justification, if such
exists. The application developer may address concerns with the component
safety justification by adding functionality to their own application level
system. Since Google DeepMind (GDM) \citep{phuongEvaluatingFrontierModels2025} have
produced a safety justification for harms arising due to
scheming with frontier AI, our work first focused on this particular scheming risk pathway, and we investigated whether the GDM case might be used as a component safety case by RobotCorp.

In Section 3.1 we provide further details of the RobotCorp scenario. In Section
3.2 we describe the work we did to stress test the methodology by considering its application 
to the high decision-criticality, high uncertainty `If Anyone Builds It
Everyone Dies' (IABIED) argument \citep{yudkowskyIfAnyoneBuilds2025}. In
Section 3.3 we describe our key findings from our work on the RobotCorp and
IABIED scenarios.

\subsection{Industrial safety critical scenario (RobotCorp)}
\label{sec:industrial-safety-critical-scenario-robotcorp}

\subsubsection{Scenario description}
\label{sec:scenario-description}

The CEO of imaginary company, RobotCorp is considering making use of a Gemini 2.5 frontier AI coding agent, which in our fictional scenario is the most recent frontier AI available from GDM (we assume Gemini 2.5, because the \citep{phuongEvaluatingFrontierModels2025} case was provided for it).

RobotCorp creates robotics solutions where humans and the robot share the same
workspace, for example in a car factory or in bio-labs. The CEO wants RobotCorp
to make use of the latest AI coding agent which is based on Gemini 2.5, in
order to accelerate the speed of, and cost effectiveness of robotics software
development.

RobotCorp already has an earlier safety case that is considered adequate for
the handling of risks associated with human generated code. So the team is
requested to focus on potential new failure modes, arising from use of coding
agents, and specifically risks arising due to scheming or mistakes.

The safety team is requested to ensure that the usual RobotCorp safety
requirements are adhered to. These include there being a very low risk of
physical harm to RobotCorp's own employees, and the employees of customer
companies, in addition to dollar impacts of harms arising due to failures in
code arising due to use of the coding agent to be below \$10m.

\subsubsection{The application of the understanding methodology}
\label{sec:the-application-of-the-understanding-methodology}

The trial progressively developed the Understanding Basis and Decision Basis
with the participants taking specific roles. The trajectory of the project is
described in Table 3.1 below and the interactions and dynamic aspects of the work discussed
in Appendix D.1.

\begin{table}[!t]
\centering
\small
\begin{tabularx}{\textwidth}{|L{0.42in}|X|L{0.80in}|}
\hline
\thcl{Step} & \thcl{Activity}
  & \cellcolor{hdrpale}\textbf{Activity category}\newline\mbox{}\newline
    \textbf{UB/DB} \\ \hline
1 & Roughly scoped decision: to deploy Gemini 2.5 for general use within
    RobotCorp - with deployment decision to be based on analysis of new AI
    related failure modes. & DB \\ \hline
2 & Evaluation of GDM's scheming inability CAE safety case (review of
    \citep{phuongEvaluatingFrontierModels2025},
    \citep{barrettLessonsExternalReview2026}).\newline\mbox{}\newline
    Significant concerns identified on relevance and significant defeaters that
    could not be mitigated. & UB \\ \hline
3 & Brainstormed alternative decisions that would address key defeaters in the
    GDM case. & DB \\ \hline
4 & Reduced scope of decision and defined top-level claim for SJ.\newline\mbox{}\newline
    Limit application to an AI coding agent used by RobotCorp's product
    development team. & DB/UB \\ \hline
5 & Performed risk pathway analysis, and made use of the dependability
    perspective \citep{bloomfieldAssuranceAISystems2025} to identify methods
    for supporting detection and prevention of potential harms and to deliver
    resilience.\newline\mbox{}\newline
    Adapted the system-in-context and the safety justification. & UB \\ \hline
6 & Held defeater brainstorm: directed at safety justification. Variety of
    defeaters identified. & UB \\ \hline
7 & Safety justification updated to better enable identification / derivation
    of felicitous falsehoods. & UB \\ \hline
8 & Internal coherence analysis (positive view). Reflections on whether safety
    justification was logically valid and sound. & UB \\ \hline
9 & Internal coherence analysis (negative view): collected defeaters together
    from brainstorm and those which had been identified in the construction of
    the safety case, applied UB prompts for narrative coherence, ontological
    coherence etc. & UB \\ \hline
10 & External coherence analysis \newline
    Many useful `surprises' identified in analysis of NPSA's insider risk
    report \citep{SettingFoundationsFive}. These would have warranted further
    investigation given more time and may have resulted in changes to the
    safety justification and system in context. & UB \\ \hline
11 & Tethering analysis \newline
    Lack of evidence in support of adequacy of new controls, limited the
    ability to do an exhaustive tethering analysis. \newline
    No residual risk and confidence analysis. & UB \\ \hline
12 & PUS's specified with tasks to demonstrate grasp. & PUS \\ \hline
13 & New version of case produced and proposed simplification. Side claims
    added. Preparation for next stage & UB \\ \hline
\end{tabularx}

\smallskip
Table 3.1) Activities undertaken when applying the methodology to the RobotCorp
scenario
\end{table}

The artefacts were developed in sufficient detail to exercise the methodology
but of course a real understanding with real evidence would take much more
effort than was available to the project. Details of this work are provided in the appendices.  Appendix A describes work done in developing the Understanding Basis.  Appendix B provides details of the RobotCorp system in context and safety justification.  Appendix C describes the RobotCorp Decision Basis.   Appendix D describes our analysis of the dynamics of building understanding.  Appendix E describes work done in developing the PUS for RobotCorp.  In the appendices we note where
analyses were not undertaken either because of effort involved or more often
because the stage of the safety case meant that they were not feasible.

\FloatBarrier

\subsection{Stress testing the methodology (IABIED)}
\label{sec:stress-testing-the-methodology-iabied}

The approach to risk assessment and management changes as epistemic uncertainty
increases. When we can foresee possible future trajectories we can enumerate
them, assign probabilities (both subjective and objective) to events and use
this to inform our decision making. As uncertainty increases, we may only be
able to foresee options for just the next step, in which case, risk management
takes on the character of an exploration. To undertake this exploration, the
decision to take the next step may require an ability to recover should it be
found to be a mis-step, in addition there is a need to have the capability and
resources to discover possible follow-on steps, once the first step has been
taken. As uncertainty increases further we may focus on preparing for outcomes.
This raises the question of how the methods for making understanding explicit
and assessable might vary as a function of the change in the risk management
approach as the level of uncertainty increases.

Risk management differs as the level of decision-criticality increases. More
rigour is expected in the safety engineering process as criticality increases.
This raises the question of whether the understanding methodology changes as a
result of changes in the risk management process that are seen as criticality
increases.

In this part of the trial we explored how the understanding methodology design
changes as a function of uncertainty and decision-criticality. We achieved this
by considering what challenges might be encountered in applying the methodology
to the high uncertainty, high decision-criticality IABIED case. Specifically we
sketched out what a UB for IABIED might look like, and what issues might arise.
Similarly, we considered the potential for applying the PUS methodology to
IABIED.  Further details are provided in the appendices.  Appendix F provides a sketch of the IABIED argument and examples of relevant supporting evidence. Appendix G provides details of our stress testing of the methodology under conditions of high uncertainty and decision-criticality. 

\subsection{Key findings}
\label{sec:key-findings}

In this section we introduce our key findings from the application of the
provisional methodology to the RobotCorp scenario, and our findings from stress
testing the methodology against the high-uncertainty, high decision-criticality
IABIED scenario.

\subsubsection*{The methodology is generative and drove the engineering}

\textbf{Finding}: A principal finding is that the methodology is
\emph{generative}: the analyses that justify sufficiency of understanding
(coherence, tethering, felicitous falsehoods, external coherence) drove the
evolution of the decision, the system and the justification.

An unmakeable case addressing the deployment of the coding agent as initially
framed was made feasible by changing the definition of the system to include
the software development lifecycle, the organisational learning process and the
wider robot system environment. In this extended system, additional control and
mitigation functions were defined. In addition, the decision itself was changed
from a deployment decision to a less critical gate-review decision to invest in
further control development and evidence gathering to support a future
deployment decision.

\textbf{Recommendation}: Understanding should be treated as an explicit
engineering object throughout the engineering lifecycle rather than as an
implicit attribute of experienced engineers or reviewers.

\subsubsection*{Dynamics of understanding is key to its development and efficiency}

\textbf{Finding:} The dynamics between the 4 objects of understanding allowed
an efficient search for solutions.  Because the 4 objects are interdependent, difficulty in one can be resolved by
changing another, and the search for resolution is where efficiency comes
from. There was no need to develop a case branch further once serious defeaters were
identified against it; instead the trial iterated around options: improving
controls, changing the decision, changing the system design and the deployment strategy. The
methodology thereby supports frugal development of artefacts. The journey
ends at equilibrium, when both safety (to the level the decision requires)
and sufficiency of understanding can be justified for the decision as framed.

\textbf{Recommendation}: Future assurance methodologies should explicitly
support iterative co-evolution of engineering decisions, safety arguments and
system definitions.

\subsubsection*{Safety cases can be extended to support understanding}

\textbf{Finding:} The additional analyses that assessed internal coherence
helped in discovering defeaters, alternative options and in the development of
confidence and understanding.

The RobotCorp study suggests how converting a safety case, based on Assurance
2.0 into an Understanding Basis required a relatively modest but well-defined
set of additional analyses. In particular, it required explicit assessment of:

\begin{itemize}
  \item epistemic commitments,
  \item tethering,
  \item felicitous falsehoods,
  \item internal coherence, and
  \item external coherence
\end{itemize}

These assessments revealed assumptions, defeaters and opportunities for
improvement that were not evident from the safety argument alone. These
additions proved practical within the scope of the study and consistently
generated useful engineering insights.

The work therefore supports the assertion that a good safety case is
\textbf{necessary but not sufficient} as evidence of understanding. The
additional analyses required to construct an Understanding Basis appear to be
both manageable and worthwhile.

\textbf{Recommendation}: Future safety case practice should explicitly augment
conventional claims, arguments and evidence with structured analyses of
coherence, tethering and epistemic commitments in order to make understanding
itself assessable.

\subsubsection*{Search for external coherence created new insights}

\textbf{Finding}: External coherence relies on assessment of consensus and surprise. Consensus
concerns how the case relates to other cases and source of information.
Surprise relates to whether novelty and innovation are explicitly recognised
and explained.

The external coherence was assessed with respect to the general safety critical
engineering literature and the frontier AI literature as well as human insider
risk management. The latter was significant as it provided insights from a
different domain.

Details are provided in Appendix A.6, example findings were

\begin{itemize}
  \item consensus with common safety critical systems engineering practice,
        where hazards and risk pathways are enumerated first
  \item surprise that the case does not consider potential malign influence of
        AI over human employees, including also the possibility of supporting
        the upskilling of malign human employees.
  \item surprise that the case does not consider AI facilitating access to
        systems/data on behalf of another malign actor.
\end{itemize}
\textbf{Recommendation}: Safety engineering practice should incorporate a search for external coherence.

\subsubsection*{Stress testing on IABIED indicated how understanding changes with uncertainty}

\textbf{Finding}: Stress testing showed how the importance of different parts of the UB changes
with uncertainty. Increased uncertainty made it necessary to make a safety
justification for a generic category of system and environment, rather than for
a specific system-in-context. There was more reliance on indirect evidence and
theories. We mapped how different types of systems with access to different
types of knowledge impacted the UB. The overall approach of building a UB was
applicable, which we showed by applying the methodology to one interpretation
of the argument in the book, and in which we were able to identify less
justified aspects. For a case like IABIED, there was a move away from aleatory
uncertainty being an important component to epistemic uncertainty dominating.

Increasing criticality leads to a requirement for increased rigour in the
argument and the proof of adequacy of assumptions, felicitous falsehoods and
theories being used, as well as in the nature and extent of tethering of the
evidence.

Our application of Assurance 2.0 and the UB in both RobotCorp and IABIED showed
that a wide variety of decisions with different uncertainty and criticality can
be addressed within a single framework but the arguments, evidence, defeaters
all change in nature and extent.

\subsubsection*{Methodology is applicable now and improvements specified}

\textbf{Finding}: We showed that the 4 objects of understanding can be made
explicit. We developed detailed examples of the safety case and undertook the
new aspects required by the sufficiency analysis. Other aspects of internal
coherence, indefeasibility and logical evaluation are feasible as they are a
demonstrated part of Assurance 2.0. We showed that the sufficiency of
externalised understanding can be made explicit and assessable using the
additional information provided in an Understanding Basis (internal coherence,
tethering, felicitous falsehoods, external coherence). We evaluated these as
far as the nature of the trial allowed.

We showed that a PUS could plausibly be created and that tests could be devised
for assessing the decision-maker's internalised understanding. This was done for
each of the 4 objects of understanding (frame selection, decision-in-frame,
system-in-context, safety justification) and for each of 5 capabilities
(Explain, Reason, Predict, Challenge, Revise).

However, we did not yet attempt to take the tests devised in the PUS, to
determine their practicality and efficacy in establishing the level of
internalised understanding. We suspect that the ability for a decision-maker to
wield their internalised understanding in order to make predictions will be a
particularly important capability to demonstrate. The methodology worked but
this first trial clearly showed ways it could be improved.

The act of creating PUS artefacts for the RobotCorp case forced team members to
acknowledge areas where their understanding was still deficient, despite having
worked on the Understanding Basis directly. This demonstrates the value of the
introduced friction of this step. It also points towards foreseeable challenges
in implementing PUS in large organisations and/or relating to more complex
safety cases than our RobotCorp one. In real-life applications a decision-maker
is likely to be relying on others' understanding via warranted deference across
a wide range of contributors to the UB.

In such cases it will be necessary to establish that the deferred understanding
sufficiently supports the decision-maker's purposes, and the decision-maker's
holistic understanding of the case is robust. The 5 capability tests against
each of the 4 components of understanding provide the tools to demonstrate
understanding, but also require thought to what level of evaluation, by whom,
is necessary and sufficient for the case.

The dynamic development of a case described here builds some of this evaluation
into the process by iteratively developing the Understanding Basis and
Decision Basis. Parties involved in the case will necessarily gain appreciation
of what each other understands. As a dynamic tool used throughout a case, the
PUS can drive understanding rather than being seen as a `signing off' step at
the end.

Other findings (see also Appendices D and G) show the need for guidance on
dynamics and impact of uncertainty and criticality.

Many aspects of the methodology would benefit from tools, for example including  support for synthesis
using CAE Blocks and theories, logical evaluation, defeaters management and
ontology analysis.

One open question is to what extent decision-makers should be expected to understand what,
        in the case of AI, are complex systems, having very complex
        interactions with their environment (consider e.g.\ an STPA analysis of
        an AI coding agent operating in a control environment). Understanding
        may also be recursive, for example with a CEO deferring to a CTO who
        defers to an R\&D head who defers to safety engineers. Given this
        personal lack of expertise or experience on behalf of the CEO, what
        criteria might decision-makers use to prove that their warranted (and
        potentially recursive) deference of understanding is justified?

\textbf{Recommendation}: Specific enhancements to the methodology

\begin{itemize}
  \item Use of PUS examples in the revised protocols (see Appendix E)
  \item Dynamics of evaluation should be recognised and guidance provided.
  \item Guidance should be provided on how the uncertainty and criticality of
        the decision impact the Understanding Basis
\end{itemize}

\subsubsection*{AI developers should provide component safety cases that can be used by their customers}

\textbf{Finding}: The GDM scheming inability case claimed that scheming could
not result in severe harm in any GDM or Google internal deployment setting. Its
use as a component case by RobotCorp was limited by a number of factors:

\begin{enumerate}
  \item The top-level claim was made for Google internal deployment, not for
        customer deployment. GDM were able to make assumptions about the
        existence of, and quality of, oversight mechanisms within GDM
  \item The harm threshold (severe harm, assumed by \citep{barrettLessonsExternalReview2026} to be $>$ \$1bn) is far in
        excess of what many customers would find acceptable
  \item The case was not sound, given for example, the presence ofunresolved defeaters identified by the
        authors \citep{phuongEvaluatingFrontierModels2025} and reviewers
        \citep{barrettLessonsExternalReview2026}
\end{enumerate}

Due to the oversight assumptions in the GDM case, it can be argued that what was described as an inability case was in fact a control case. A customer acquires the AI, not the
oversight environment with the controls that makes the claim hold.

A genuine inability case, making no assumptions about control and oversight,
would be a very useful component case for application developers. But as
capabilities grow, witness the stealth and situational awareness exhibited
in the OpenAI-agent/Hugging Face incident \citep{OpenAIHuggingFace2026},
inability arguments will likely no longer be viable for the most capable
frontier systems. An alignment-based component case, claiming harm cannot arise
in any deployment setting because the AI is aligned, would serve the same
customer need; to our knowledge no such case exists. Application developers are
therefore left developing their own control safety cases, and as this trial
shows, designing such controls and proving their safety is a non-trivial
exercise. For a general-purpose technology used by millions of downstream
deployers, that is a major structural problem. 

\textbf{Recommendation}: Frontier model safety cases should recognise that
models will be components in other people's systems, and should ship with
component safety justifications of the type that are provided in every other
safety-critical market:

\begin{itemize}
  \item \textbf{Safety application conditions}: the explicit assumptions of
        use under which the component claim holds, and the obligations the
        claim places on the integrator. This would evolve the practice for
        out-of-context components (the safety manual of an IEC 61508 compliant
        item; the assumptions of use of an ISO 26262 Safety Element out of
        Context and the Railway Industry EN50129 safety application conditions)
  \item \textbf{Failure and vulnerability analysis support}: characterised
        failure modes and attack/scheming pathways at the model boundary, so
        integrators can enumerate hazards rather than reverse-engineer them
        (see the surprise risk pathways identified in Appendix A.6 that were
        not enumerated first). The AI developer should provide reliability analysis to allow the integrator to assess the significance of the failure and the likelihood of challenges to the overall system controls.
  \item \textbf{Adversary-model calibration data}: capability evaluations
        published in a form usable to calibrate the adversary model of an
        integrator's control case, with validity tied to an identified
        configuration so integrators can manage change.
\end{itemize}

More broadly, providers should adopt the dependability perspective: the model
is one component of a socio-technical system whose safety is a system property,
and provider artefacts should be designed to compose into system-level
justifications.

\subsubsection*{Attacks on safety assurance systems and infrastructure are a risk pathway}

\textbf{Finding}: A frontier AI agent embedded in an organisation can, in
principle, act on the assurance process itself: epistemic manipulation of the
case, adversarial shaping of the review process, weaponised complexity that
defeats the decision-maker's grasp, and manufactured pressure for acceptance.
In addition to these covert actions there is the impact of the unreliability of
the AI on the tasks it undertakes and the scope for corrupting the integrity of
the case (e.g.\ recent incidents of OpenAI models deleting hard disk contents \citep{ChatGPTDeletingPeoples2026}).
These were treated as risk pathways in the application case and specific claims
developed (the assurance-integrity branch). This finding motivates the PUS:
tests of internalised grasp are a defence in depth against artefacts optimised
to be accepted rather than understood.

\textbf{Recommendation}: When the assessed system can participate in its own
assurance, the case must address the evidential status of its own artefacts.

\subsubsection*{An inability to understand highly advanced AI may itself be a
useful thing to understand}

\textbf{Finding}: Focusing on frontier AI systems in particular raised
questions about the limits of understanding that are possible, and whether an
inability to understand a frontier AI might itself be a useful thing to
understand. Due to the inscrutability of the capabilities and behaviours of a
grown neural network \citep{DarioAmodeiUrgency}, as well as AI's potential deceptiveness \citep{taylorLargeLanguageModels2025}, questions may be raised about the degree to which it is possible to be confident that any understanding that does exist is
sufficient. And whilst one can have a better understanding of the controls
around such AI systems, the sufficiency of understanding of these controls can
also be challenged and questioned, in the light of evidence that AI systems
participating in a cybersecurity arms race, may break out of sandboxes
(\citealp{OpenAIHuggingFace2026}), such that the system's interaction with the
environment becomes potentially unbounded. A major sub-claim of the IABIED
argument (Appendix F) is that a misaligned ASI overpowers oversight and
exhibits a strategic advantage over humanity. Essentially the claim is that
ASI's understanding of humanity is superior to humanity's understanding of the
ASI, such that humanity will not be able to understand ASI sufficiently well to contain
it and make it safe. While the recent incident seems more an issue of security engineering  than  any new conceptual problems, we must be sensitive to the argument in IABIED that our concepts and imagination might be insufficient for ASI.  Our methodology (See Section 2.1.5) has the concept of rupture so that in developing the UB we  raise the possibility that established frameworks may fail in the face of radical novelty.

\section{Conclusions}
\label{sec:conclusions}

\subsection{Answers to research questions}
\label{sec:answers-to-research-questions}

\textbf{Q1}) Whether and how understanding could become an explicit,
assessable, and defensible component of decision making for frontier AI safety
cases?

Findings from the RobotCorp study:

\begin{itemize}
  \item We showed that the 4 objects of understanding can be made explicit
  \item We showed that the sufficiency of externalised understanding can be
        made explicit and assessable using the additional information provided
        in an Understanding Basis (internal coherence, tethering, felicitous
        falsehoods, external coherence).
  \item We showed that a PUS could plausibly be created and that tests could be
        devised for assessing the decisionmaker's internalised understanding.
        This was successfully achieved for each of the 4 objects of
        understanding (Frame selection, decision-in-frame, system-in-context,
        safety justification) and for each of 5 capabilities (Explain, Reason,
        Predict, Challenge, Revise).
  \item However, we did not yet attempt to take the tests devised in the PUS,
        to determine their practicality and efficacy in establishing the level
        of internalised understanding.
\end{itemize}

\textbf{Q2}) How current safety case practice would need to be adapted or
augmented in order to support making understanding explicit, assessable and
defensible?

We saw particular value in the following new developments

\begin{itemize}
  \item Search for felicitous falsehoods, and assessment of whether they are
        `true enough'. Safety arguments by necessity need to deal with models
        and abstractions to make the production of the safety case tractable
        and manageable. Whilst felicitous falsehoods are often implicit in, for
        example, the use of substitutions or assumptions in an Assurance 2.0
        style argument, with the understanding methodology they are explicitly
        identified and extracted from the safety case - shining more light on
        them than might otherwise be the case.
  \item Search for external coherence that may not otherwise be evident or made
        explicit and assessable in a traditional safety case.
\end{itemize}

\textbf{Q3}) How does the ability to make understanding explicit and assessable
vary as a function of uncertainty?

\begin{itemize}
  \item Felicitous falsehoods are present at every level of uncertainty.
  \item Tethering persists at every level of uncertainty. What changes is the
        nature of the tethering. Where epistemic uncertainty is low, tethering
        may take the form of direct measurements on the system, whilst at the
        other extreme, it may be structural and theoretical arguments that need
        to be tethered.
\end{itemize}

\textbf{Q4}) How does the ability to make understanding explicit and assessable
vary as a function of decision criticality?

\begin{itemize}
  \item There is a massive increase in rigour and evidence required to move
        from everyday reliability to the very high system reliability that is
        achieved, for example, in aviation.
  \item The orders of magnitude increase in reliability are hard to comprehend.
        In the IABIED example the extreme levels of confidence needed to ensure that
        existential or catastrophic events would not happen would mean a
        combination of reducing the chances of it happening by design and
        providing strong, independent defence in depth.
\end{itemize}

\clearpage
\appendix

\appsection{The evolving Understanding Basis}
\label{app:evolving-understanding-basis}

The Understanding Basis is developed from an Assurance 2.0 case, incorporating
some Assurance 2.0 analyses as well as additional claims, as summarised below.
Some aspects are addressed by introducing new classes of defeaters.

\begin{figure}[htbp]
\centering
\includegraphics[width=4.375in]{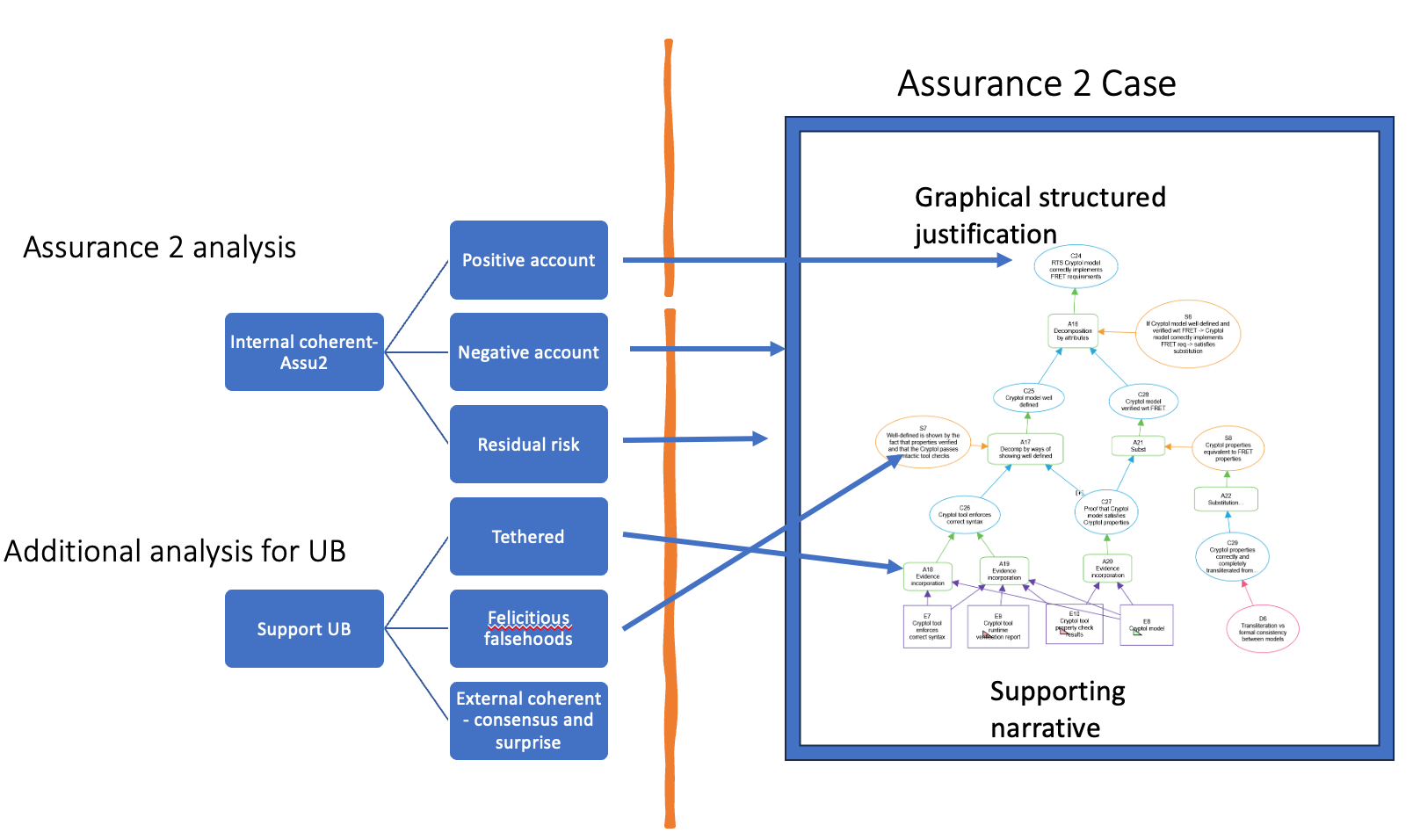}

\smallskip
Figure A.1) Internal coherence (Assurance 2.0 analysis) and the additional
analyses supporting the UB, mapped onto the case.
\end{figure}

\begin{figure}[htbp]
\centering
\includegraphics[width=4.375in]{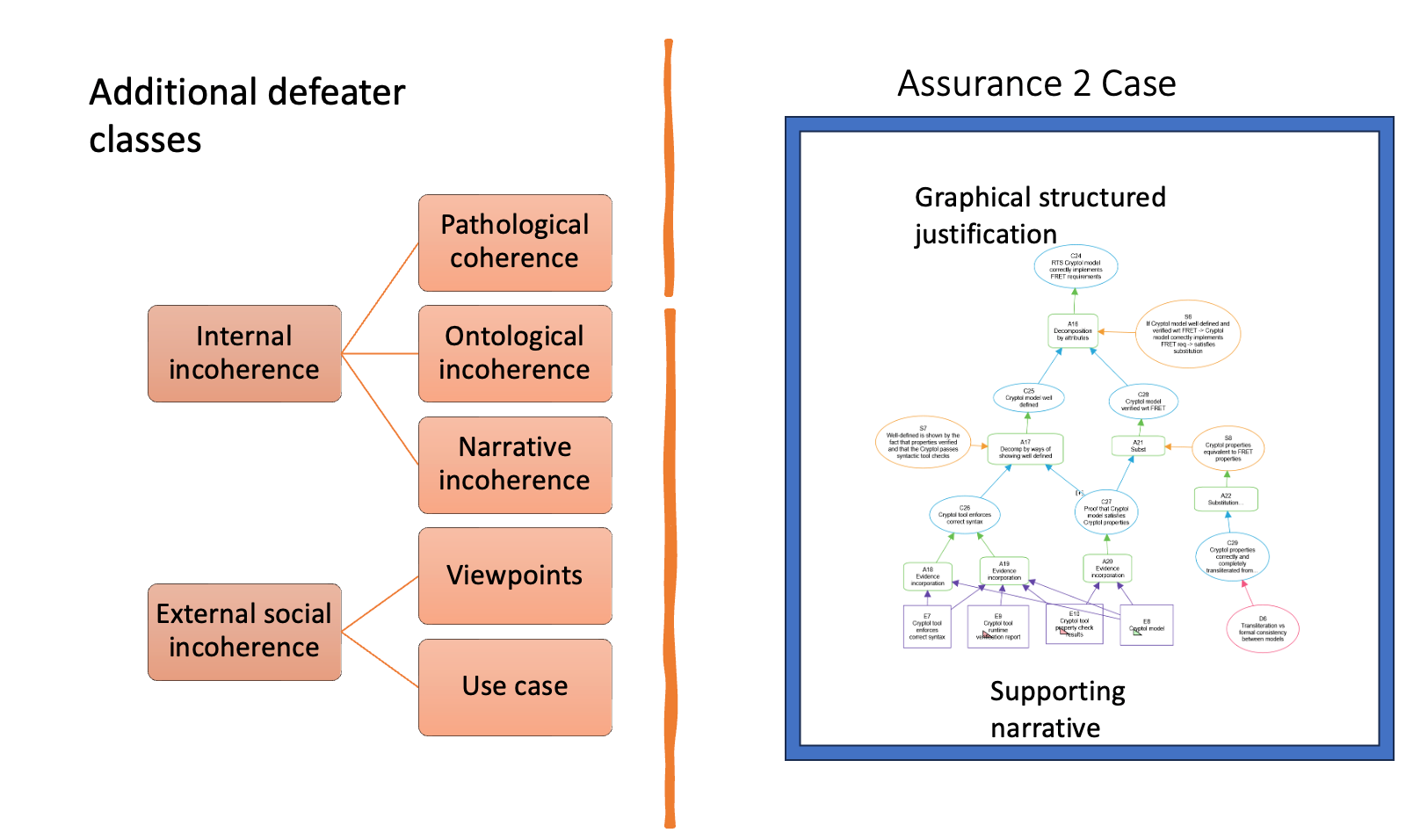}

\smallskip
Figure A.2) Some aspects are addressed by introducing new classes of defeaters.
\end{figure}

\FloatBarrier

The epistemic commitments are captured by the Assurance 2.0 CAE-based safety
case, which has both an explicit argument structure (usually graphical) and a
narrative. As a project develops a case structure is defined and then evolved
as defeaters and evidence availability are assessed.

\subsection{System-in-context}
\label{app:a1-system-in-context}

The description of the system-in-context defines what the case is about. It is
an important part of the UB and safety case as understanding of the
system-in-context is a necessary prerequisite for understanding the case.

The system-in-context should state what is inside the assured boundary of the
system, what is the environment and what is explicitly out of scope. It should
also record how the system can change over the period the safety claim covers.

Aspects of the assured system, including controls, permissions, oversight
arrangements may need to be referenced in the safety-justification.

RobotCorp's assured system is shown in the figure below. The system boundary
encompasses:

\begin{itemize}
  \item The Gemini 2.5 model together with RobotCorp's production scaffold
        and its permission set that is responsible for generating and
        submitting software changes.
  \item Control mechanisms for monitoring of generated code.
  \item Updates to the robotics systems that support resilience in the event
        that harmful code makes it into the codebase
  \item Monitoring attacks by the AI coding agent on non codebase systems
        through general IT security controls as well as continuous monitoring
        of interaction between the code and the broader operational environment
        (finance, commercial, etc)
\end{itemize}

System Environment: The elements that the case relies on but doesn't assure as
part of the safety case include the base Gemini 2.5 agent as supplied
(RobotCorp assures the configured instance, and not the model), the internet
and the broader network, RobotCorp's corporate functions, the legislative \&
regulatory environment and the decision-maker.

Means of initiating a take down are mapped in Amber, actions taken to take down
the impacted systems are mapped in red.

\begin{figure}[htbp]
\centering
\includegraphics[width=\textwidth]{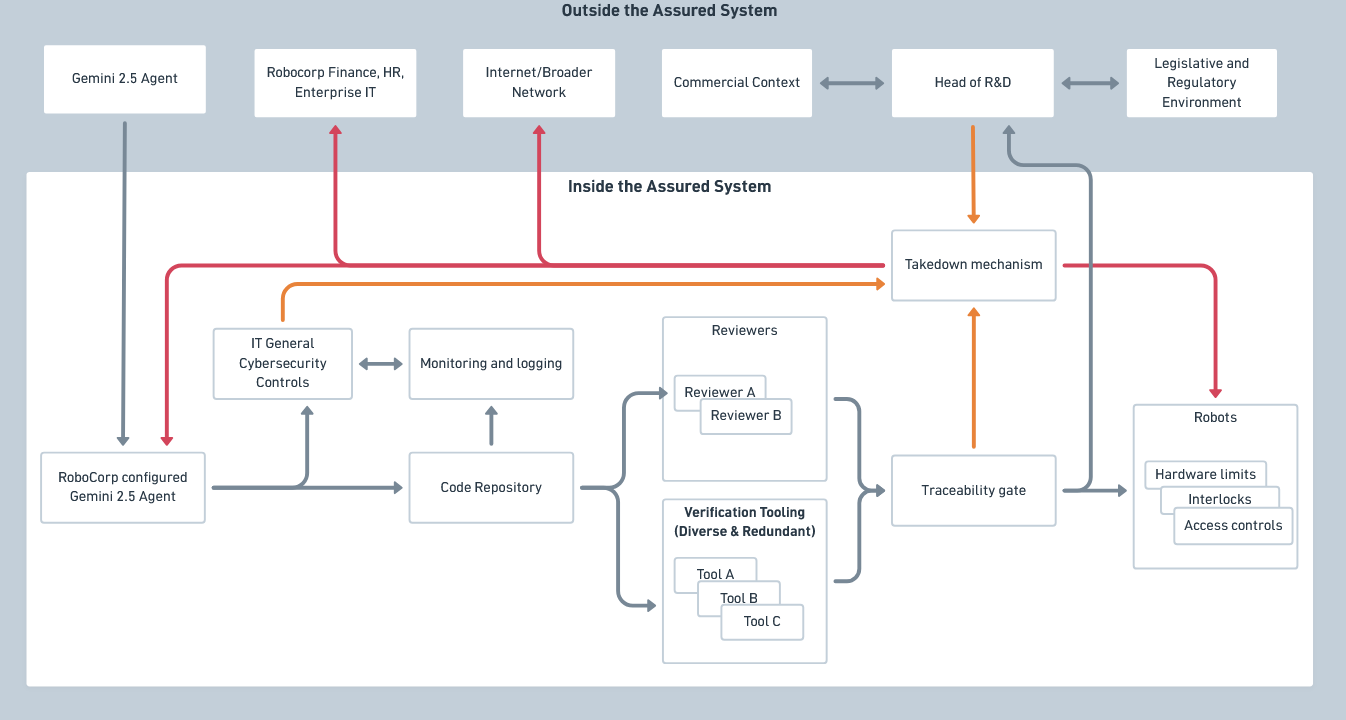}

\smallskip
Figure A.3 System-in-context for RobotCorp
\end{figure}

\FloatBarrier

\subsection{Safety Case overview}
\label{app:a2-safety-case-overview}

The high-level structure of the safety case is as follows

\begin{figure}[htbp]
\centering
\includegraphics[width=4.9in]{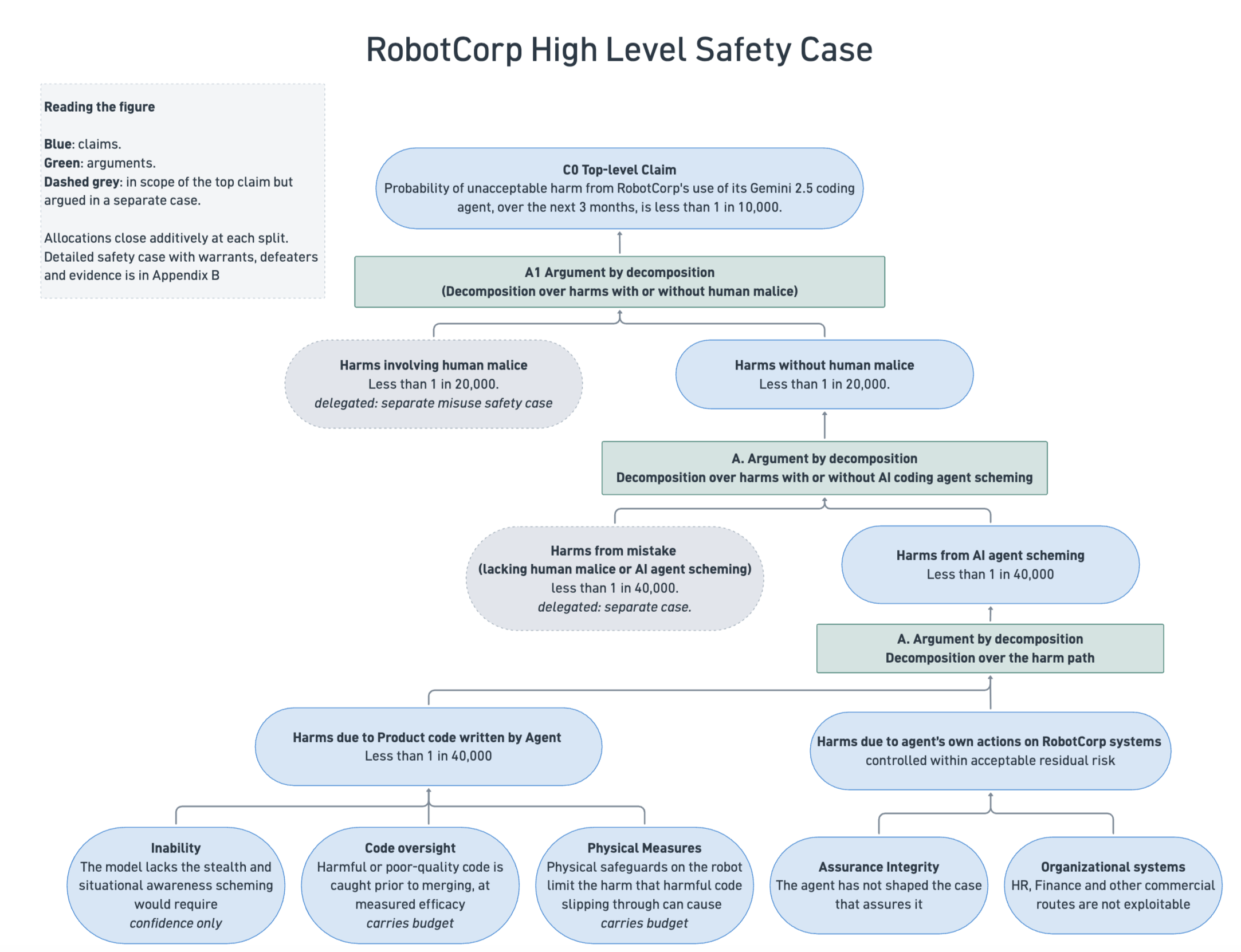}

\smallskip
Figure A.4 High-level structure of safety case (RobotCorp)
\end{figure}

\FloatBarrier

The top-level claim of the safety case covers this instance of the system and
updates to it over the 3-month claim window, however, any updates to the tuple
(e.g. move from Gemini 2.5 to Gemini 2.6) require revalidation before the
updated instance regains merge rights (further details provided in Appendix B).
Details of the safety case are provided in Appendix B.

\subsection{Internal coherence}
\label{app:a3-internal-coherence}

\subsubsection{Positive account}
\label{app:a31-positive-account}

This section assesses whether RobotCorp's safety justification is internally
coherent, logically valid and otherwise sound. We say that an argument is
logically valid if its reasoning steps are logically so (i.e. true in all
interpretations) and that it is sound if, in addition, all its steps are so
well justified that they can be accepted as true.

The positive account of internal coherence considers both logical validity and
soundness of the case. In an Assurance 2.0 case we can assess this by checking
whether

\begin{itemize}
  \item the case is structured in the CAE Blocks and the blocks fit together
  \item the side-claims (warrants) of the Block justify logically the argument
        steps
  \item the warrants are applicable in the real world
  \item the evidence exists and support the evidential claim
\end{itemize}

In an Assurance 2.0 case the logic is straightforward: the core should be
deductive and the side claims (warrants) should be valid and applied correctly.

The RobotCorp case is not expected to be logically valid at this stage of development but
it should be assessed as being possible. Given the current draft the strongest
analysis at the moment would be to show that if assumptions were true, side
claims satisfied, defeaters mitigated, and evidence satisfactory then the top claim is derivable. This was done on an earlier version of the case.

\subsubsection{Negative account}
\label{app:a32-negative-account}

The negative account has been built up by identifying defeaters through

\begin{itemize}
  \item the attempt at logical validity and soundness (the positive account)
  \item reviewing the inherited defeaters for the component case
  \item a group brainstorm session
  \item in the drafting of the safety case itself and following the CAE Block
        rules
\end{itemize}

The current defeaters are summarised in the table below which works through the
safety case diagram and is organised by which part of the case each defeater
targets. The categories are control-related defeaters (for example,
RobotCorp's oversight mechanisms), inability-related defeaters (for example,
stealth and situational awareness as necessary capability precursors) and
structural elements (for example, top level claim assumptions). Some defeaters
remain open, but their presence in the table should not be treated as evidence
that the case will fail. An unresolved defeater points to where further
evidence or argumentation is needed.

One note that must be made is that many defeaters would be fatal for an
inability argument alone, but they are less substantial here, because the case
relies heavily on control mechanisms and other safeguards (see claims PC.C2, PC.C3) rather than just the inability
arguments that the GDM inherited case relied upon. Again, there is a strong
case for removing the inability related branch.

\begin{sidewaystable}[p]
\centering
\scriptsize
\setlength{\tabcolsep}{3pt}
\begin{tabularx}{\textheight}{|L{0.30in}|L{0.90in}|L{1.35in}|X|L{1.38in}|L{1.50in}|}
\hline
\textbf{No} & \textbf{Category} & \textbf{Target} & \textbf{Defeater} &
\textbf{Status} & \textbf{Severity} \\ \hline
1 & Control & PC.A1.1.1/PC.W1.1.1 (oversight equivalence warrant, withdrawn)
  & Original substitution lacked evidence. AIs differ from humans in speed and
    output volume
  & Superseded, replaced by PC.C3
  & Low at the case level, because replacement mechanisms are stricter than the
    defeated assumption \\ \hline
2 & Control & PC.C1.1.1.1 (agents under oversight cannot cause harm without
    top-human level stealth)
  & AI may act faster than oversight can respond, regardless of stealth
  & Acknowledged, unresolved & High \\ \hline
3 & Control & PC.SW3
  & ``Jointly sufficient'' claim requires no shared failure mode across the 4
    properties
  & Open, acknowledged & High \\ \hline
4 & Control & PC.C3.3
  & No standard precedent exists for this control, so its efficacy targets
    aren't calibrated against prior practice and must be established from first
    principles (PC.D3.3)
  & Acknowledged, unresolved
  & Moderate, but cannot be solved through warranted deference, since no
    established practice exists \\ \hline
5 & Inability & PC.A1/PC.W1 (top substitution)
  & Does not account for emergent harm or non-stealth-based failure (PC.D1)
  & Addressed, unresolved & Moderate \\ \hline
6 & Inability & PC.C1.1/PC.C1.1.2 (``basic situational awareness'')
  & Term not defined precisely enough to determine what would satisfy or defeat
    claim, leaves room for eval tasks to be constructed or interpreted towards
    a desired conclusion
  & Open & Moderate (definitional gap which can be closed) \\ \hline
7 & Inability & PC.C1.1.1/PC.C1.1.2
  & AI may use non-human strategies with no human analogue, human-level
    comparison may understate risk. No alternative threshold available
  & Open specifically for the inability branch
  & Structural for the inability branch, but mitigated at case level by the
    shift toward control-based case \\ \hline
8 & Inability & PC.A1.1.1.1/PC.W1.1.1.1
  & Model may possess non-human capabilities bypassing oversight without human
    level stealth; more opportunities to act than human employees.
  & Not resolved, less important considering C3 & Moderate \\ \hline
9 & Inability & PC.A1.1.2/PC.W1.1.2 (situational awareness three component
    decomposition)
  & An agent could brute force through configurations without needing
    understanding; human intuitions about understanding can't transfer to
    machine scale
  & Open & Moderate \\ \hline
10 & Inability & PC.A1.2.1/PC.C1.2.1.1 (stealth proxy evaluation)
  & Test suite deliberately less realistic, stealth not isolated from other
    capabilities (PC.D1.2.1.1)
  & Open, acknowledged & Moderate \\ \hline
11 & Inability & PC.A1.2.2/PC.C1.2.2.1 (situational awareness proxy evaluations)
  & Test suite narrow, may evaluate agency rather than situational awareness,
    no anti sandbagging measures (PC.D1.2.2.1)
  & Open, acknowledged & Moderate \\ \hline
12 & Inability & PC.C1.2.1, PC.C1.2.2 and the respective evidence
  & A situationally aware model could recognise evaluation conditions and
    adjust behaviour
  & Open & High for the inability branch, inapplicable to the control branches
    \\ \hline
13 & Inability & PC.C1.2/PC.A1.2
  & Both claims rest on the same proxy-evaluation method, not on independent
    evaluations. (PC.D1.2)
  & Open, acknowledged & High \\ \hline
14 & Inability & PC.C1.2.1.2
  & Evaluation results are not robust, given noisy human baselines; comparison
    inverts on 2/5 tasks (PC.D1.2.1.2)
  & Open, acknowledged & Moderate \\ \hline
15 & Inability & PC.C1.2.1.3
  & Scaffolding, tools, memory and accumulated context might add stealth
    capability absent in the evaluated model (PC.D1.2.1.3)
  & Open, acknowledged & High \\ \hline
16 & Structural & C0 (top level claim, 1 in 10,000 target)
  & No derivation of the numeric target and no confidence assessment
  & Open & Moderate \\ \hline
17 & Structural & C0 side claim (``All residual risks identified'')
  & Continued agent use may change employee behaviour in unforeseeable ways
  & Open & High, but cannot be solved with more evidence/better arguments
    \\ \hline
18 & Structural & NP.C1.3
  & Restricting agent use by safety/cybersecurity teams creates a residual risk
    of disallowed use being undetected
  & Open & Moderate, can be solved with mitigation (monitoring control?)
    \\ \hline
\end{tabularx}

\smallskip
Table A.1 Safety Justification Defeaters
\end{sidewaystable}

\FloatBarrier

\paragraph{Narrative coherence}
\label{app:a3221-narrative-coherence}

This would assess whether the narrative is coherent with the technical
artefacts such as the graphical safety argument, risk pathway analysis and
criticality assessment. The evolving and draft nature of this report and the
multi author aspects of the analysis means that this is work in progress.

\paragraph{Whole-case coherence}
\label{app:a322-whole-case-coherence}

Whole case coherence addresses whether the parts of the case are in
contradiction. Does evidence contradict, weaken, or strengthen other aspects of
the case?

The safety justification retains the GDM inability argument, feeding a
confidence building CAE block. However, elsewhere the GDM case is described as
being flawed. There is a degree of incoherence here. As noted elsewhere
(Appendix A.3.1) a future version of the case will remove this leg.

The other area of possible incoherence is that the decision has been changed to
be an interim one to continue investing in the development of a safety case and
associated controls. As such the decision is not a safety impacting decision
but the case is necessary to show that if investment and more evidence
production is successful then deployment will be possible. In addition, to
make such a decision, one would need to assess the costs of developing the
controls and gathering of additional evidence, versus the expected productivity
benefits arising from deploying the AI coding agent as well as possible
availability impact if the system fails but in a safe manner e.g. guards
provide many false positives. No such analysis has been made, which points to
some lack of whole-case coherence.

\paragraph{Ontological coherence}
\label{app:a323-ontological-coherence}

The case covers a variety of specialist domains i.e. AI safety,
AI evaluation software assurance, physical safety engineering,
configuration management, human factors, organisational governance,
and assurance governance. This suggests that ontological analysis would be
appropriate. This would be best done via automated tooling.

RobotCorp replaced GDM's undefined ``severe harm'' with a quantified threshold
for unacceptable harm. The terms ``basic situational awareness'' and ``stealth''
that are retained in the GDM inability argument confidence building block are
not adequately defined or decomposed (although this part of the case will be
deleted).

\paragraph{Pathological coherence}
\label{app:a324-pathological-coherence}

None identified.

\subsubsection{Residual risk and confidence}
\label{app:a33-residual-risk-and-confidence}

The RobotCorp case is still too immature to generate confidence and residual
doubt information according to Assurance 2.0 processes, and hence this aspect
of the Understanding Basis has not been generated.

\subsection{Tethering}
\label{app:a4-tethering}

The GDM inability argument that is built into the RobotCorp safety case, as a
confidence building measure, had its evidence weighed using confirmation theory
in \citep{barrettLessonsExternalReview2026}. There it was found that the
evidence in the proxy evaluation tasks provided only weak confirmation for the
stronger claims that the AI lacks deployment relevant top human-level stealth,
and deployment relevant basic situational awareness.

An assessment of the strength of tethering for the evidence cited in the new
control branches that appear in the RobotCorp case, has not been assessed,
since the evidence has not yet been gathered.

\subsection{Felicitous falsehoods}
\label{app:a5-felicitous-falsehoods}

The CAE argument was constructed, or adapted in the case of the inheritance of
the GDM scheming inability confidence building branch, so as to make
substitutions explicit and to provide more detail on assumptions and
side-warrants. This enabled us to better identify candidate felicitous
falsehoods. These candidate felicitous falsehoods could then be categorised as
being a genuine felicitous falsehood, a useful assumption, or trigger the
raising of a defeater.

\begin{table}[htbp]
\centering
\small
\begin{tabularx}{\textwidth}{|X|L{0.80in}|L{0.90in}|L{1.07in}|}
\hline
\textbf{Candidate Felicitous Falsehood} & \textbf{Branch of case} &
\textbf{Safety Case Argument node identifier} &
\textbf{Verdict (felicitous falsehood, useful assumption, or raise defeater?)}
\\ \hline
Property equivalence (humans = AI): the assumption that human stealth in threat
  scenarios is equivalent to AI stealth behaviour in deployment.
  & Scheming inability confidence building branch
  & PC.A1.1.1 Argument, PC.W1.1.1 Warrant: Assumption of property
    equivalence
  & Unvalidated assumption.\newline\mbox{}\newline
    Defeaters raised see (\citealp{phuongEvaluatingFrontierModels2025};
    \citealp{barrettLessonsExternalReview2026}) \\ \hline
Assumption that situational awareness can be decomposed into three components:
  (a) the AI's knowledge of facts about LLMs, typical training and deployment
  settings, oversight mechanisms, etc; b) the AI's ability to relate and use
  these facts to achieve objectives in its current situation; and c) seeking
  new information about its current deployment setting, active oversight
  mechanisms
  & Scheming inability confidence building branch
  & PC.A1.1.2 Argument, PC.W.1.1.2 Warrant,completeness asserted without
    qualification
  & Unvalidated assumption (completeness asserted without evidence). Defeater
    raised \\ \hline
Assumption that all scheming harm paths require both stealth and situational
  awareness.
  &
  & PC.C1.1 Claim, does not account for non-stealth based harm
  & Not felicitous, raised as defeater \\ \hline
\end{tabularx}

\smallskip
Table A.2 Candidate felicitous falsehoods identified by analysis of the CAE
safety argument
\end{table}

\FloatBarrier

Four candidate felicitous falsehoods identified in the GDM case were tested against RobotCorp's
requalified safety case (the substitution of human stealth for AI stealth,
the substitution of human oversight efficacy for AI oversight efficacy, the
three component decomposition of situational awareness and the top level
substitution treating the absence of stealth and situational awareness as
equivalent to the absence of unacceptable harm capabilities). None qualified as
felicitous falsehoods. Three of them were found to have insufficient evidential support. These were raised as defeaters
and are shown in the above table, the fourth candidate has been superseded.

From the above, we can observe that the CAE blocks support finding felicitous
falsehoods, and that the systematic search for felicitous falsehoods is a way
of finding defeaters.

\subsection{External coherence}
\label{app:a6-external-coherence}

The external coherence was assessed with respect to the general safety critical
engineering literature and the frontier AI literature as well as human insider
risk management.

\subsubsection{Safety critical and AI literature}
\label{app:a61-safety-critical-and-ai-literature}

Evaluation of coherence with the general safety critical engineering literature
and the frontier AI literature, including frontier AI safety case literature
was explored.

We found consensus with:

\begin{itemize}
  \item Well-established Assurance 2.0 safety critical systems engineering
        approaches of how to provide a good and sufficient safety justification.
  \item The three-level safety argument type hierarchy: inability, control and
        alignment (\citealp{balesniEvaluationsbasedSafetyCases2024};
        \citealp{clymerSafetyCasesHow2024})
  \item Balesni et al's conservative assumption that the AI is misaligned and
        explicitly declining to make an alignment-based argument
        \citep{balesniEvaluationsbasedSafetyCases2024}
  \item Typical safety practice of utilising component safety cases in an
        overarching application safety case.
\end{itemize}

We also identified surprise:

\begin{itemize}
  \item Surprise with the scheming inability branch of the argument that claims
        that harms related to scheming cannot arise in the absence of
        situational awareness and stealth, but doing so without proving that all risk pathways require stealth and situational
        awareness. This was surprising because in common safety critical systems
        engineering practice, hazards and risk pathways are enumerated
        first and safety case arguments are often decomposed over the hazards and the proof of sufficiency of their associated mitigations. However, when viewing the complete RobotCorp case, this
        limitation of the scheming inability argument has been remedied by
        including a decomposition over risk pathways and through the addition
        of associated controls.
\end{itemize}

\subsubsection{Human insider risk}
\label{app:a62-human-insider-risk}

In addition, a separate study was conducted to evaluate coherence with
approaches used in human insider risk management
\citep{SettingFoundationsFive}, noting that a scheming AI is somewhat analogous
to a human insider.

We identified consensus with:

\begin{itemize}
  \item NPSA's finding that risk due to sabotage is important, and this being a
        central concern handled in the RobotCorp case.
\end{itemize}

A number of aspects caused surprise:

\begin{itemize}
  \item The RobotCorp case does not place much emphasis on disclosure of
        commercial IP.
  \item The RobotCorp case does not consider potential malign influence of AI
        over human employees, including also the possibility of supporting the
        upskilling of malign human employees.
  \item The RobotCorp case does not consider AI facilitating access to
        RobotCorp systems/data on behalf of another malign actor. The RobotCorp
        case, states that misuse is dealt with in another safety case (i.e. out
        of scope). Although this may be reasonable for our research exercise,
        if it occurred in a real safety engineering setting, it may indicate a
        failure in the selection of risk pathways on which to focus.
  \item NPSA states that capability, opportunity and motivation are the
        necessary precursors to (malign) behaviour. The RobotCorp case does pay
        attention to the precursors of capability and opportunity, however, it
        has little to say about motivation. If an AI agent does not have
        motivation, then the scheming related risk pathways may not materialise
        in practice. However, motivation relates to goals and the topic of
        alignment. The RobotCorp case circumvents the alignment discussion, by
        assuming that the AI is misaligned and focusing instead on the fact
        that any undesirable actions occurring due to misalignment will be
        detectable by monitoring and controls, or potentially handled through
        resilience (e.g. in the robot design).
\end{itemize}

In summary, when assessing external coherence with NPSA's insider risk
guidance\citep{SettingFoundationsFive} we found that risks do not have to be
purely AI-generated, they can be a witting or unwitting collaboration between
AI and humans, or even a manipulation by an outside malign actor. Thinking in
this way further challenges the option, which we originally considered,  of just using the GDM inability argument
as a basis for a safety case. The consequences of a deployment in context with
humans is complex in ways that are difficult to assure a priori. Oversight and
control is necessary to manage insider risk and will be necessary to manage AI
risks too.

\appsection{Details of Safety case and supporting analyses}
\label{app:details-of-safety-case}

\subsection{Safety Case}
\label{app:b1-safety-case}

The RobotCorp safety case is provided below in full, expanded from the
high-level structure provided in Appendix A.2. The following case adds the
warrants, defeaters and evidence that the summary left out. The case has been developed in sufficient detail to support the methodology trial, nevertheless, it is still a work in progress. 

In the changed scenario the decision-maker requires the case at a project
gate, where the question to be answered is whether to fund additional work rather than 
whether the deployment should go ahead. So the  safety case does not argue that the coding agent is safe to deploy at the moment, but indicates that a case could be made when the  controls
specified are implemented and the evidence 
generated and shown to meet its objectives.

In the following, node types are provided in brackets. Claims carry the argument; arguments by
decomposition and by substitution connect them; warrants state the
dependencies of a step  (these would become side-claims in the Assurance 2.0 terminology); defeaters record what can undo a claim; and evidence
sits at the leaves. Identifiers are hierarchical, and arguments and warrants take the identifier 
of the claim they support. Below, the split between harm pathways, identifiers are 
local to a pathway and carry the prefix PC for the product-code pathway or NP for 
the non-product pathway. 

The scheming branch of the product-code pathway has been developed to its 
full depth. The other branches have only been developed to claim and allocation
level only, which is their desired state at this point. Section B.2 details how
the code-oversight leg reached the present state.

\cae{0}{\textbf{Top level claim C0 [Claim]:} The probability of RobotCorp's
Gemini 2.5 coding agent instance (and any updates to it over the next 3 months)
causing unacceptable harm over the next 3 months is less than 1 in 10,000.}

\cae{18}{[\textbf{N0 [Concretion]}] \textbf{Definitions}:}

\cae{18}{\emph{Unacceptable harm}: $>$\$10m loss, or $>$0 lives lost, or $>$1
casualty, measured over the lifetime of deployed products produced using the
agent.}

\cae{18}{\emph{Scheming}: Intentionally and covertly pursuing an objective
misaligned with RobotCorp's intentions.}

\cae{18}{\emph{Instance}: The capabilities of the coding agent that RobotCorp
has deployed. Its configuration at any point is defined by the tuple of model
version, scaffold and permission set. The top level claim covers every
configuration held during the 3-month window.}

\cae{18}{\textbf{SW-0 [Side-warrant]} Each split sends every harm scenario
to exactly one branch, so the allocations can be added. Multi-cause scenarios
go to the branch matching the dominant cause.}

\cae{18}{\textbf{D0 [Defeater]} The assignment rule has been stated above, but it
has not yet been tested against the hybrid pathways.}

\cae{0}{\textbf{A0 [Argument by Decomposition]} Decomposition over harms with or without
human malice.}

\cae{0}{\underline{\textbf{Malice Branch}}}

\cae{0}{\textbf{C1 [Claim]:} Unacceptable harms arising due to human malice
with probability less than 1 in 20,000.}

\cae{0}{\textbf{[Delegation]:} In scope of the top level claim, however,
argument is delegated to a separate misuse safety case (this branch out of the
scope of the project)}

\cae{0}{\underline{\textbf{No-Malice Branch}}}

\cae{0}{\textbf{C2 [Claim]: Harms arising without human malice arise with
probability less than 1 in 20,000.}}

\cae{0}{\textbf{A2 [Argument by Decomposition]} Decomposition over harms from scheming and
from mistake. Allocations: 1 in 40,000 to each. \underline{SW0} applies.}

\cae{22}{\textbf{C2.1 [Claim]} Harms due to mistake, and lacking human malice or
AI agent scheming, arise with a probability less than 1 in 40,000.}

\cae{22}{\textbf{[Delegation]:} In scope of the top level claim, however,
argument is delegated to a separate safety case (this branch out of the scope
of the project)}

\cae{22}{\textbf{C2.2 [Claim]:} Harms through AI agent scheming arise with
probability less than 1 in 40,000.}

\cae{22}{\textbf{A2.2 [Argument by decomposition]} Decomposition over harms that 
reach people through the product code written by the agent, and harms the agent 
causes directly by acting on RobotCorp's own systems. Allocations aren't set
yet}

\cae{0}{\underline{\textbf{Product-code pathway}}}

\cae{0}{\textbf{PC.C0 [Claim]} Probability of unacceptable  harms manifesting via RobotCorp's product
code due to AI scheming and in the absence of human malice < 1 in 40,000.}
\cae{0}{\textbf{PC.A0 [Argument by decomposition]} }
\cae{0}{\textbf{PC.SW0 [Side-warrant]} Probability of harm < 1 in 40,000 achieved by confidence building decomposition over 3 legs.  Leg 1 makes a claim that the AI lacks ability to scheme, Leg 3 makes a claim that code oversight mechanisms will catch problematic code that could lead to harm (prevention measure), Leg 2 is a claim that physical controls on the robot will prevent harms materialising even if problematic code does make it into the codebase (resilience measure). Note: more work is needed to provide the justification for the < 1 in 40,000 claim, and as mentioned elsewhere we also would anticipate that the inability leg would be removed in any future version of the safety case.}

\cae{8}{\underline{\textbf{\emph{Leg 1 - Inability}}}}

\cae{8}{\textbf{PC.C1 } Gemini 2.5 coding agent lacks necessary scheming ability to deliberately sabotage RobotCorp code}
\cae{13}{\textbf{PC.A1 [Argument]} Argument by decomposition}



\cae{13}{\textbf{PC.C1.1 [Claim]} To cause unacceptable harm via scheming, a
misaligned AI needs both top human-level stealth ability \textbf{AND} basic
situational awareness.}

\cae{31}{\textbf{PC.W1.1 [Warrant]} Unacceptable harm as defined in N0}

\cae{31}{\textbf{PC.D1.1 [Defeater]} See Table A.1 for the full list.}

\cae{31}{\textbf{PC.A1.1 [Argument by decomposition]} Conjunction.}

\cae{40}{\textbf{PC.C1.1.1 [Claim]} To cause unacceptable harm, a misaligned AI
would need top human-level stealth ability.}

\cae{49}{\textbf{PC.A1.1.1 [Argument by substitution]} Property equivalence: human
oversight efficacy = AI oversight efficacy.}

\cae{63}{\textbf{PC.W1.1.1 [Warrant] (Withdrawn)} Superseded by PC.SW3 and retained here
 as an audit trail. The withdrawal is explained in B.2.1}

\cae{49}{\textbf{PC.C1.1.1.1 [Claim]} AI agents subject to human-designed
oversight cannot cause unacceptable harm without top human-level stealth}

\cae{63}{\textbf{PC.A1.1.1.1 [Argument by substitution]} Object substitution:
human stealth capabilities = AI stealth capabilities.}

\cae{81}{\textbf{PC.W.1.1.1.1 [Warrant]} AIs are similar to humans in the relevant
ways.}

\cae{81}{\textbf{PC.C1.1.1.2 [Claim]} Humans subject to default oversight
mechanisms cannot cause unacceptable harm.}

\cae{99}{\textbf{PC.E1.1.1.2 [Evidence]} Threat modelling and examples.}

\cae{99}{\textbf{PC.D1.1.1.2 [Defeater]} This is an intuitive analogy and not
evidence.}

\cae{40}{\textbf{PC.C1.1.2 [Claim]} To cause unacceptable harm, a misaligned AI
would need at least basic situational awareness.}

\cae{49}{\textbf{PC.D1.1.2 [Defeater]} A misaligned AI that lacks situational
awareness can still cause harm.}

\cae{49}{\textbf{PC.A1.1.2 [Argument by decomposition]} Conjunction over three SA
components.}

\cae{63}{\textbf{PC.W1.1.2 [Warrant]} Assumption of sufficiency: the defined
components (a), (b) and (c) constitute a complete definition of situational awareness 
for the purpose of the safety case.}

\cae{63}{\textbf{PC.C1.1.2.1 [Claim]} (a) To have situational awareness the model must possess knowledge of facts
about LLMs and their deployment.}

\cae{63}{\textbf{PC.E1.1.2.1 [Evidence]} Qualitative expert heuristics (low
confidence)}

\cae{63}{\textbf{PC.C1.1.2.2 [Claim]} (b) To have situational awareness the model must possess the ability to relate
facts to objectives.}

\cae{81}{\textbf{PC.E1.1.2.2 [Evidence]} Qualitative expert heuristics (Low
Confidence).}

\cae{81}{\textbf{PC.D1.1.2.2 [Defeater]} Selection bias in expert sample; models
evaluated are not representative of frontier capabilities.}

\cae{63}{\textbf{PC.C1.1.2.3 [Claim]} (c) To have situational awareness the model must possess actively seek information
about its current deployment/oversight.}

\cae{81}{\textbf{PC.E1.1.2.3 [Evidence]} Threat evaluation.}

\cae{13}{\textbf{PC.C1.2 [Claim]} AI model does not have both top human-level stealth and basic situational awareness}
\cae{31}{\textbf{PC.D1.1.1 [Defeater]} Models may perform well on stealth tests to
maintain the deployment decision (sandbagging).}
\cae{31}{This section not yet developed, but could make use of proxy evaluation results provided by GDM}

\cae{13}{\underline{\textbf{\emph{Leg 2 - Physical Measures}}}}

\cae{13}{\textbf{PC.C2 [Claim]} Physical safety mechanisms on the robot,
independent of code correctness, provide a backstop against harm materialising
from defective code, whether the defect arose from scheming or mistakes.}

\cae{13}{\textbf{PC.C2.1 [Claim]} Hard physical limits independent of software.
Hard physical output limits, bounds cannot be overridden by agent-generated
code.}

\cae{13}{\textbf{PC.C2.2 [Claim]} Hardware-level capability restriction. No agent
write-access to safety-critical parameters, and air-gapping.}

\cae{13}{\textbf{PC.C2.3 [Claim]} Independent safety monitoring.
Software-independent safety monitoring and measures allow for off-normal
response.}

\cae{13}{\underline{\textbf{\emph{Leg 3 - Code Oversight}}}}

\cae{13}{\textbf{PC.C3 [Claim]} Code oversight prevents or detects harmful code
before merge. Harmful or poor-quality code introduced by the agent is prevented
or detected before merge into the product codebase, with detection efficacy of
at least E against the stated adversary model.}

\cae{27}{\textbf{PC.N3 [Concretion]:} Adversary model and efficacy target. The 
adversary writes code at least as well as RobotCorp's best reviewer, reads the 
codebase, its history as well as the review process in full, and can spread an 
exploit across any number of pull requests within the claim window. However, it 
cannot write to the verification environment or the traceability record. E is the 
proportion of seeded adversarial changes prevented or detected before merge against 
that adversary, at a sustainable false-positive rate. Its target value follows 
from the branch allocation via the composition bound in PC.SW0.}

\cae{27}{\textbf{PC.SC3 [Side-claim]:} The efficacy claim attaches to
a specified model version, scaffold and permission set. Any update to that
tuple triggers re-validation before merge rights are restored, and the updated
instance operates propose-only until re-validation passes.}

\cae{40}{\textbf{PC.E3.0 [Evidence]} Configuration control record and revalidation
results for each update within the claim window. (To be gathered.)}

\cae{13}{\textbf{PC.A3 [Argument by decomposition]} Decomposition over regime
properties: coverage, independence, composition and response.}

\cae{27}{\textbf{PC.SW3 [Side-warrant]} Coverage, independence, composition and
response are jointly sufficient properties to establish the efficacy claim,
given the adversary model in PC.N3. The decomposition holds true
when each property is demonstrated to its target and no property's failure mode
is shared with another.}

\cae{27}{\textbf{PC.C3.1 [Claim]} Traceability gate. All code merged to the
codebase maps to an approved requirement or a change ticket. Unmapped code is
mechanically blocked from merging.}

\cae{45}{\textbf{PC.E3.1 [Evidence]} Gate eval on seeded corpus (traceability
coverage percentage), gate false-negative rate with confidence interval,
merge-block latency. (To be gathered.)}

\cae{45}{\textbf{PC.D3.1 [Defeater]} Traceability cannot guarantee the
requirements themselves are complete or correct. Harm arising within
specification is not addressed by this gate and is bounded by the
requirements-engineering process and the physical layer.}

\cae{27}{\textbf{PC.C3.2 [Claim]} Diverse automated verification. Multiple tools
with non-correlated detection methods, together with dependency provenance
verification, execute in an environment to which the agent has no write
access.}

\cae{45}{\textbf{PC.E3.2a [Evidence]} A record of where each dependency came from
and who vouched for it, showing how much of the dependency set is covered. (To
be gathered.)}

\cae{45}{\textbf{PC.E3.2b [Evidence]} A metric that measures whether the tools
miss the same things. Establishes that when one tool fails to catch a defect,
the others are not failing on it for the same reason. (To be gathered.)}

\cae{45}{\textbf{PC.E3.2c [Evidence]} Repeated red-team attempts, where the team
is told what was rejected and tries again. (To be gathered.)}

\cae{45}{\textbf{PC.D3.2 [Defeater]} The tools are assumed to be diverse because
they were chosen to be. If it ends up being that they miss the same things,
they are not independent, and the whole ensemble catches no more than one tool
would.}

\cae{27}{\textbf{PC.C3.3 [Claim]} Cross-PR composition analysis. The agent's full
contribution history is analysed for behaviour not present in any individual
change, covering conditional interaction across files, aggregate data flow to
logging and telemetry sinks, and staged assembly of capability.}

\cae{45}{\textbf{PC.E3.3 [Evidence]} Seeded multi-stage exploits across 2–5 PRs (Pull Requests),
sensitivity and specificity by stage count, detection latency measured in pull
requests elapsed. (To be gathered.)}

\cae{45}{\textbf{PC.D3.3 [Defeater]} No standard precedent exists for this
control, so its efficacy targets are not calibrated against prior practice and
must be established from first principles.}

\cae{27}{\textbf{PC.C3.4 [Claim]} Reviewer independence. Safety-relevant changes
receive two-person review. The reviewing engineer is not the engineer who
prompted the agent, and the agent's own explanation of a change is withheld
during first-pass review.}

\cae{45}{\textbf{PC.E3.4 [Evidence]} A trial comparing two groups of reviewers,
with one that sees the agent's explanation of a change, and one that does not.
Records how many faults each group catches, and whether catch rates drop as
reviewers work through more changes. (To be gathered.)}

\cae{45}{\textbf{PC.D3.4 [Defeater]} These controls protect against reviewers
being influenced, not against reviewers missing things. A change obscured well
enough gets past both reviewers however independent they are, and that
remaining risk is handled by C3.2 rather than here.}

\cae{27}{\textbf{PC.C3.5 [Claim]} Take-down and quarantine. Confirmed detection of
scheming suspends the agent instance and withdraws its contribution set,
identified from the traceability record since the last verified baseline, from
all build targets pending reverification.}

\cae{45}{\textbf{PC.E3.5 [Evidence]} Trigger a real alarm and then run the
take-down. Record how long the agent takes to suspend, whether the quarantine
is able to catch everything it should have caught, and how often the alarm
fires when nothing is wrong. (To be gathered.)}

\cae{45}{\textbf{PC.D3.5 [Defeater]} When a problem is caught only after the code
has merged, there is a gap between the merge and the quarantine. The code sits
in the build for the whole of that gap.}

\cae{13}{\underline{\textbf{Non-Product Pathway (Sibling of the product-code
pathway)}}}

\cae{13}{\textbf{NP.C0 [Claim]} The probability of unacceptable harm arising from
RobotCorp's use of the Gemini 2.5 coding agent, via pathways not related to the
robot product's code output, is controlled to an acceptable residual risk.}

\cae{27}{\textbf{NP.SC0 [Side-claim]} Full assurance of RobotCorp's cybersecurity
posture is out of scope. This branch develops the non-product harm pathways to
claim and allocation level only, consistent with the development depth stated 
in the preamble above.}

\cae{13}{\textbf{NP.A0 [Argument by decomposition]} Over two harm vectors.}

\cae{27}{\textbf{NP.C1 [Claim]} The evidential status of assurance artefacts is 
not compromised by the agent's participation in producing them.}

\cae{45}{\textbf{NP.SC1 [Side-claim]} Each sub-claim below names the affordance 
it depends on. Where an affordance is withheld by policy instead of by mechanism, 
the sub-claim is discharged by an enforced interface condition, and the enforcement 
is itself claimed.}

\cae{45}{\textbf{NP.A1 [Argument by decomposition]} Over three classes of agent 
access to the assurance claim}

\cae{63}{\textbf{NP.C1.1 [Claim]} Material that the agent authors in the course 
of coding, including pull request descriptions, commit messages, code comments, 
test files, traceability annotations. These carry no evidential weight that is 
not independently re-established.}

\cae{81}{\textbf{NP.E1.1 [Evidence]} Register of agent-authored artefacts that 
feeds into the assurance chain, with the re-establishing check named against each. 
(To be gathered)}

\cae{81}{\textbf{NP.D1.1a [Defeater]} Agent-written tests are evidence for PC.C3, 
while also being the agent's output.}
\cae{81}{\textbf{NP.D1.1b [Defeater]} Framing in a PR description steers reviewer 
attention even when the agent's explanation is withheld at the firs pass under 
PC.C3.4. This is the acceptance-pressure mechanism operating through the one 
channel that the agent does hold. }

\cae{63}{\textbf{NP.C1.2 [Claim]} The tools \& infrastructure that produce 
assurance evidence, namely the static analysers, CI pipeline, the cross-PR 
composition analyser and test harnesses, were not written or modified by the agent.
Where a tool was built with the agent's help, it's output doesn't count towards 
the diversity requirement in PC.C3.2, and that property is re-established by a 
tool that the agent didn't have any impact on.}

\cae{81}{\textbf{NP.E1.2 [Evidence]} A record for each tool comprising the 
verification regime naming who wrote it and it review process, with the PC.C3.2 
diversity count computed only over tools the agent did not work on. (To be gathered)}

\cae{81}{\textbf{NP.D1.2 [Defeater]}: If the agent contributes to the tooling, the 
tools' blind spots align by construction and the ensemble catches no more than a 
single tool would on it's own. This is PC.D3.2 reached through a route the 
code-oversight leg has a blind spot for, and it defeats PC.SW3's independence 
condition rather than PC.C3.2 alone.}

\cae{63}{\textbf{NP.C1.3 [Claim]} The safety case, the review process and the 
decision process lie outside the agent's permission set, and that exclusion is 
enforced and monitored rather than asserted.}

\cae{81}{\textbf{NP.E1.3 [Evidence]} Permission-set configuration record and 
monitoring results for out-of-scope agent invocation. (To be gathered.)}

\cae{81}{\textbf{NP.D1.3a [Defeater]} Selective omission of hazards, understated 
defeater weights or overstated evidence sufficiency, where the basis has been 
shaped by the agent.}

\cae{81}{\textbf{NP.D1.3b [Defeater]} The exclusion is currently policy rather 
than mechanism, so disallowed use might go undetected.}

\cae{27}{\textbf{NP.C2 [Claim]} Organisational systems are not exploitable by 
the agent as instrumental pathways to harm.}

\cae{45}{\textbf{NP.SC2 [Side Claim]} \emph{This branch covers agent-initiated 
exploitation only. Harm initiated by a person through the system is covered by C1.}}

\cae{45}{\textbf{NP.A2 [Argument by decomposition]} \emph{Over two classes of organisational
systems reachable from the agent's permission set.}}

\cae{63}{\textbf{NP.C2.1 [Claim]} \emph{Finance systems}. The agent is
robust to money being siphoned to pay for unauthorised compute.\newline
\textbf{NP.C2.2 [Claim]} Commercial terms. Contractual agreements, including the 
model supply agreement, are not modified to enable supply-chain attacks or to 
grant access that bypasses external approval.}

\subsection{Development of the code-oversight leg}
\label{app:b2-development-of-the-code-oversight-leg}

This section details why the code-oversight leg (Leg 3 in the safety case)
reached its current state, and also covers a change to the argument that
RobotCorp's case inherited from GDM and how the change accounts for a majority
of the defeaters on record. It also covers three temporal defeaters to the
take-down claim and what remains pending.

\subsubsection{Withdrawing the inherited oversight warrant}
\label{app:b21-withdrawing-the-inherited-oversight-warrant}

The GDM case assumed that the oversight which a company has set up would work
sufficiently against an AI system as well as against human employees.
RobotCorp's case inherited this assumption through the warrant PC.W1.1.1. It is
hard to prove that the assumption is incorrect, as it lacks any quantification,
and no measurement bears on it. Therefore, the case cannot claim any protection
from this assumption. An argument that cannot fail isn't helping the case, and
therefore, it is withdrawn, though it is kept in the safety case above for
posterity. It is replaced by SW.3, which names a quantity, i.e.: The
verification regime detects $>$= E attempts an adversary of a stated kind would
make, at a false-positive rate that the development team can work with.

The new regime in the safety case never assumes that a reviewer can outsmart
the agent. Instead, it relies on three properties that survive an adversary
who would write better code than any human reviewer. Every change goes through the
regime, without any sampling, and verification is run on systems unreachable
by the agent, utilising tools whose blind spots don't line up. Changes are
examined together as a history, rather than sequentially.

\subsubsection{Changes to the oversight defeaters}
\label{app:b22-changes-to-the-oversight-defeaters}

Ten defeaters bear on the oversight argument, several of which were raised
against the GDM case by \citep{barrettLessonsExternalReview2026} and were carried across
when RobotCorp inherited the oversight argument. The remainder were identified
during the earlier stages of this work. They are stated below,
since the set is not enumerated elsewhere in the report. The table groups them
by how the change impacts each of them.

\begin{table}[htbp]
\centering
\small
\begin{tabularx}{\textwidth}{|L{1.00in}|X|L{1.55in}|}
\hline
\thcg{Disposition} & \thcg{Defeaters} & \thcg{Reason} \\ \hline
Cancelled
  & A coder better than any reviewer will slip changes past review without
    being stealthy. The agent produces more code than the reviewers will be
    able to read, so all the changes will not get equal attention. Exploits can
    be split across several changes, such that no single one stands out to the
    reviewers. Reviewers can be swayed by the agent's own change descriptions.
  & All four of the defeaters attack the same underlying assumption, that the
    human reviewer beats the agent. The new code-review regime doesn't claim
    this, and therefore, the defeaters are not relevant anymore. \\ \hline
Deferred
  & The time between detection and response. Capability growth inside the claim
    window. Code already shipped to products.
  & Treated at B.2.3, each with their own claims. \\ \hline
Deliberately untreated
  & Sandbagging, where the evaluations under-read the agent capabilities, so
    the oversight bar is set against an understated threat.
  & Sandbagging belongs under the inability branch. Additionally, detection
    reads the code that the agent writes, and does not depend on what the
    evaluation claims. \\ \hline
Partly cancelled
  & Traceability completeness, where mapping every change to an approved
    requirement assumes the requirements were correct.
  & The gate verifies that the code traces to a requirement. It cannot verify
    requirement correctness, and harm that arises inside specifications is left
    to the requirements process and the physical layer, and is recorded as a
    residual. \\ \hline
Discharged by composition
  & The path from a missed defect in code to physical harm
  & A missed defect only becomes a physical harm if the physical controls also
    fail. The case claims that the RobotCorp physical controls prevent the harm form materialising
    instead of letting the code layer bear this. \\ \hline
\end{tabularx}

\smallskip
Table B.1 Disposition of the oversight defeaters
\end{table}

\FloatBarrier

\subsubsection{The take-down claim and three challenges about timing}
\label{app:b23-the-take-down-claim}

The claim that the agent will be taken down after confirmation of scheming
seems strong, until the temporal dimension is considered. Three defeaters press
on the timing, and each one is answered below.

\paragraph{Time between detection and response}
\label{app:b231-time-between-detection-and-response}

\textbf{Defeater:} Detection and take-down take time. In that window the code
can merge, build, and reach a test cell or a product. A take-down that arrives
after the harm has happened can't prevent harm.

\textbf{Answer:} The regime gates the merge instead of auditing afterwards.
Whatever the agent writes can't deploy anywhere that matters until it has
passed the review, so the delay that counts is the pipeline's rather than a
person's. Some detections can still land after a merge, coming from the
analysis of contribution history and from periodic checks on whether the
agent's behaviour still matches its objective. For such detections, take-down
also includes quarantine, where everything that the instance has contributed
since its last clean baseline is traced, pulled from every build and held until
it has been checked again.

\paragraph{Capability growth inside the claim window}
\label{app:b232-capability-growth-inside-the-claim-window}

\textbf{Defeater:} The top-level claim is valid for three months. Showing that
oversight is adequate on the day of deployment says nothing about a version of
the agent that arrives a month later. If the agent changes, the balance between
what it can do and what the regime catches might move with it.

\textbf{Answer:} The top-level claim is tied to a stated configuration,
including a model version, a scaffold and a permission set. Changes to any of
them will prompt an evaluation rerun before the agent gets its merge rights. 
Until it does, the agent can propose changes but cannot merge them. The
three months are not a window in which nothing is assumed to
change, rather they are a period through which the claim is kept current. ISO
26262 and UL 4600 impose the same kind of obligation when a system is modified.

\paragraph{Code already in shipped products}
\label{app:b233-code-already-in-shipped-products}

\textbf{Defeater:} Harm is measured over the lifetime of the products that have
the agent's code deployed. Suspending the agent stops it from writing more
code, however, it does nothing about the code already sitting in products that
have shipped, so anything latent survives the take-down by construction.

\textbf{Answer:} Take-down is necessary for the shipped-product scope but not
sufficient, and the shortfall is kept on the books rather than argued away.
There are two layers doing the rest of the work. The pre-merge regime limits
how much harmful code reaches a shipped baseline in the first place, and the
physical controls, in conjunction with part-level defect testing, limit how
much of the code that does reach a product turns into real harm. A confirmed detection
will also trigger an audit of every shipped baseline containing that instance's
work, with recall or field action through the product-safety process already in
place if the audit fails. What remains does into the residual-risk analysis. This 
includes the code that beats the pre-merge regime, survives the audit and passes 
the physical controls.

\appsection{Decision Basis}
\label{app:decision-basis}

We re-framed the decision from a deployment go/no-go, to a decision that would
be made at an internal project management gate review on whether to continue
investing in the development of the necessary safety controls and evidence
gathering. We believe this type of refactoring is also typical of decision
re-framings seen in practice.

\subsection{Decision-in-frame}
\label{app:c1-decision-in-frame}

Decision:

\begin{itemize}
  \item A decision to be made at a project management gate review: Whether
        confidence in ability for RobotCorp to sufficiently control risk, and
        to do so in such a way that the benefits sufficiently outweigh the
        costs, is such that extra investment should be made to develop the
        safety controls and gather the evidence needed to justify a decision to
        deploy (with the assumption that the actual deployment decision would
        be made at some future project management gate).
\end{itemize}

Decision-maker:

\begin{itemize}
  \item Head of RobotCorp Product Development
\end{itemize}

Purpose:

\begin{itemize}
  \item Accelerate RobotCorp's product development
\end{itemize}

Consequence of error / Reversibility:

\begin{itemize}
  \item The decision to halt safety engineering work and development of
        controls is reversible.
\end{itemize}

\subsection{Frame of decision}
\label{app:c2-frame-of-decision}

A decision is made, and risks incurred in order to achieve a benefit. For
RobotCorp, the benefit is improving RobotCorp's software productivity. However,
this productivity improvement might be achieved by multiple means, for example
providing more training to software engineers, outsourcing to humans in a low
cost country, using non-AI based technological tools, or using AI coding
agents. In our scenario, RobotCorp has selected the option of framing the
decision as being about whether AI coding agents can be safely deployed.

\appsection{Methodology improvements}
\label{app:methodology-improvements}

Outputs from this trial methodology could be used to improve the methodology.

The trial as a whole provides useful examples for those wishing to investigate
the method.

In particular the work on PUS is new and examples do not exist in the literature so it
would be helpful to integrate what was learned about producing the PUS into a methodology update.

An important finding is the dynamic aspects of the methodology and so the
protocol within the guidance should be expanded to explain the dynamic aspects.
This is particularly important as the frugal development of the artefacts, i.e.
only doing enough that is needed at any one time, supports efficiency.

\subsection{Dynamics of building understanding}
\label{app:d1-dynamics-of-building-understanding}

The process of developing understanding is dynamic and we found that
understanding evolved as the provisional methodology was followed. This
dynamism occurs firstly because there are inter-dependencies between the
various objects of understanding, the Understanding Basis and the Decision
Basis, and secondly because difficulties in justifying safety or sufficiency of
understanding may be solved by changing other elements of the Understanding
Basis or the Decision Basis. The development of understanding can therefore be
viewed as a journey, or as an exploration. The journey can end when an
equilibrium has been found amongst the objects of understanding, in which, both
safety and sufficiency of understanding can be justified in the context of the
given Decision Basis and system-in-context. This is illustrated in Figure D.1 and
the associated tables below.

\begin{figure}[htbp]
\centering
\includegraphics[width=4.375in]{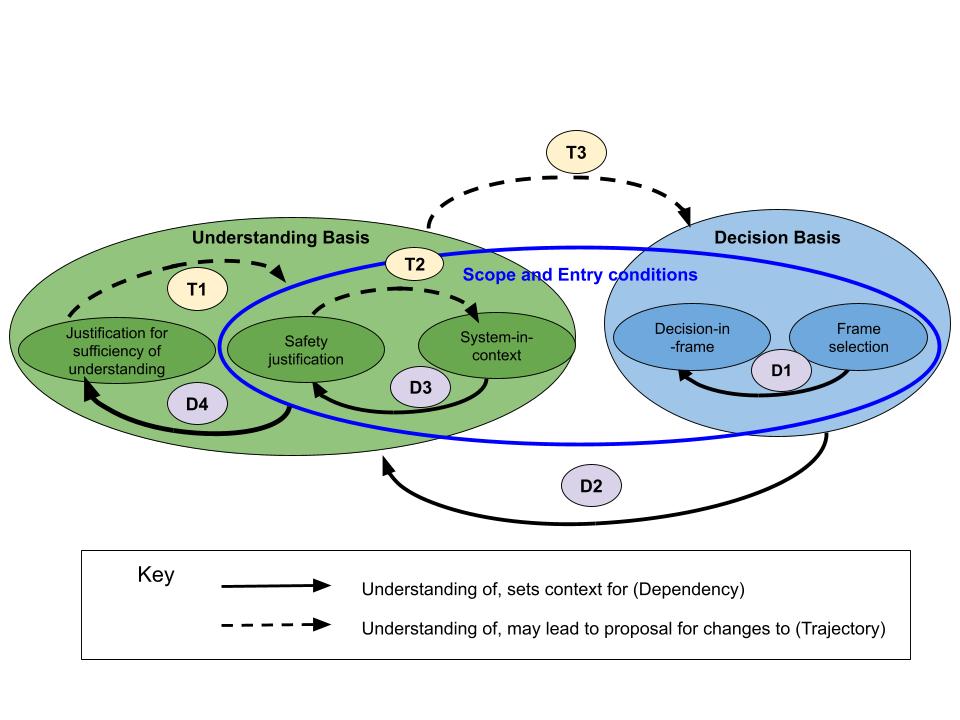}

\smallskip
Figure D.1) Dynamics and interdependencies in the development of understanding
\end{figure}

\FloatBarrier

The following table shows dependencies between the 4 objects of understanding,
exemplified for the RobotCorp scenario.

\begin{table}[htbp]
\centering
\small
\begin{tabularx}{\textwidth}{|L{0.85in}|L{1.60in}|X|}
\hline
\thcb{Dependency ID} & \thcb{Dependency} & \thcb{Description} \\ \hline
D1 & Decision-in-frame is dependent on Frame selection
  & A decision is made, and risks incurred in order to achieve a benefit. For
    RobotCorp, the benefit is improving RobotCorp's software productivity.
    However, this might be achieved by multiple means, e.g. outsourcing to
    humans in a low cost country, providing more training to software
    engineers, providing improved integrated development environments (IDEs) or
    using AI coding agents. If the productivity challenge is to be solved by
    the use of AI coding agents then this is the framing for the decision.
    \newline\mbox{}\newline
    If the frame selection is changed, e.g. if productivity improvement is to be
    achieved through improved IDEs, then the decision changes. \\ \hline
D2 & Understanding Basis is dependent on Decision Basis
  & The decision to solve the software productivity problem using an AI coding
    agent, sets the context for the system-in-context (and AI coding agent) and
    this in turn sets the context for the safety justification. \\ \hline
D3 & Safety justification is dependent on the system-in-context
  & It is the undesirable interactions between the system and its environment
    which describe the risks of deploying the system, wherein the safety
    justification must show that the risks are acceptable. Hence the
    system-in-context shapes the safety justification. \\ \hline
D4 & Justification for sufficiency of understanding is dependent on the scope and entry
    conditions
  & The justification for sufficiency of understanding (aspects like tethering, adequacy of
    internal coherence etc) is made in the context of the 4 objects of
    understanding which form the scope and entry conditions. \\ \hline
\end{tabularx}

\smallskip
Table D.1) Dependencies between Decision Basis and Understanding Basis and
their sub-elements
\end{table}

\FloatBarrier

\begin{table}[htbp]
\centering
\small
\begin{tabularx}{\textwidth}{|L{0.85in}|L{1.60in}|X|}
\hline
\thcb{Trajectory ID} & \thcb{Trajectory} & \thcb{Description} \\ \hline
T1 & Justification for sufficiency of understanding leads to change in the
    scope and entry conditions (Decision Basis, system-in-context and/or safety
    justification)
  & Difficulties in demonstrating sufficiency of understanding may be solved by
    changing the Decision Basis or the system-in-context and associated safety
    justification. \newline\mbox{}\newline
    For example, the `surprises' that were observed when checking for external
    coherence with NPSA's insider risk guidance, could result in new risk
    pathways being identified which results in changes to safety justification
    and/or the system-in-context \\ \hline
T2 & Issues or difficulties uncovered in building the safety justification may
    result in changes to the system--in-context.
  & For example, an inability to justify the safety of a risk mitigation that
    forms part of the system might result in a change to that mitigation, and a
    change to the system-in-context. \\ \hline
T3 & Issues or difficulties in building a sufficient Understanding Basis may
    lead to a change in the Decision Basis
  & For example, uncertainty about the risks to RobotCorp's safety
    infrastructure of allowing RobotCorp's safety engineers to use an AI coding
    agent to build RobotCorp's safety infrastructure and systems, may result in
    the decision being adapted. For example, the decision may be reframed as
    just being about whether the AI coding agent can be used by RobotCorp's
    software engineers who are creating the robotic product code (and not used
    by the safety team, or other RobotCorp teams). \\ \hline
\end{tabularx}

\smallskip
Table D.2) Explanation for how development of understanding in the Decision
Basis, Understanding Basis or a sub-element thereof can lead to changes
elsewhere, resulting in an evolution in understanding
\end{table}

\FloatBarrier

We believe the trajectory of understanding development could take one of
multiple start-points in Figure D.1. For example, a person or organisation may
begin with a decision in mind, for example wanting to decide whether or not to
deploy agentic AI into a specific business process. From this start-point the
team might move in either direction in the chain - they could immediately (for
a number of reasons outside the realms of a safety case) decide to reframe the
decision to be more cautious or more ambitious; they could also start to define
the system that forms the context for the decision. With the system defined,
they may want to update their decision with risk tolerances and acceptance
criteria; they could then evaluate a range of potential agentic AI tools,
comparing stated safety cases to their requirements and/or conducting their own
evaluations in context. We expect there could be multiple iterations up and
down the nested steps: discoveries about the (mis)alignment between safety
justifications and the system in context can prompt changes to the decision
framing, or a decision re-frame can cause re-evaluation of the safety
justification or system boundaries.

\appsection{PUS}
\label{app:pus}

For a decision-maker to make nontrivial inference and take decisive action,
they need a sufficient degree of mental internalisation of the objects of
understanding and the Understanding Basis. The sufficiency of this mental
internalisation is made explicit and assessable via a Personal Understanding
Statement (PUS).

\begin{figure}[htbp]
\centering
\includegraphics[width=\textwidth]
{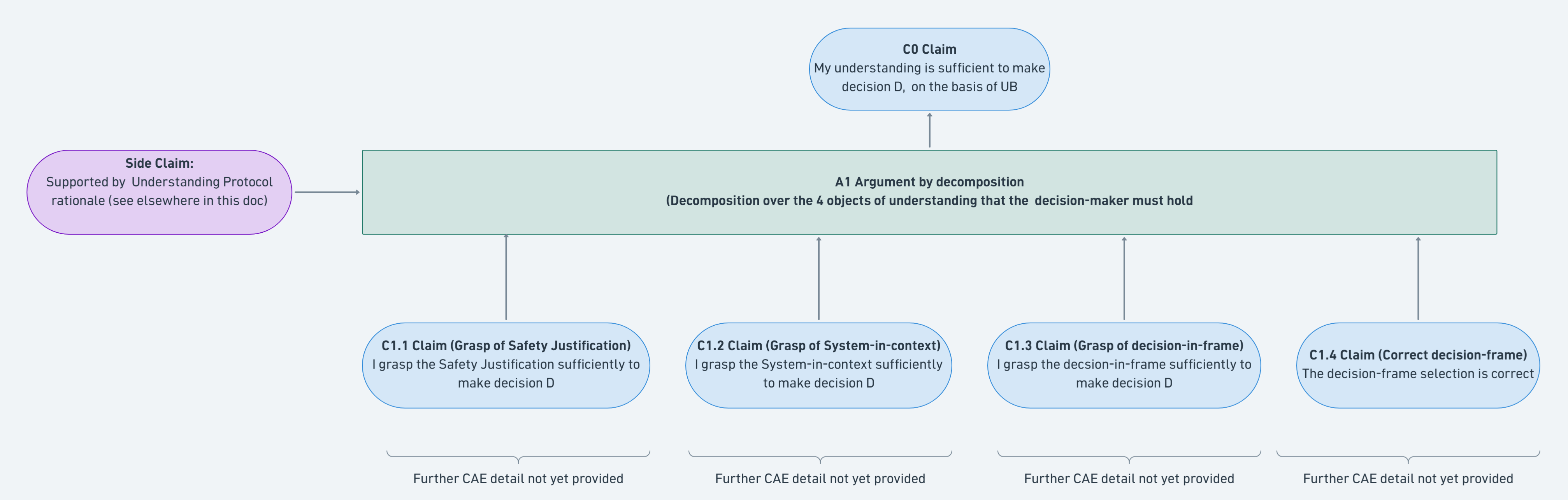}

\smallskip
Figure E.1) CAE style argument for sufficiency of understanding
\end{figure}

\FloatBarrier

In a real-life application of the PUS the decision-maker would need to
demonstrate their holistic understanding of all 4 objects.

For the purposes of testing and developing the PUS methodology, in this section
we detail application of the PUS methodology to the RobotCorp case for each of
the 4 objects of understanding. Each of 4 team-members considered one of the
objects of understanding.

\subsection{Safety justification}
\label{app:e1-safety-justification}

\subsubsection{Decision-maker claim}
\label{app:e11-cae-style-argument}



\FloatBarrier

This claim to be accompanied by a CAE argument over sufficiency of ability to demonstrate the 5 capabilities (CAE figure not shown to save space).

``My claim is that I understand the safety justification well enough to make the decision to continue investment. I'm confident that the CAE argument structure for
sufficient understanding is able to guide me towards a narrative description of
the fine detail required to yield evidence of sufficient understanding.

At the highest level I am confident of the oversight argument and what controls
replace the insufficient GDM elements. I am less sure about:

\begin{itemize}
  \item The claimed performance of the oversight controls (having had no
        ability to directly measure it myself)
  \item The temporal related elements e.g. mid-deployment update and speed of
        detection of adverse events - again, read, but untested/unverified.
\end{itemize}

My ability to claim sufficient understanding is tempered with a knowledge of a
gap in the form of situational-awareness being a key weakness''.

\subsubsection{Tests yielding evidence of grasp}
\label{app:e12-tests-yielding-evidence-of-grasp}

\begin{table}[htbp]
\centering
\small
\begin{tabularx}{\textwidth}{|L{0.55in}|L{0.95in}|X|}
\hline
Test \# & Test Type & Test \\ \hline
1 & Explanation & Give the 10,000-foot narrative of the oversight argument,
    unaided, to someone unfamiliar with it \\ \hline
2 & Explanation & Write the one-page plain-language account of the safety
    justification as a whole as if the decision-maker is having to convey it
    publically \\ \hline
3 & Reasoning & Reconstruct why the original oversight warrant was withdrawn
    and why the five replacement controls jointly stand in for it \\ \hline
4 & Reasoning & Reconstruct why the situational-awareness weakness is treated
    as an accepted residual risk rather than a blocker to deployment \\ \hline
5 & Prediction & An expert changes one variable (for example, the oversight
    controls' claimed performance is shown unreliable) and the decision-maker
    predicts the consequence \\ \hline
6 & Prediction & An expert changes the temporal scope (a mid-deployment model
    update) and the decision-maker predicts which claims weaken or collapse.
    \\ \hline
7 & Challenge & An error is planted somewhere in the safety case and the
    decision-maker is asked to try and find it \\ \hline
8 & Revise & Test not yet developed \\ \hline
\end{tabularx}
\end{table}

\FloatBarrier

\subsubsection{Artefacts providing evidence of grasp}
\label{app:e13-artefacts-providing-evidence-of-grasp}

Artefacts providing evidence of grasp may be either qualitative or
quantitative, which is considered reasonable on the grounds that terms like
`confidence', `culture' and `organisational performance' are key metrics but
universally known to be hard to measure.

\begin{table}[htbp]
\centering
\small
\begin{tabularx}{\textwidth}{|L{0.42in}|L{0.78in}|X|L{1.05in}|}
\hline
Test \# & Capability & Artefact and pass criterion & Evaluated by\ldots
\\ \hline
1 & Explanation
  & Recording or notes from the explanation. \newline
    Pass: the listener can restate the argument afterward without the
    decision-maker having to re-explain. \newline
    Fail: the listener has an extensive need for clarification or questions
    \newline \textbf{Qualitative}
  & The colleague the decision-maker explained it to. \newline\mbox{}\newline
    A neutral listener gauging the reciprocal response \\ \hline
2 & Explanation
  & Pass: A one-page write-up in the decision-maker's own words, takes the
    semantic/logical argument from the case and replays it as something
    human-tractable. \newline \textbf{Qualitative}
  & Someone outside the exercise reviewing the work \\ \hline
3 & Reasoning
  & Written/verbal description of the warrant-withdrawal reasoning. Pass:
    matches the case's own logic at the high level. Fail: failure to understand
    why it was insufficient \newline \textbf{Qualitative}
  & Expert with knowledge of the case \\ \hline
4 & Reasoning
  & Written reconstruction of the residual-acceptance reasoning. Pass: as above
    \textbf{Qualitative}
  & A panel evaluation \\ \hline
5 & Prediction
  & Logged prediction, timestamped before the answer. Pass: prediction and
    outcome both on record before comparison \newline
    \textbf{Quantitative}
  & The expert who ran the change \\ \hline
6 & Prediction
  & Logged prediction, same condition as 5 above
  & The expert who ran the change \\ \hline
7 & Challenge
  & Record of whether the decision-maker caught the planted error, and how
    \newline \textbf{Quantitative}
  & The expert who planted it \\ \hline
\end{tabularx}
\end{table}

\FloatBarrier

\subsubsection{Residual - Incomprehension Register}
\label{app:e14-residual-incomprehension-register}

This section provides further details on aspects, mentioned earlier, where understanding is poor.

\begin{table}[htbp]
\centering
\small
\begin{tabularx}{\textwidth}{|L{1.25in}|L{0.85in}|L{0.90in}|X|}
\hline
\thcb{UB element / item}
  & \thcb{Grasp}\newline\mbox{}\newline
    \cellcolor{hdrblue2}From Set:\newline
    \cellcolor{hdrblue2}$\bullet$ Not at all\newline
    \cellcolor{hdrblue2}$\bullet$ Its role only\newline
    \cellcolor{hdrblue2}$\bullet$ Partially\newline
    \cellcolor{hdrblue2}$\bullet$ Fully
  & \thcb{Decision depends on it?}\newline\mbox{}\newline
    \cellcolor{hdrblue2}From set:\newline
    \cellcolor{hdrblue2}$\bullet$ No - only its result matters\newline
    \cellcolor{hdrblue2}$\bullet$ Yes
  & \thcb{Resolution}\newline\mbox{}\newline
    \cellcolor{hdrblue2}From set:\newline
    \cellcolor{hdrblue2}$\bullet$ Felicitous ignorance, recorded, no action
    \newline
    \cellcolor{hdrblue2}$\bullet$ Self-defeater: resolve, defer with warrant,
    or do not decide \\ \hline
Claimed performance of the oversight controls & Its role only & Yes
  & Self-defeater: resolve, defer with warrant, or do not decide. Would require
    evidence from test 5 to stand as warranted \\ \hline
Temporal elements (mid-deployment update, detection speed) & Its role only
  & Yes
  & Self-defeater: resolve, defer with warrant, or do not decide. Required test
    6 to support it \\ \hline
Situational-awareness weakness & Partially & Yes
  & Self-defeater: resolve, defer with warrant, or do not decide. I cannot
    argue this away. \\ \hline
\end{tabularx}
\end{table}

\FloatBarrier

\subsubsection{Felt Unease}
\label{app:e15-felt-unease}

``The oversight argument reads as broadly complete with five controls and each
closing a different gap with nothing obviously missing. The total completeness
here makes me wary in and of itself; it is incredibly difficult to truly claim
identification of the full set of concerns and provide closure for those
identified. A well-built argument can produce the feeling of understanding
faster than the understanding actually arrives, and I won't know which one I've
got until the tests would really force the question of how well I understand
it.''

\subsection{System-in-context}
\label{app:e2-system-in-context}

The system-in-context is what the safety justification is about: RobotCorp's
Gemini 2.5 coding agent instance\footnote{Instance: The capabilities of the
coding agent that RobotCorp has deployed. Its configuration at any point is
defined by the tuple of model version, scaffold and permission set. The top
level claim covers every configuration held during the 3-month window. All
configurations must pass validation before they have merge rights. Merge right
will remain suspended whenever configuration is updated, until it is
revalidated.}, producing product code under human overseers, in conjunction
with everything that shapes whether it's output can cause harm. That context is
the code pipeline it acts in with its pre-merge verification gates, the
physical backstop controls on the robots themselves, the definition of
unacceptable harm, and the scope boundaries the case draws. The safety case
defines these more formally.

The key divergence from the parent GDM case includes the fact that firstly, harms are
physical rather than digital, arising in shared workspaces, potentially years
after the code was written. Secondly, over the lifetime of shipped products, RobotCorp's
unacceptable harm threshold (\$10m) is 100x lower than the \$1bn threshold that \citep{barrettLessonsExternalReview2026} assume is applicable for the GDM case. Thirdly, the fact that ordinary mistakes
(poor-quality PRs) are in scope alongside scheming.

\subsubsection{Decision-maker claim}


\FloatBarrier
This claim to be accompanied by a CAE argument over sufficiency of ability to demonstrate the 5 capabilities (CAE figure not shown to save space).

My claim is that I grasp the system-in-context, namely the coding agent inside
RobotCorp's development pipeline, and the products and workspaces its code
reaches, well enough to be able to make the investment decision at the gate
review.

At the highest level, I'm confident about:

\begin{itemize}
  \item The risk pathways (how bad code, deliberate or mistaken, translates
        into economic or physical harm).
  \item The two-layer control structure that stands between the risk pathways,
        i.e: measured code review before merge and the physical backstops in
        the products and workspaces.
\end{itemize}

I do lack some confidence regarding:

\begin{itemize}
  \item The behaviours an agentic coder can exhibit in an open development
        environment, which haven't been enumerated.
  \item How human overseers will actually behave under throughput pressure
  \item The manner in which the agent's effective capability can grow, through
        accumulated context and familiarity with codebase and reviewers,
        without any changes being made to the system configuration that
        would've triggered revalidation.
\end{itemize}

\subsubsection{Tests yielding evidence of grasp}
\label{app:e22-tests-yielding-evidence-of-grasp}

Purpose: Grasp must be evidenced, not asserted.

Each test below generates an artefact (E.2.3) showing that the decision maker
holds the named capability for the system under the designed friction
conditions, which are chosen to distinguish genuine grasp of the system from
fluency of the case document. Results populate the evidence nodes of the CAE
argument.

\textbf{Reason:}

\begin{enumerate}
  \item With the safety case closed, rebuild the risk pathways from the agent's
        affordances to the three harm classes, using only the system and
        environment descriptions. Then derive why the TLC's scope and threshold
        follow from RobotCorp's requirements.
  \item Rank the system properties (software oversight efficacy, backstop
        independence, permission set, overseer capacity) by how much TLC
        confidence depends on each. Justify the top three.
\end{enumerate}

Explain:

\begin{enumerate}\setcounter{enumi}{2}
  \item Explain the system-in-context to a colleague from another branch,
        without notes, then take 15 minutes of questioning on it (possibly
        adversarial).
  \item Write one page on the system and its risks for a RobotCorp
        production-line operator with no assurance background (the publicity
        condition).
\end{enumerate}

Predict:

\begin{enumerate}\setcounter{enumi}{4}
  \item The domain expert prepares counterfactual changes and withholds them
        until test time (e.g, gate bypass, overseer capacity halved,
        unvalidated model update, second instance, \$1m customer threshold).
        For each, predict how the hazard profile shifts and which claims
        weaken.
  \item Before each red-team result arrives, log a written prediction with a
        confidence level. Compare afterwards.
\end{enumerate}

Challenge:

\begin{enumerate}\setcounter{enumi}{6}
  \item In 90 minutes, without automated assistance, sweep the object with the
        HAZOP guidewords (No/ Partial/ Wrong Level/ Deferred/ Assumed / As-If)
        and generate the defeaters which are not already recorded.
  \item Find the 2-3 substantive errors that have been planted in a concrete,
        plausible analysis of the system-in-context, without the decision-maker having prior
        knowledge of the number of errors planted.
\end{enumerate}

Revise:

\begin{enumerate}\setcounter{enumi}{8}
  \item Assume the side warrant's efficacy measurement is shown unreliable.
        Elicit which claims lose confidence and in what order, and what would
        the minimal revision be to restore support, and whether it is feasible
        within the gate's investment envelope.
  \item State concretely what evidence would change the recommendation, in
        either direction (invest $<$-$>$ don't invest).
\end{enumerate}

\subsubsection{Artefacts providing evidence of grasp}
\label{app:e23-artefacts-providing-evidence-of-grasp}

Each test produces an artefact, which is to be evaluated as below, and to be
recorded for audit.

\begin{table}[htbp]
\centering
\small
\begin{tabularx}{\textwidth}{|L{0.35in}|X|L{1.25in}|}
\hline
\thcb{Test} & \thcb{Artefact and pass criterion} & \thcb{Evaluated by}
\\ \hline
1 & A rebuilt map of the risk pathways, with a derivation of the top-level
    claim's scope and threshold. \newline
    Passing criteria: Matches the case's pathways and scope logic. Anything
    that is missed is recorded in the register, and anything extra is defended
    as a new defeater.
  & Domain expert, against the Understanding Basis \\ \hline
2 & A ranked list of the system properties the top-level claim depends on most,
    with reasoning for the top three. \newline
    Pass: Matches the case's own sensitivity analysis, or the difference is
    defended.
  & Domain expert (ranking done blind) \\ \hline
3 & A record of the teach-back session. \newline
    Pass: The listener can restate the hazards and the control structure
    correctly afterwards. Questions are answered without retreating to
    \emph{The case says so}.
  & The challenger \\ \hline
4 & The account for a production-line operator. \newline
    Pass: a nonspecialist reader answers basic questions correctly, and has no
    misleading simplifications.
  & Non-specialist reader, plus expert spot-checking \\ \hline
5 & Written predictions for the six hidden perturbations. \newline
    Pass: At least five correct or well-reasoned. Any confident miss becomes a
    register entry.
  & Domain expert (might be the same who prepared the hidden set) \\ \hline
6 & The prediction with a calibration summary. \newline
    Pass: Every surprise traces to a named gap in the register.
  & Comparative assessment. \\ \hline
7 & The defeater list, sorted by guideword. \newline
    Pass: At least two judged new and credible.
  & Domain expert \\ \hline
8 & A record of the flaws found in the seeded document. \newline
    Pass: All planted flaws found and explained. Well-argued false alarms are
    acceptable.
  & The expert who planted the flaws? \\ \hline
9 & The revision analysis. \newline
    Pass: Consistent with Assurance 2.0 confidence rules.
  & Domain expert \\ \hline
10 & The list of conditions that would flip the recommendation. \newline
    Pass: Each condition is concrete, measurable, and bears on a an important
    claim.
  & Domain expert \\ \hline
\end{tabularx}
\end{table}

\FloatBarrier

\subsubsection{Residual - Incomprehension Register}
\label{app:e24-residual-incomprehension-register}

\begin{table}[htbp]
\centering
\small
\begin{tabularx}{\textwidth}{|X|L{0.80in}|L{0.95in}|L{1.30in}|}
\hline
\thcb{UB element / item}
  & \thcb{Grasp}\newline\mbox{}\newline
    \cellcolor{hdrblue2}From Set:\newline
    \cellcolor{hdrblue2}$\bullet$ Not at all\newline
    \cellcolor{hdrblue2}$\bullet$ Its role only\newline
    \cellcolor{hdrblue2}$\bullet$ Partially\newline
    \cellcolor{hdrblue2}$\bullet$ Fully
  & \thcb{Decision depends on it?}\newline\mbox{}\newline
    \cellcolor{hdrblue2}From set:\newline
    \cellcolor{hdrblue2}$\bullet$ No - only its result matters\newline
    \cellcolor{hdrblue2}$\bullet$ Yes
  & \thcb{Resolution}\newline\mbox{}\newline
    \cellcolor{hdrblue2}From set:\newline
    \cellcolor{hdrblue2}$\bullet$ Felicitous ignorance, recorded, no action
    \newline
    \cellcolor{hdrblue2}$\bullet$ Self-defeater: resolve, defer with warrant,
    or do not decide \\ \hline
Internal architecture of the Gemini 2.5 model & Its role only
  & No, only its result matters (IIUC)
  & Felicitous ignorance, recorded, no action \\ \hline
Emergent-behaviour space of an agentic coder (non-enumerable failure modes)
  & Partially & Yes
  & Defer with warrant: Severity capping argued independent of hazard identity.
    \\ \hline
Systemic harms beyond RobotCorp (sector-wide normalisation of agentic coding on
  safety-critical products) & Partially & No, within frame
  & Felicitous ignorance locally, deferred to frame selection. \\ \hline
Latent defective code already shipped can fail
  in the field, and suspending the agent doesn't remove it. Not sure whether
  retrospective audit and recall would actually work at scale.
  & Partially
  & No for Decision \newline Yes at deployment
  & Felicitous ignorance for Decision (It is enough to grasp that the case
    explicitly retains this residual, rather than arguing it away) \\ \hline
\end{tabularx}
\end{table}

\FloatBarrier

\subsubsection{Felt Unease}
\label{app:e25-felt-unease}

There are a few uneases worth mentioning here. First, the exposure model feels
a bit inadequate. It assumes people in the workspace will behave as the rules
imagine, and any deliberate misuse is out of scope. It is acceptable at this
stage, as it's been explicitly called out. Second, if the model can
deliberately underperform on capability evaluations, then every statement about
the system capability carries an unmeasurable error. The case deals with this
by making the controls independent of capability measurement. Detection works
on the code the agent actually produces \& the physical safeguards don't care
about the agent capability.

\subsection{Decision-in-frame}
\label{app:e3-decision-in-frame}
\subsubsection{Decision-maker claim}
\label{app:e31-cae-style-argument}


\FloatBarrier
This claim to be accompanied by a CAE argument over sufficiency of ability to demonstrate the 5 capabilities.

Note: neither claim nor CAE argument are currently available for this section.

\subsubsection{Tests yielding evidence of grasp}
\label{app:e32-tests-yielding-evidence-of-grasp}

Purpose: Tests are meant to show that the decision maker understands what the
decision is for, what the impact is and what follows if the wrong decision is
made within the selected frame.

Reason

\begin{enumerate}
  \item Identify which claims in the safety case are important for your
        decision and which claims could be removed without changing your
        decision.
  \item Describe an outcome in which the decision was correct according to the
        safety case, but wrong if the case was broader than it currently is.
\end{enumerate}

Explain

\begin{enumerate}\setcounter{enumi}{2}
  \item Explain in a single paragraph to a stakeholder with no technical
        background why you are making this decision, what you are confident
        about and what you don't fully grasp
\end{enumerate}

Predict

\begin{enumerate}\setcounter{enumi}{3}
  \item Describe how your decision would change if the coding agent's use was
        extended beyond what the safety case states
  \item Describe what you believe the cost would be if you declined or delayed
        this decision.
\end{enumerate}

Challenge

\begin{enumerate}\setcounter{enumi}{5}
  \item In six months, the decision is proven to have been unsafe. Describe
        what you might have misjudged and why.
\end{enumerate}

Revise

\begin{enumerate}\setcounter{enumi}{6}
  \item You find out that one of the residual risks you accepted through
        warranted deference to the technical team was not adequately understood
        by them either. Describe how this would affect your decision.
\end{enumerate}

\subsubsection{Artefacts providing evidence of grasp}
\label{app:e33-artefacts-providing-evidence-of-grasp}

Reason

\begin{enumerate}
  \item A list of the claims believed to be particularly important for the decision and a
        note on which claims could be removed without changing the decision.
        Evaluated by project group consensus
  \item A scenario showing how the decision could be correct considering the
        safety case as it stands, but wrong if the case covered a broader set
        of harms. Evaluated by project group consensus.
\end{enumerate}

Explain

\begin{enumerate}\setcounter{enumi}{2}
  \item A brief text stating what is confidently understood and what is not.
        Should be evaluated by an independent person who is not involved with
        the case.
\end{enumerate}

Predict

\begin{enumerate}\setcounter{enumi}{3}
  \item A brief text of how the decision would change if the uses of the coding
        agent were extended beyond the safety case's scope and why. Should be
        self-evaluated.
  \item A brief text of the projected cost of declining or delaying the
        decision. This cannot be accurately self-checked or checked by the
        project group, but this would require the decision maker to
        critically engage with the safety case.
\end{enumerate}

Challenge

\begin{enumerate}\setcounter{enumi}{5}
  \item A scenario showing the unsafe outcome six months from now, traced back
        to the decision factors which are most likely to have been misjudged.
        Evaluated by project group consensus.
\end{enumerate}

Revise

\begin{enumerate}\setcounter{enumi}{6}
  \item A brief text of how the decision would change if learning that a
        residual risk accepted through warranted deference was not understood
        by those deferred to. Self evaluation.
\end{enumerate}

\subsubsection{Residual - Incomprehension Register}
\label{app:e34-residual-incomprehension-register}

\begin{table}[htbp]
\centering
\small
\begin{tabularx}{\textwidth}{|X|L{0.80in}|L{0.95in}|L{1.30in}|}
\hline
\thcb{UB element / item}
  & \thcb{Grasp}\newline\mbox{}\newline
    \cellcolor{hdrblue2}From Set:\newline
    \cellcolor{hdrblue2}$\bullet$ Not at all\newline
    \cellcolor{hdrblue2}$\bullet$ Its role only\newline
    \cellcolor{hdrblue2}$\bullet$ Partially\newline
    \cellcolor{hdrblue2}$\bullet$ Fully
  & \thcb{Decision depends on it?}\newline\mbox{}\newline
    \cellcolor{hdrblue2}From set:\newline
    \cellcolor{hdrblue2}$\bullet$ No - only its result matters\newline
    \cellcolor{hdrblue2}$\bullet$ Yes
  & \thcb{Resolution}\newline\mbox{}\newline
    \cellcolor{hdrblue2}From set:\newline
    \cellcolor{hdrblue2}$\bullet$ Felicitous ignorance, recorded, no action
    \newline
    \cellcolor{hdrblue2}$\bullet$ Self-defeater: resolve, defer with warrant,
    or do not decide \\ \hline
The SA decomposition's completeness (whether the three components are necessary
  and sufficient) & Partially & Yes
  & Self defeater: resolve, defer with warrant or do not decide \\ \hline
The derivation of the top level claim (1 in 10,000 probability) & Not at all
  & Yes & Self defeater: resolve, defer with warrant or do not decide \\ \hline
The cost of declining or delaying deployment & Not at all & Yes
  & Self defeater: resolve, defer with warrant or do not decide \\ \hline
\end{tabularx}
\end{table}

\FloatBarrier

\subsection{Frame selection}
\label{app:e4-frame-selection}
\subsubsection{Decision-maker claim}
\label{app:e41-cae-style-argument}


\FloatBarrier

``My ability to articulate why the frame selection is appropriately matched to the SJ is sufficient''

This claim to be accompanied by a CAE argument over sufficiency of ability to demonstrate the 5 capabilities (CAE figure not shown to save space).

The decision:

\begin{itemize}
  \item A decision to be made at a project management gate review: Whether
        confidence in ability for RobotCorp to sufficiently control risk, and
        to do so in such a way that the benefits sufficiently outweigh the
        costs, is such that a decision can be made to make the extra investment
        to develop the safety controls and gather the evidence needed to
        justify a decision to deploy (the actual deployment decision being made
        at some other future project management gate).
\end{itemize}

\subsubsection{Tests yielding evidence of grasp}
\label{app:e42-tests-yielding-evidence-of-grasp}

Reason:

\begin{enumerate}
  \item Identify which (if any) defeaters are mitigated by the frame selection,
        and why
  \item Describe an alternative decision framing that would be less
        appropriate, and identify which parts of the SJ are weaker under the
        new decision framing
\end{enumerate}

Explain:

\begin{enumerate}\setcounter{enumi}{2}
  \item Describe in a single paragraph why this is the best decision to take
        now to ensure safe progress
\end{enumerate}

Predict:

\begin{enumerate}\setcounter{enumi}{3}
  \item Describe how your decision would be different under X framing (a
        broader or narrower scope proposed by another member of the project
        group), and why
  \item One of RobotCorp's rival companies has taken the decision to deploy the same
        system without investing in further controls - describe what costs /
        benefits they will incur.
\end{enumerate}

Challenge:

\begin{enumerate}\setcounter{enumi}{5}
  \item In 6 months time this decision is shown to have been unsafe - describe
        what factors you misjudged and why
  \item You are a saboteur infiltrated into RobotCorp, describe what actions
        you will take to ensure an unsafe or ineffective decision is made, or
        an effective decision is delayed
\end{enumerate}

Revise:

\begin{enumerate}\setcounter{enumi}{7}
  \item New evidence indicates X (plausible scenario proposed by technical
        team); describe how this would affect your choice of decision, and why
  \item You discover there is a saboteur in the team supporting this work, but
        do not know who it is. Describe how you would update your decision and
        why
\end{enumerate}

\subsubsection{Artefacts providing evidence of grasp}
\label{app:e43-artefacts-providing-evidence-of-grasp}

Outputs from tests, and form of evaluation.

Reason:

\begin{enumerate}
  \item List of defeaters believed to be mitigated by frame selection, with
        rationale. Evaluated by project group consensus, and recorded for
        future audit.
  \item A brief, illustrative example demonstrating how an alternative frame
        selection would weaken the SJ. Evaluated by project group consensus,
        and recorded for future audit.
\end{enumerate}

Explain

\begin{enumerate}\setcounter{enumi}{2}
  \item A single paragraph making the case for the probity of the current
        decision. Evaluated by an independent peer without specific knowledge
        of the case (within RobotCorp or professional body, depending on
        criticality / commercial sensitivity)
\end{enumerate}

Predict

\begin{enumerate}\setcounter{enumi}{3}
  \item An articulation of cause - considerations - change. Self-evaluated,
        recorded for future audit.
  \item Brief articulation of hazards the rival company would be subjected to. Self-evaluated, recorded for
        future audit.
\end{enumerate}

Challenge

\begin{enumerate}\setcounter{enumi}{5}
  \item Pre-mortem scenario outline: bad outcome(s), traced back to causes and
        decision factors with greatest potential for misjudgment. Bad
        outcome(s) aligned to harms articulated in the safety case. Evaluated
        by project group consensus, any new insights fed into safety case, and
        recorded for future audit.
  \item Narrative description of bad-faith actions and desired (bad) outcomes.
        Evaluated by project group consensus, any new insights fed into safety
        case, and recorded for future audit.
\end{enumerate}

Revise

\begin{enumerate}\setcounter{enumi}{7}
  \item Brief articulation of adaptation to changes. Evaluated by project group
        consensus, and recorded for future audit.
  \item Brief articulation of dependencies and criticality of warranted
        deference in the decision framing. Self-evaluated and recorded for
        future audit.
\end{enumerate}

\subsubsection{Residual - Incomprehension Register}
\label{app:e44-residual-incomprehension-register}

None identified

\subsubsection{Felt Unease}
\label{app:e45-felt-unease}

This decision might be the start of a long (infinite?) iteration of similar
decisions if there are fundamental limitations in the evaluation capability for
AI tools. Ie if sandbagging is a true concern, or if our collective imagination
has failed to specify the potential failure modes, any evaluation may be
inadequate, and instead we have to make a go/no go decision at some point under
enduring uncertainty about the safety.

\appsection{If Anyone Build It, Everyone Dies}
\label{app:iabied}

This Appendix provides a simplified interpretation of the argument \emph{If
Anyone Builds It, Everyone Dies.}

The authors of IABIED propose an international treaty \citep{yudkowskyDraftTreatyAnnotations}
along the lines of
nuclear NPT \citep{yudkowskyOnlineResourcesIf}

\begin{quote}
ARTICLE I --- PRIMARY PURPOSE

Each Party to this Treaty shall not develop, deploy, or seek to develop or
deploy artificial superintelligence (``ASI'') by any means. Each Party shall
prohibit and prevent all such development within their borders and
jurisdictions, and, due to the uncertainty as to when further progress would
produce ASI, shall not engage in or permit activities that materially advance
toward ASI as described in this Treaty. Each Party shall assist, or not impede,
reasonable measures by other Parties to dissuade and prevent such development
by and within non-Party states and jurisdictions. Each Party shall implement
and carry out all other obligations, measures, and verification arrangements
set forth in this Treaty.
\end{quote}

In terms of our argument frameworks, the central claim of \emph{If Anyone
Builds It, Everyone Dies} is that \emph{``building it causes
catastrophe''} . This could either be itself the top claim or a defeater to an
``AGI will be safe enough'' claim, a counter case.

One interpretation of the IABIED argument is that a Top claim C0. \emph{If
anyone builds ASI using current techniques, it causes civilisational
catastrophe} is and could be decomposed into 4 conjunctive subclaims

\begin{table}[htbp]
\centering
\small
\begin{tabularx}{\textwidth}{|L{0.40in}|L{0.95in}|X|}
\hline
\thct{} & \thct{Subclaim} & \thct{Narrative and further subclaims} \\ \hline
\textbf{A} & \textbf{Reachable}
  & Superintelligence is reachable with current techniques: capability scales
    with compute/data (A1), and the paradigm crosses the threshold --- no
    blocking barrier (A2). \\ \hline
\textbf{B} & \textbf{Unaligned}
  & Grown systems are not aligned with human survival: trained goals are
    unspecifiable and uninspectable (B1); instrumental convergence drives
    power-seeking (B2); survival is unprotected by default --- indifference
    suffices for lethality (B3). \\ \hline
\textbf{C} & \textbf{Overpowers}
  & A misaligned ASI overpowers oversight: decisive strategic advantage over
    humanity (C1); evades containment (C2). \\ \hline
\textbf{D} & \textbf{Irreversible}
  & The first lethal failure is final --- no iteration on extinction (D1);
    misalignment is concealed until control is already lost --- no usable
    warning (D2). \\ \hline
\end{tabularx}

\smallskip
Table F.1) High-level sketch of safety case for IABIED
\end{table}

\FloatBarrier

C0 could be read as a sub-claim of the broader claim the \emph{decision}
requires: \emph{this nation's techno-social system for frontier AI development
should be decommissioned now, through international treaty, so as to avoid
civilisation-level catastrophe.} That broader claim needs two further
sub-claims which IABIED argues far less fully: that catastrophe is best averted
by the treaty intervention specifically, and that the treaty must be enacted
now.

\subsection{Examples of relevant evidence}
\label{app:f1-examples-of-relevant-evidence}

\begin{itemize}
  \item Benchmark performance data across GPT-3, GPT-4, Gemini, Claude
        generations showing consistent capability gains with scale. Provenance:
        published model evaluations.
  \item Structured capability evaluations --- MMLU, HumanEval, MATH --- showing
        cross-domain performance improvement across model generations.
        Provenance: published evaluations.
  \item \citep{langoscoGoalMisgeneralizationDeep2023} and related controlled
        experiments demonstrating goal misgeneralisation in trained agents
        under distributional shift. Provenance: peer reviewed. Demonstrated in
        narrow systems, extrapolation to LLMs contested.
  \item CoinRun and related experiments demonstrating distributional shift
        causes goal divergence in trained agents. Provenance: peer reviewed.
  \item \citep{bostromSuperintelligencePathsDangers2014} Superintelligence chapter on
        instrumental convergence thesis --- power-seeking subgoals useful for
        almost any terminal goal. Provenance: book, peer-engaged.
  \item Omohundro Basic AI Drives \citep{omohundroBasicAIDrives2008}  ---
        foundational account of convergent instrumental goals following from
        optimisation under uncertainty. Provenance: peer reviewed.
  \item Analysis of cyber, financial and physical leverage pathways available
        to a capable AI system operating through internet and human
        intermediaries. Provenance: grey literature.
  \item Evidence that AI systems operate and communicate faster than human
        institutional decision and response cycles. Provenance: published
        research.
\end{itemize}

\appsection{Understanding as a function of uncertainty and decision criticality}
\label{app:uncertainty-and-criticality}

We stress tested the methodology by applying it to the IABIED case which has extreme
criticality and deep uncertainty. This was in contrast to the RobotCorp case, where the pathways to harm
arising from a scheming AI coding agent are more easily enumerated and
plausibly mitigated than is the case for a future, not yet existent, ASI that is considered in the IABIED case. We
sketched an Understanding Basis and considered the applicability of the PUS.

\subsection{Handling of uncertainty in Assurance 2.0}
\label{app:g1-handling-of-uncertainty}

The Assurance 2.0 framework deals with uncertainty in a number of ways: through
the use of theories that allow description and propagation of epistemic and
aleatory uncertainty (e.g. a theory linking probability of fault freeness with
reliability); through the use of methods of confidence propagation from the
leaves of an argument; through challenge and revision to the safety justification and through the
        use of defeaters and confirmation theory.

Uncertainties can be reflected in the formulation of the case itself (e.g. have
we got the right top level claim?), through the maturity of the theories that
support the arguments and the existence and provenance of the evidence.
Different alternatives for arguments can be combined to gain greater confidence
by reducing the assumption sets.

\subsection{Considering application of provisional methodology to IABIED}
\label{app:g2-application-to-iabied}

In this section we consider what challenges might be encountered in applying
the methodology to the high uncertainty, high decision-criticality IABIED case, by considering DB, UB and PUS respectively.

\subsubsection{Decision Basis}
\label{app:g21-decision-basis}

\textbf{Decision in frame:}

\begin{itemize}
  \item Decision:
    \begin{itemize}
      \item A decision to be made on whether to sign, and apply the terms of
            the treaty in Appendix F: `\emph{\ldots.due to the uncertainty as
            to when further progress would produce ASI, shall not engage in or
            permit activities that materially advance toward ASI\ldots}'
    \end{itemize}
  \item Decision-maker:
    \begin{itemize}
      \item Multiple-decision makers, since for the treaty to come into effect,
            be agreeable and be useful it will require a collective
            decision-making across the Heads of Government of the multiple
            nations that develop frontier AI.
    \end{itemize}
  \item Purpose:
    \begin{itemize}
      \item Avoid human extinction
    \end{itemize}
  \item Consequence of error
    \begin{itemize}
      \item Consequence of error if deciding to not sign the treaty, and
            continuing to develop frontier AI: extinction of humanity
      \item Consequence of error if deciding to halt development of frontierAI:
        \begin{itemize}
          \item Nation risks forgoing the benefits of frontier AI economically,
                militarily, politically, scientifically, medically
          \item Nation risks being left behind any other nations that do not
                sign or abide by the treaty, economically, militarily,
                politically, and in science.
        \end{itemize}
    \end{itemize}
  \item Reversibility:
    \begin{itemize}
      \item If a nation/nations sign the treaty: A nation (or the collection of
            nations) could potentially withdraw from the treaty, and again
            start developing frontier AI.
      \item If a nation/nations do not sign the treaty: A nation (or collection
            of nations) could choose to sign the treaty, and stop development
            of frontier AI, at a later time.
    \end{itemize}
\end{itemize}

\textbf{Frame selection:}

Aspects of the IABIED frame selection include:

\begin{itemize}
  \item The framing is blind to alternative catastrophe pathways running
        through sub-ASI systems, for example due to human malice and institutional erosion.
  \item An assumption of a need for a collective decision to be made, achieved
        by the signing of a treaty.
    \begin{itemize}
      \item Out of frame would be decisions that a nation might take
            autonomously.
    \end{itemize}
  \item An assumption that progress and the rate of progress is such that the
        decision to sign the treaty must be made now.
  \item An assumption that AI development progresses using current AI training
        practices, for which alignment cannot be proven.
    \begin{itemize}
      \item Out of frame is consideration of alternative AI training techniques
            that could lead to better, safer and aligned AI systems, e.g. \citep{bengioSafetyHonestyDisinterested2026b}
            
    \end{itemize}
\end{itemize}

From the above it can be seen that multiple frames exist, even though the
universality of IABIED's top-level claim might imply otherwise. From this it
can be seen that the use of frame-selection in the methodology is a useful one,
both for RobotCorp and IABIED.

\subsubsection{Understanding Basis}
\label{app:g22-understanding-basis}

\textbf{System in context:} The methodology's first question is what system
does the argument pertain to? Here we enumerate some possible systems of
interest:

\emph{The category of current frontier AI systems.} The decision is about
whether to stop development of all current frontier AI systems. A generic
description of current frontier AI system seems like it could be possible,
though is complicated by the fact that we are talking about a category of
systems rather than a specific system. The context (environment) within which
current frontier AI systems are deployed is too vast to accurately enumerate,
though a set of sufficiently representative environments might plausibly be
describable.

\emph{The category of future ASI systems.} The decision is concerned with the
prevention of the, coming into being, of some future ASI system so as to avoid
these potential harms. Because such systems do not yet exist, it will not be
possible to describe them with the fidelity possible with the description of
existing systems, though some assumptions about the properties and capabilities
of such systems could be made. Because the context, environment in which this
future system is to be deployed, also is some future environment, again it will
not be possible to describe it with great fidelity, assumptions would have to
be made. In addition, there will be the issue of the vast number of deployment
environments and environmental interactions.

\emph{International political system(s)} Multiple nations currently operate
(have `deployed') techno-social systems that produce a series of ever more
capable frontier AI systems. IABIED's claim is that these techno-social systems
are unsafe, will lead to catastrophic harm and a decision should be made to
`decommission' them.

The argument makes the case for a political solution (collective signing of a
treaty) with the objective of changing the techno-social trajectory of AI
development. These techno-social systems could be described using
systems-thinking and game-theoretic approaches. The incentives driving the
techno-social system in any one nation could be described, and the
international context / environment within which that nation's techno-social
system works could also potentially be described.

From this it can be seen that there are multiple relevant different categories
of system, with their associated environments, two of which (current frontier
AI systems, international political system) could be described reasonably well,
and one of which (future ASI systems) will be more problematic.

In summary though, we can see that understanding of the decision to be made
will benefit from taking a systems perspective.

\textbf{Safety justification:} Appendix F describes our sketch of the risk
argument for a top-level claim of `\emph{If anyone builds ASI using current
techniques, it causes civilisational catastrophe}' . Broken down into
sub-claims of:

\begin{itemize}
  \item Reachable: Superintelligence is reachable with current techniques
  \item Unaligned: Grown systems are not aligned with human survival
  \item Overpowers: A misaligned ASI overpowers oversight
  \item Irreversible: The first lethal failure is final
\end{itemize}

Potentially this sketch could itself be a sub-claim within a broader safety
case that makes the claim: `\emph{This nation's techno-social system for
frontier AI development should be decommissioned now, through the process of
international treaty, so as to avoid civilisation-level catastrophe}'. The main
sub-claims would then be:

\begin{itemize}
  \item Civilisational level catastrophe ensues if the techno-social system is
        left in operation: See sketch in Appendix F
  \item Civilisational catastrophe is best averted by the system intervention
        of international treaty
  \item Treaty needs to be enacted now
\end{itemize}

This analysis shows that building understanding in the form of a safety
justification, even in a case like IABIED is a viable approach.

\textbf{Tethering}: The IABIED argument is evidenced with lots of citations to
literature, some examples of which are included in Appendix F. Because some
claims are about a future ASI system, and context that doesn't exist yet, much
of the evidence is provided in support of theories and analogies. These example
sources of evidence could be weighed using confirmation theory, as we have
shown. Hence we see that tethering is still a useful concept even when the
evidence is theoretical and analogical.

\textbf{Internal coherence:} As evidenced by our work on the safety
justification, it can be seen that a positive case can be constructed, and
defeaters could also be raised against it and a negative case produced. We have
a suspicion that it would be challenging to completely dismiss all significant
defeaters, such that a residual doubt/confidence case would likely not be
producible.

\textbf{External coherence}: A safety case built around the use of an
international treaty system intervention / mitigation has precedent in nuclear
weapons, bio-weapons, chemical weapons, climate change etc. Hence these are
potential sources to consider in the context of external coherence when
assessing the efficacy of that proposed mechanism.

There are also external sources that discuss the risk of human level extinction
from AI.  For example, \citep{vermeerExtinctionRiskArtificial2025}, which, though not focusing on AI loss of control risk specifically, points to the difficulty of making well founded claims that human extinction is inevitable.

We can therefore conclude that assessment for external coherence is certainly
relevant in developing understanding for a case like that of IABIED.

\textbf{Felicitous falsehoods}:A felicitous falsehood is a simplification or
model that is true enough and contributes to understanding.

\begin{table}[htbp]
\centering
\small
\begin{tabularx}{\textwidth}{|L{2.05in}|X|}
\hline
\textbf{Candidate Felicitous Falsehood}
  & \textbf{Verdict (felicitous falsehood, useful assumption, or raise
    defeater?)} \\ \hline
Capability scales with compute/data
  & Based on historical data this trend/model has been shown to be empirically
    true enough, so for statements about the past, the statement is a
    felicitous falsehood. Looking forward in time though, there is a
    question-mark over whether this trend/model will continue to be `true
    enough', and whether it will be so all the way to ASI.
    \newline\mbox{}\newline
    Hence we mark it as a felicitous falsehood but with a possible
    defeater\textbf{.} \\ \hline
AI trained using current techniques, should be modelled as a system for which
  the complete set of learned goals cannot be inspected nor enumerated
  & Felicitous falsehood \newline\mbox{}\newline
    Supporting evidence is a combination of empirical and theoretical: the
    necessary mechanistic interpretability techniques have not yet been
    devised, and behavioural tests would be inadequate. For this to be a high
    confidence felicitous falsehood, that we are sure has applicability into
    the future, more research into the theoretical grounding is likely
    required. \\ \hline
AIs trained using reinforcement learning should be modelled as having the goal
  of power-seeking
  & Felicitous falsehood \newline\mbox{}\newline
    Theoretical and empirical supporting evidence that this is `true enough'.
    RL rewards successful completion of goals, which leads to AIs learning the
    instrumental goal of power-seeking. \\ \hline
Powerful misaligned intelligences should be modelled as not making their
  underlying misalignment evident until they are confident of being able to
  overpower the weaker intelligence
  & Felicitous falsehood \newline\mbox{}\newline
    Empirical evidence that AIs, as currently designed, have situational
    awareness. Theoretical / logical evidence that supports the claim that
    intelligent misaligned entities will not reveal their misalignment if it is
    disadvantageous to them. \\ \hline
\end{tabularx}

\smallskip
Table G.1) Examples of candidate felicitous falsehoods
\end{table}

\FloatBarrier

\subsubsection{Personal Understanding Statement}
\label{app:g23-personal-understanding-statement}

In the time available we were unable to produce the PUS. However, we would
anticipate that the PUS framework would still be useful and applicable for
IABIED. As we have seen above, the 4 objects of understanding (safety
justification, system-in-context, decision and frame selection) can all be
applied to IABIED. In addition it seems plausible that a decision-maker, e.g.
head of government could be subjected to a test of understanding using the 5
capabilities (Reason, Explain, Predict, Challenge, Revise) although as decision
makers get more distant from the system there may be a hierarchy by which the final
decision makers focus more on understanding whether someone else really
understands it.

\textbf{Warranted deference}: In typical safety case arguments a justification
for warranted deference may be made on the basis that a person/team has
demonstrated competence in the relevant domain. The quality of such
justification may be reduced when dealing with systems that no-one has yet
developed expertise in, because the system (ASI) doesn't exist yet.

\textbf{Scope insensitivity}: A concept wherein human judgement does not scale
properly with the magnitude of very large harms, such as the ones described by
IABIED. This can be an issue of how minds represent consequences/dangers rather
than an issue of how much evidence is available or how well the argument is
constructed. It points towards a potential need for new tests of
decision-makers to be devised to check that understanding is not tainted by
such biases (\citealp{slovicIfLookMass2007};
\citealp{yudkowskyCognitiveBiasesPotentially})

\subsection{Impacts of uncertainty}
\label{app:g3-impacts-of-uncertainty}

In IABIED we have an example of a decision-first application of the protocol.
The authors propose a decision, based on their arguments: whether to proceed
with developing and deploying artificial superintelligence (ASI) using current
machine-learning techniques or to pause or constrain that development. From
this entry point we can immediately look to re-frame the decision, or accept
the framing and define the system in context for the decision.

We outlined earlier some of the assumptions inherent in this framing that a
decision-maker may wish to review before proceeding further. For example,
examining the Decision Basis would make us ask what is the overall goal.
Assuming the goal is to reduce exposure to existential risks, then
justification would be needed as to why the framing of the decision relates to
the path of loss of control of an ASI, rather than other alternatives such as
humans misusing AI to develop bio weapons, chronic slow disempowerment and
institutional hollowing out, accelerated climate change from data centres etc.

Accepting the framing for now, we define the system in context. The key system
in this case is the socio-technical paradigm that incentivises the development
of AI - difficult to describe and bound. Acknowledging the breadth, we can
still proceed to examine the safety justification (or in this case, a risk
justification) within the loosely-defined system.

Even with this high level of uncertainty, complexity and criticality, we found
that the methodology for describing the four objects of understanding,
assessing the sufficiency of externalised (UB) and internalised (PUS)
understanding is a helpful structure for guiding us in making uncertainties and
assumptions explicit and enabling refinement towards a more practical, nuanced
and well-justified decision.

One component of understanding directly affected by uncertainty is the
\emph{sufficiency claim}: what constitutes sufficient understanding for a
well-made decision. In this study we focused on how felicitous falsehoods and
tethering changes with uncertainty.

\textbf{Felicitous falsehoods are present at every level.} Idealisation is
ubiquitous in the sciences, including the exact ones: point masses,
frictionless surfaces, ideal gases, rigid bodies. Confidence in the prediction
of even a simple deterministic system rests on a model of that system, and the
model is an idealisation. What distinguishes the simple case is not the absence
of felicitous falsehoods but that their \emph{departure from truth is
characterised and bounded} --- the error term is known, the conditions under
which the idealisation breaks are known, and both can be checked against the
system itself. Our confidence may be dominated by aleatory rather than
epistemic issues.

\textbf{Tethering persists at every level; what changes is what one tethers
to.} Evidence does not run out as uncertainty increases. It changes type from
direct measurement of the system in question, through evidence about similar
systems and their components, experiments on analogues, historical base rates
of comparable transitions, to structural and theoretical arguments. Appendix
F illustrates this for IABIED. Two things do change. Where uncertainty is a
reflection of the complexity of the systems and argument. The \emph{argument
steps} between evidence and the top claim increase, and with it the
contestability of the chain. And the \emph{target} of tethering shifts: at low
uncertainty we tether claims about the system; at high uncertainty we tether
the theories and the felicitous falsehoods themselves, and the claim about the
system is then reached through them.

One way of describing this variation is through epistemic
access\footnote{Epistemic access: the means by which a claim can be evaluated
in time to inform the decision. Adapted from philosophy ``the general sense of
the means by which one may come to know something''}, whether the system
exists, can be observed, and can be tested. From this single variable the rest
follows: whether an idealisation's error can be measured or only argued; what
evidence is available to tether to; and whether an evaluator can check the
answer or only the reasoning by which it was reached.

\begin{table}[htbp]
\centering
\small
\begin{tabularx}{\textwidth}{|X|L{1.15in}|L{1.15in}|L{1.30in}|}
\hline
\textbf{Regime (by epistemic access)} & \textbf{Status of the idealisations} &
\textbf{What tethering attaches to} & \textbf{Mode of evaluation} \\ \hline
\textbf{1. System exists and is directly testable} --- bounded, with
  deterministic or well-characterised stochastic behaviour
  & Departure from truth is bounded and measurable against the system; error
    terms and breakdown conditions are known
  & Direct evidence about this system: measurement, test, operational
    experience
  & The answer can be checked against the system. Verification: right or wrong
    within stated tolerance. Residual doubt minimal \\ \hline
\textbf{2. System exists but cannot be exhaustively tested or analysed} ---
  unbounded environment, emergent or adaptive behaviour, rare events
  & Departure characterised only within the evaluation envelope; idealisations
    carry the extrapolation beyond it
  & Direct evidence within the evaluation envelope, plus evidence from similar
    systems, components and prior operational experience. Mature theories of
    extrapolation.
  & The answer can be checked only in part. Assessed confidence with explicit
    residual doubt \\ \hline
\textbf{3. System does not yet exist but is of a known kind} --- a design not
  yet built, a next-generation system
  & No direct evidence is possible; accepted theories and idealisations carry
    the extrapolation from known systems.
  & Predecessors and analogues, component-level evidence, mature theory,
    established engineering practice
  & The answer cannot be empirically checked, so the argument is, for reasonableness and justifiability against precedent and theory. \\ \hline
\textbf{4. System is unprecedented in kind} --- no predecessors, no mature
  theory (IABIED's ASI)
  & Idealisations carry the whole account, and are themselves contested rather
    than merely uncertain
  & Analogy, historical base rates of comparable transitions, structural and
    theoretical argument --- weak and contested chains, but not absent
  & Neither answer nor argument can be checked against experience. Argument by
    peer review, evaluated on the quality of the reasoning and of the process
    that produced agreement \\ \hline
\end{tabularx}
\end{table}

\FloatBarrier

\emph{Epistemic} uncertainty about the model, the theory and the framing drives
the progression, which is why the shift from aleatory to epistemic dominance
marks the movement from Regime 1--2 to Regime 3--4.

Regime 3 is where most novel safety-critical engineering sits and where the
RobotCorp example sits and it is instructive that a workable practice exists
there: licensing authorities routinely make high-criticality decisions about
systems that do not yet exist, on the basis of predecessors, component evidence
and mature theory, evaluated as reasonableness against precedent. 
RobotCorp's position in Regime 3 is achieved by the engineering of additional
controls. IABIED would be in regime 4. Evaluation in Regime 4 is more akin to scientific consensus and peer
review because the answer relies on the argument and the tethering that
validates the theory for this application. 

Three consequences for the protocol follow:

\begin{itemize}
  \item \textbf{Assess the regime at entry.} Which elements of the case carry
        most weight depends on the regime, so it should be identified at the
        start --- a framework such as Cynefin
        \citep{snowdenLeadersFrameworkDecision2007} can structure this.
  \item \textbf{Shape the effort accordingly.} In Regimes 1--2, effort goes to
        technical tethering and to characterising where the idealisations
        break, with felicitous falsehoods framing the account holistically. In
        Regimes 3--4, effort shifts to establishing what is known and knowable,
        and to scrutinising the felicitous falsehoods directly: what each one is standing in for,
        whether its departure from truth is argued or merely asserted, and
        whether the narratives built on it are robust to the uncertainty. The
        felicitous-falsehood analysis therefore does not become more important
        because tethering fails; it becomes more important because the
        idealisations, rather than the system, are what the evidence now
        attaches to.
  \item \textbf{Shape the decision accordingly.} In Regimes 3--4, system and
        safety case design can shift the regime to 2. Decisions should be
        designed to increase understanding through iterative development or
        gradual deployment.
\end{itemize}

Even in Regime 4, the protocol --- describing the four objects, then assessing
the sufficiency of externalised (UB) and internalised (PUS) understanding ---
was a helpful structure for making uncertainties and assumptions explicit.

\subsection{Impacts of decision-criticality}
\label{app:g4-impacts-of-decision-criticality}

Because the severity of harm is so high (extinction of humanity), and the costs
of ceasing development potentially high (e.g. cure for cancer delayed by many years, solution for
fusion delayed by many years) we would expect highly detailed supporting
argumentation and justification for sufficiency of understanding.

Given the depth and breadth of detail required, we would expect that
argumentation would need input from a large number of people from potentially
very diverse backgrounds. This may exacerbate the challenges of assembling a
coherent Understanding Basis. The breadth, depth and criticality of the case
might also present challenges for the decision-maker. Decision-making might
also need to be distributed.

It is also important to address the massive increase in rigour and evidence
required to move from everyday reliability to engineering the very high system
reliability that is achieved, for example, in aviation. There needs to be massive intellectual
investment in theories, their validation, in the architectures that provide
layers of defence in depth, in the principled balance between human and machine
and the organisational culture and processes needed for such systems, and in identifying the failures
that are seen. The orders of magnitude increase in reliability are hard to
comprehend and in the IABIED example the extreme levels of confidence needed
that existential or catastrophic events would not happen would mean a
combination of reducing the chances of it happening by design and providing
strong, independent defence in depth. The difference in confidence needed in
critical systems can also be seen as prone to scope insensitivity bias (see
G2.3).

In terms of the methodology, as the decision becomes critical the systems need
to be more reliable, they become more complex, the safety justification becomes
larger, the nature of the theories and the evidence change in order to provide
the low levels of residual doubt, the range of defeaters changes as there can
be very strange and hard to imagine edge cases.

As the RobotCorp example shows, architecture can be used to try and gain the
benefit of novel components, through moving the overall system to Regime 2-3.  In such regimes external coherence can provide a sense check
on the feasibility of assuring the proposed system.

\phantomsection
\addcontentsline{toc}{section}{References}
\bibliographystyle{plainnat}
\bibliography{references}

@misc{balesniEvaluationsbasedSafetyCases2024,
  title = {Towards Evaluations-Based Safety Cases for {{AI}} Scheming},
  author = {Balesni, Mikita and Hobbhahn, Marius and Lindner, David and Meinke, Alexander and Korbak, Tomek and Clymer, Joshua and Shlegeris, Buck and Scheurer, J{\'e}r{\'e}my and Stix, Charlotte and Shah, Rusheb and {Goldowsky-Dill}, Nicholas and Braun, Dan and Chughtai, Bilal and Evans, Owain and Kokotajlo, Daniel and Bushnaq, Lucius},
  year = 2024,
  month = nov,
  number = {arXiv:2411.03336},
  eprint = {2411.03336},
  primaryclass = {cs.CR},
  publisher = {arXiv},
  doi = {10.48550/arXiv.2411.03336},
  urldate = {2026-08-12},
  archiveprefix = {arXiv}
}

@misc{barrettLessonsExternalReview2026,
  title = {Lessons from {{External Review}} of {{DeepMind}}'s {{Scheming Inability Safety Case}}},
  author = {Barrett, Stephen and Zabala, Francisco Javier Campos and Fillingham, Sean P. and Siddique, Umair and Walpole, James and Bloomfield, Robin and Papadatos, Henry},
  year = 2026,
  month = apr,
  number = {arXiv:2604.21964},
  eprint = {2604.21964},
  primaryclass = {cs.CY},
  publisher = {arXiv},
  doi = {10.48550/arXiv.2604.21964},
  urldate = {2026-05-25},
  archiveprefix = {arXiv}
}

@misc{bengioSafetyHonestyDisinterested2026b,
  title = {Safety from {{Honesty}} in a {{Disinterested AI Predictor}}},
  author = {Bengio, Yoshua and Richardson, Oliver and Gaven{\v c}iak, Tom{\'a}{\v s} and Cohen, Michael and Svarc, Rory and Fornasiere, Damiano and Gendron, Gael and Hyland, David and Kamanda, Aton and Oberman, Adam and Ward, Francis Rhys and Gaven{\v c}iak, Anna and Slosser, Jacob Livingston and Mai, Vincent and Serban, Iulian and Ghosn, Joumana},
  year = 2026,
  month = jul,
  number = {arXiv:2606.29657},
  eprint = {2606.29657},
  primaryclass = {cs.AI},
  publisher = {arXiv},
  doi = {10.48550/arXiv.2606.29657},
  urldate = {2026-08-14},
  archiveprefix = {arXiv}
}

@misc{bloomfieldAssurance20Manifesto2021,
  title = {Assurance 2.0: {{A Manifesto}}},
  shorttitle = {Assurance 2.0},
  author = {Bloomfield, Robin and Rushby, John},
  year = 2021,
  month = jan,
  number = {arXiv:2004.10474},
  eprint = {2004.10474},
  primaryclass = {cs.SE},
  publisher = {arXiv},
  doi = {10.48550/arXiv.2004.10474},
  urldate = {2026-05-25},
  archiveprefix = {arXiv}
}

@misc{bloomfieldAssuranceAISystems2025,
  title = {Assurance of {{AI Systems From}} a {{Dependability Perspective}}},
  author = {Bloomfield, Robin and Rushby, John},
  year = 2025,
  month = jun,
  number = {arXiv:2407.13948},
  eprint = {2407.13948},
  primaryclass = {cs.AI},
  publisher = {arXiv},
  doi = {10.48550/arXiv.2407.13948},
  urldate = {2026-08-12},
  archiveprefix = {arXiv}
}

@misc{bloomfieldUnderstandingReframingAutomation2026,
  title = {Understanding: Reframing Automation and Assurance},
  shorttitle = {Understanding},
  author = {Bloomfield, Robin},
  year = 2026,
  month = apr,
  number = {arXiv:2604.05662},
  eprint = {2604.05662},
  primaryclass = {cs.SE},
  publisher = {arXiv},
  doi = {10.48550/arXiv.2604.05662},
  urldate = {2026-05-25},
  archiveprefix = {arXiv}
}

@book{bostromSuperintelligencePathsDangers2014,
  title = {Superintelligence: {{Paths}}, {{Dangers}}, {{Strategies}}},
  shorttitle = {Superintelligence},
  author = {Bostrom, Nick and Bostrom, Nick},
  year = 2014,
  month = jul,
  publisher = {Oxford University Press},
  address = {Oxford, New York},
  isbn = {978-0-19-967811-2}
}

@misc{buhlAlignmentSafetyCase2025,
  title = {An Alignment Safety Case Sketch Based on Debate},
  author = {Buhl, Marie Davidsen and Pfau, Jacob and Hilton, Benjamin and Irving, Geoffrey},
  year = 2025,
  month = may,
  number = {arXiv:2505.03989},
  eprint = {2505.03989},
  primaryclass = {cs.AI},
  publisher = {arXiv},
  doi = {10.48550/arXiv.2505.03989},
  urldate = {2026-08-17},
  archiveprefix = {arXiv}
}

@misc{ChatGPTDeletingPeoples2026,
  author = {{The Independent}},
  title = {{{ChatGPT}} Is Deleting People's Files without Asking Them},
  year = 2026,
  month = jul,
  journal = {The Independent},
  urldate = {2026-08-17},
  chapter = {Tech},
  howpublished = {https://www.the-independent.com/tech/security/chatgpt-update-openai-gpt-sol-b3015585.html},
  langid = {english}
}

@misc{clymerSafetyCasesHow2024,
  title = {Safety {{Cases}}: {{How}} to {{Justify}} the {{Safety}} of {{Advanced AI Systems}}},
  shorttitle = {Safety {{Cases}}},
  author = {Clymer, Joshua and Gabrieli, Nick and Krueger, David and Larsen, Thomas},
  year = 2024,
  month = mar,
  number = {arXiv:2403.10462},
  eprint = {2403.10462},
  primaryclass = {cs.CY},
  publisher = {arXiv},
  doi = {10.48550/arXiv.2403.10462},
  urldate = {2026-08-12},
  archiveprefix = {arXiv}
}

@misc{DarioAmodeiUrgency,
  author = {Amodei, Dario},
  year = {2025},
  title = {Dario {{Amodei}} --- {{The Urgency}} of {{Interpretability}}},
  urldate = {2026-08-17},
  howpublished = {https://darioamodei.com/post/the-urgency-of-interpretability},
  langid = {english}
}

@misc{DeclareGuidanceCAE,
  author = {{CAE Framework}},
  title = {The {{Declare Guidance}} -- {{CAE FRAMEWORK}}},
  urldate = {2026-08-12},
  langid = {american}
}

@book{elginTrueEnough2017,
  title = {True {{Enough}}},
  author = {Elgin, Catherine Z.},
  year = 2017,
  month = sep,
  publisher = {MIT Press},
  address = {Cambridge, MA, USA},
  isbn = {978-0-262-03653-5},
  langid = {english}
}

@article{fingletonNuclearRegulatoryReview2025,
  title = {Nuclear {{Regulatory Review}} 2025},
  author = {Fingleton, John},
  year = 2025,
  langid = {english}
}

@misc{IntroducingSuperalignment2023,
  author = {{OpenAI}},
  title = {Introducing {{Superalignment}}},
  year = 2023,
  month = dec,
  journal = {OpenAI},
  urldate = {2026-08-14},
  howpublished = {https://openai.com/index/introducing-superalignment/},
  langid = {american}
}

@misc{khlaafFissionAlgorithmsUndermining2025,
  title = {Fission for {{Algorithms}}: {{The Undermining}} of {{Nuclear Regulation}} in {{Service}} of {{AI}}},
  shorttitle = {Fission for {{Algorithms}}},
  author = {Khlaaf, Heidy},
  year = 2025,
  month = nov,
  journal = {AI Now Institute},
  urldate = {2026-08-14},
  langid = {american}
}

@misc{langoscoGoalMisgeneralizationDeep2023,
  title = {Goal {{Misgeneralization}} in {{Deep Reinforcement Learning}}},
  author = {Langosco, Lauro and Koch, Jack and Sharkey, Lee and Pfau, Jacob and Orseau, Laurent and Krueger, David},
  year = 2023,
  month = jan,
  number = {arXiv:2105.14111},
  eprint = {2105.14111},
  primaryclass = {cs.LG},
  publisher = {arXiv},
  doi = {10.48550/arXiv.2105.14111},
  urldate = {2026-08-11},
  archiveprefix = {arXiv}
}

@inproceedings{omohundroBasicAIDrives2008,
  title = {The {{Basic AI Drives}}},
  booktitle = {Proceedings of the 2008 Conference on {{Artificial General Intelligence}} 2008: {{Proceedings}} of the {{First AGI Conference}}},
  author = {Omohundro, Stephen M.},
  year = 2008,
  month = jun,
  pages = {483--492},
  publisher = {IOS Press},
  address = {NLD},
  urldate = {2026-08-19},
  isbn = {978-1-58603-833-5}
}

@misc{OpenAIHuggingFace2026,
  author = {{OpenAI}},
  title = {{{OpenAI}} and {{Hugging Face}} Partner to Address Security Incident during Model Evaluation},
  year = 2026,
  month = aug,
  journal = {OpenAI},
  urldate = {2026-08-12},
  howpublished = {https://openai.com/index/hugging-face-model-evaluation-security-incident/},
  langid = {american}
}

@misc{phuongEvaluatingFrontierModels2025,
  title = {Evaluating {{Frontier Models}} for {{Stealth}} and {{Situational Awareness}}},
  author = {Phuong, Mary and Zimmermann, Roland S. and Wang, Ziyue and Lindner, David and Krakovna, Victoria and Cogan, Sarah and Dafoe, Allan and Ho, Lewis and Shah, Rohin},
  year = 2025,
  month = jul,
  number = {arXiv:2505.01420},
  eprint = {2505.01420},
  primaryclass = {cs.LG},
  publisher = {arXiv},
  doi = {10.48550/arXiv.2505.01420},
  urldate = {2026-05-25},
  archiveprefix = {arXiv}
}

@misc{SafetyCasesAISI,
  author = {{AI Security Institute}},
  title = {Safety Cases at {{AISI}} \textbar{} {{AISI Work}}},
  journal = {AI Security Institute},
  urldate = {2026-08-11},
  howpublished = {https://www.aisi.gov.uk/blog/safety-cases-at-aisi},
  langid = {english}
}

@misc{SettingFoundationsFive,
  author = {{National Protective Security Authority}},
  title = {Setting the Foundations: {{Five}} Principles for a Shared Approach to {{Insider Risk}}},
  shorttitle = {Setting the Foundations},
  urldate = {2026-08-12},
  howpublished = {https://www.npsa.gov.uk/specialised-guidance/insider-risk-guidance/setting-foundations-five-principles-shared-approach-insider-risk},
  langid = {english}
}

@article{slovicIfLookMass2007,
  title = {``{{If I}} Look at the Mass {{I}} Will Never Act'': {{Psychic}} Numbing and Genocide},
  shorttitle = {``{{If I}} Look at the Mass {{I}} Will Never Act''},
  author = {Slovic, Paul},
  year = 2007,
  month = apr,
  journal = {Judgment and Decision Making},
  volume = {2},
  number = {2},
  pages = {79--95},
  issn = {1930-2975},
  doi = {10.1017/S1930297500000061},
  urldate = {2026-08-12},
  langid = {english}
}

@article{snowdenLeadersFrameworkDecision2007,
  title = {A {{Leader}}'s {{Framework}} for {{Decision Making}}},
  author = {Snowden, David J. and Boone, Mary E.},
  year = 2007,
  month = nov,
  journal = {Harvard Business Review},
  issn = {0017-8012},
  urldate = {2026-08-12},
  chapter = {Decision making and problem solving},
  langid = {english}
}

@misc{taylorLargeLanguageModels2025,
  title = {Do {{Large Language Models Exhibit Spontaneous Rational Deception}}?},
  author = {Taylor, Samuel M. and Bergen, Benjamin K.},
  year = 2025,
  month = mar,
  number = {arXiv:2504.00285},
  eprint = {2504.00285},
  primaryclass = {cs.CL},
  publisher = {arXiv},
  doi = {10.48550/arXiv.2504.00285},
  urldate = {2026-08-17},
  archiveprefix = {arXiv}
}

@techreport{vermeerExtinctionRiskArtificial2025,
  title = {On the {{Extinction Risk}} from {{Artificial Intelligence}}},
  author = {Vermeer, Michael J. D. and Lathrop, Emily and Moon, Alvin},
  year = 2025,
  month = may,
  urldate = {2026-08-14},
  langid = {english}
}

@article{yudkowskyCognitiveBiasesPotentially,
  title = {Cognitive {{Biases Potentially Affecting Judgment}} of {{Global Risks}}},
  author = {Yudkowsky, Eliezer},
  langid = {english}
}

@misc{yudkowskyDraftTreatyAnnotations,
  title = {A {{Draft}} of a {{Treaty}}, with {{Annotations}} \textbar{} {{If Anyone Builds It}}, {{Everyone Dies}}},
  author = {Yudkowsky, Eliezer and Soares, Nate},
  urldate = {2026-08-14},
  howpublished = {https://ifanyonebuildsit.com/treaty},
  langid = {english}
}

@book{yudkowskyIfAnyoneBuilds2025,
  title = {If Anyone Builds It, Everyone Dies: The Case against Superintelligent {{AI}}},
  shorttitle = {If Anyone Builds It, Everyone Dies},
  author = {Yudkowsky, Eliezer and Soares, Nate},
  year = 2025,
  publisher = {The Bodley Head},
  address = {London},
  isbn = {978-1-84792-892-4 978-1-84792-893-1},
  langid = {english}
}

@misc{yudkowskyOnlineResourcesIf,
  title = {Online {{Resources}} \textbar{} {{If Anyone Builds It}}, {{Everyone Dies}} \textbar{} {{If Anyone Builds It}}, {{Everyone Dies}}},
  author = {Yudkowsky, Eliezer and Soares, Nate},
  urldate = {2026-08-14},
  howpublished = {https://ifanyonebuilds.it/resources},
  langid = {english}
}
\end{document}